\documentclass[12pt]{article}
\pdfoutput=1

\usepackage{amsmath}
\usepackage{amssymb}
\usepackage{mathrsfs}
\usepackage{graphicx}
\usepackage{cite}
\usepackage{multirow}
\usepackage{longtable}
\usepackage{lscape}
\usepackage{subfigure}

\newcommand{\beq}{\begin{equation}}
\newcommand{\eeq}{\end{equation}}
\newcommand{\be}{\begin{eqnarray}}
\newcommand{\ee}{\end{eqnarray}}

\newcommand{\ms}{\Delta m^2_{21}}
\newcommand{\ma}{\Delta m^2_{31}}

\newcommand{\sss}{\sin^2 \theta_{12}}
\newcommand{\sch}{\sin^2 \theta_{13}}
\newcommand{\stch}{\sin^2 2\theta_{13}}
\newcommand{\sa}{\sin^2 \theta_{23}}
\newcommand{\sta}{\sin^22 \theta_{23}}

\newcommand{\dcp}{\delta_{\mathrm{CP}}}
\newcommand{\tmt}{\theta_{23}}
\newcommand{\tet}{\theta_{13}}
\newcommand{\tem}{\theta_{12}}

\def\nue{{\nu_e}}
\def\anue{{\bar\nu_e}}
\def\numu{{\nu_{\mu}}}
\def\anumu{{\bar\nu_{\mu}}}
\def\nutau{{\nu_{\tau}}}

\makeatletter
\renewcommand{\fnum@table}{\textbf{\tablename~\thetable}}
\renewcommand{\fnum@figure}{\textbf{\figurename~\thefigure}}
\makeatother

\begin{document}

\begin{titlepage}

\renewcommand{\thefootnote}{\alph{footnote}}

\vspace*{0.2cm}

\renewcommand{\thefootnote}{\fnsymbol{footnote}}
\setcounter{footnote}{-1}

{\begin{center} 

{\large\bf

Physics Potential of Long-Baseline Experiments

}

\end{center}} 

\renewcommand{\thefootnote}{\alph{footnote}}

\vspace*{.6cm}

{\begin{center} {
               \large{\sc
                 Sanjib Kumar Agarwalla\footnote[1]{\makebox[1.cm]{Email:}
                 sanjib@iopb.res.in}}
                 }
\end{center}
}
\vspace*{0cm}
{\it
\begin{center}

\footnotemark[1]
       Institute of Physics, Sachivalaya Marg, Sainik School Post, \\
                              Bhubaneswar 751005, Orissa, India
       
\end{center}}

\vspace*{0.5cm}

{\Large \bf
\begin{center} Abstract \end{center}  }

The discovery of neutrino mixing and oscillations over the past decade provides firm evidence for new physics beyond the Standard Model.
Recently, $\theta_{13}$ has been determined to be moderately large, quite close to its previous upper bound. 
This represents a significant milestone in establishing the three-flavor oscillation picture of neutrinos. 
It has opened up exciting prospects for current and future long-baseline neutrino oscillation 
experiments towards addressing the remaining fundamental questions, in particular the type of the neutrino mass 
hierarchy and the possible presence of a CP-violating phase. Another recent and crucial development is the indication of non-maximal 
2-3 mixing angle, causing the octant ambiguity of $\theta_{23}$. 
In this paper, I will review the phenomenology of long-baseline neutrino oscillations with a special emphasis on 
sub-leading three-flavor effects, which will play a crucial role in resolving these unknowns.
First, I will give a brief description of neutrino oscillation phenomenon. Then, I will discuss our present global
understanding of the neutrino mass-mixing parameters and will identify the major unknowns in this sector.
After that, I will present the physics reach of current generation long-baseline experiments. Finally, I will 
conclude with a discussion on the physics capabilities of accelerator-driven possible future long-baseline 
precision oscillation facilities.

\vspace*{.5cm}

\end{titlepage}
\newpage

\renewcommand{\thefootnote}{\arabic{footnote}}
\setcounter{footnote}{0}

\section{Introduction and motivation}
\label{sec:intro-motiv}

We are going through an exciting phase when the light of new findings is breaking apart our long-held understanding 
of the Standard Model of particle physics. This revolution started in part with the widely confirmed claim that neutrinos have mass, 
and it will continue to be waged by currently running and upcoming neutrino experiments. Over the last fifteen years or so, fantastic 
data from world-class experiments involving neutrinos from 
the sun~\cite{Cleveland:1998nv,Altmann:2005ix,Hosaka:2005um,Ahmad:2002jz,Aharmim:2008kc,Aharmim:2009gd,Arpesella:2008mt},
the Earth's atmosphere~\cite{Fukuda:1998mi,Ashie:2005ik}, 
nuclear reactors~\cite{Araki:2004mb,:2008ee,An:2012eh,An:2012bu,Ahn:2012nd,Abe:2011fz,Abe:2012tg}, and 
accelerators~\cite{Ahn:2006zza,Adamson:2008zt,Adamson:2011qu,Adamson:2013ue,Abe:2011sj,Abe:2013xua} 
have firmly established the phenomenon of neutrino flavor oscillations~\cite{Pontecorvo:1967fh,Gribov:1968kq}. 
This implies that neutrinos have mass and they mix with each other, providing an exclusive example of experimental 
evidence for physics beyond the Standard Model. 

\begin{figure}[tp]
\begin{center}
\includegraphics[width=10cm, height=7.0cm]{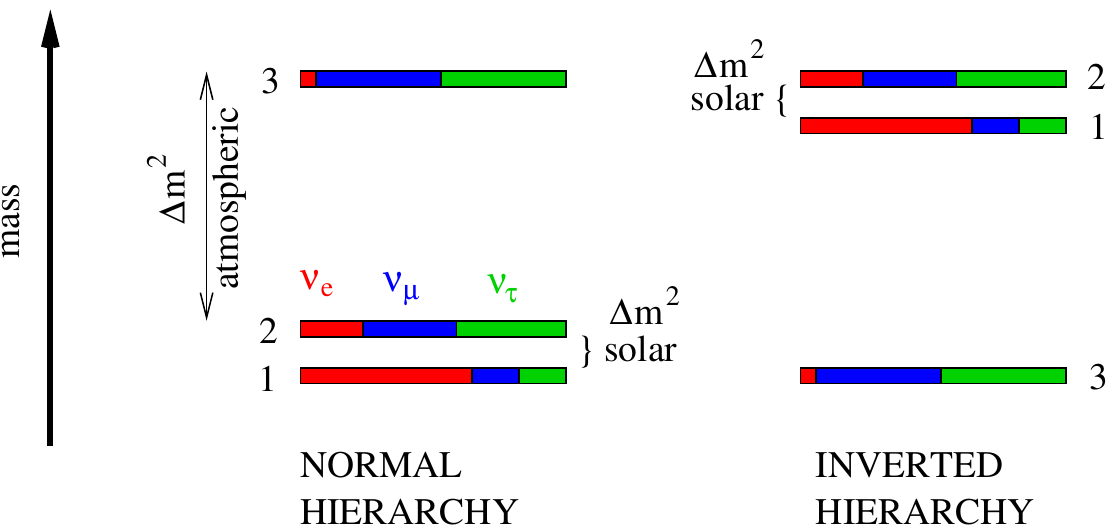}
\caption{\footnotesize{
The sign of $\Delta m^2_{31} \equiv m_3^2 - m_1^2$ is unknown.
The hierarchy of the neutrino mass spectrum can be normal or inverted.}}
\label{fig:hierarchy_scheme}
\end{center}
\end{figure}

The most recent development in the field of neutrinos is the discovery of the smallest lepton mixing angle $\tet$.
Finally it has been measured to be non-zero with utmost confidence by the reactor neutrino experiments 
Daya Bay~\cite{An:2012bu} and RENO~\cite{Ahn:2012nd}. They have found a moderately large value of $\tet$
\begin{center}
\hskip0.5cm$\stch|_{\mathrm{\bf Daya~Bay~(rate-only)}}$ = $0.089 \pm 0.010 \,({\mathrm{stat}}) \pm 0.005 \, ({\mathrm{syst}})$ \cite{An:2012bu}, and
\vskip0.2cm$\left.\stch\right|_{\mathrm{\bf RENO~(rate-only)}}$ = $0.113 \pm 0.013 \, ({\mathrm{stat}}) \pm 0.019 \, ({\mathrm{syst}})$~\cite{Ahn:2012nd},
\end{center}
in perfect agreement with the data provided by the another reactor experiment Double Chooz~\cite{Abe:2011fz,Abe:2012tg}   
and the accelerator experiments MINOS~\cite{Adamson:2013ue} and T2K~\cite{Abe:2013xua}.
All the three global fits~\cite{Tortola:2012te,Fogli:2012ua,GonzalezGarcia:2012sz} of all the world neutrino oscillation data available  
indicate a non-zero value of $\tet$ at more than $10\sigma$ and suggest a best-fit value of $\sch \simeq 0.023$ with a 
relative $1\sigma$ precision of 10\%. Daya Bay experiment is expected to reduce this uncertainty to a level of 5\% by 2016 when
they will finish collecting all the data~\cite{dayabay_NF12}. These recent high precision measurement of a moderately large value 
of $\tet$ signifies an important breakthrough in validating the standard three-flavor oscillation picture of neutrinos~\cite{Hewett:2012ns}. 
It has created exciting opportunities for current and future neutrino oscillation experiments to address the remaining fundamental
unknowns. This fairly large value of $\tet$ has provided a {\it `golden'} avenue to directly probe the 
neutrino mass hierarchy\footnote{Two possibilities are there: it can be either normal (NH) if $\Delta m^2_{31} \equiv m^2_3 -m^2_1 > 0$, 
or inverted (IH) if $\Delta m^2_{31} < 0$, as described in Figure~\ref{fig:hierarchy_scheme}.} using the Earth matter effects, and to search for 
leptonic CP violation\footnote{If the Dirac CP phase, $\dcp$ differs from $0^\circ$ or $180^\circ$.} in accelerator based
long-baseline neutrino oscillation experiments~\cite{Minakata:2012ue,Machado:2013kya}.
Another recent and important development related to neutrino mixing parameters is the hint of non-maximal
2-3 mixing by the MINOS accelerator experiment~\cite{Nichol:2013caa,Adamson:2013whj}. 
However, the maximal value of $\tmt$ is still favored by the atmospheric neutrino data, dominated by 
Super-Kamiokande~\cite{Itow:2012}. Combined analyses of all the neutrino oscillation data 
available~\cite{Tortola:2012te,Fogli:2012ua,GonzalezGarcia:2012sz} also prefer the deviation from maximal 
mixing for $\tmt$ i.e., $(0.5 - \sa) \ne 0$. In $\numu$ survival probability, the dominant term mainly depends on $\sta$.
Now, if $\sta$ differs from 1 as indicated by the recent neutrino data, then we get two solutions for $\tmt$:
one $< 45^\circ$, named as lower octant (LO) and the other $> 45^\circ$, named as higher octant (HO).
In other words, if the quantity $(0.5 - \sa)$ is positive (negative) then $\tmt$ belongs to LO (HO).
This leads to the problem of octant degeneracy of $\tmt$~\cite{Fogli:1996pv} 
which is a part of the overall eight-fold degeneracy~\cite{Barger:2001yr,Minakata:2002qi}, where the other two degeneracies 
are $(\tet,\dcp)$ intrinsic degeneracy~\cite{BurguetCastell:2001ez} and the (hierarchy, $\dcp$) degeneracy~\cite{Minakata:2001qm}.
The resolution of the three fundamental issues in the neutrino sector: neutrino mass hierarchy, octant of $\tmt$
and CP violation is possible only by observing the impact of three-flavor effects in neutrino oscillation 
experiments~\cite{Pascoli:2013wca,Agarwalla:2013hma}. 

The information on neutrino mass hierarchy is very important in order to settle the structure of neutrino mass matrix 
which in turn can give crucial piece of information towards the underlying theory of neutrino masses and mixing~\cite{Albright:2006cw}. 
This is also a vital ingredient for neutrinoless double beta decay searches investigating the Majorana nature of neutrinos. 
If $\Delta m^2_{31}< 0$, and yet no neutrinoless double beta decay is observed even in the very far future experiments, that would be 
a strong hint that neutrinos are not Majorana particles~\cite{Pascoli:2005zb}.
Another fundamental missing link that needs to be addressed in long-baseline neutrino oscillation experiments is to measure 
$\dcp$ and to explore leptonic CP violation. This new source of low energy CP violation in neutral lepton sector has drawn 
tremendous interest because of the possibility of leptogenesis leading to baryogenesis and the observed baryon asymmetry 
in the universe~\cite{Fukugita:1986hr}. Leptogenesis demands the existence of CP violation in the leptonic sector,
for a recent review, see~\cite{DiBari:2012fz}. The possible link between leptogenesis and neutrino oscillations has been 
studied in~\cite{Joshipura:2001ui,Endoh:2002wm,Pati:2003gv}. 
It is likely that the CP violating phase in neutrino oscillations is not directly responsible to generate the CP violation
leading to leptogenesis. But, there is no doubt that a demonstration of CP violation in neutrino
oscillations will provide a crucial guidepost for models of leptonic CP violation and leptogenesis.
Precise measurement of $\tmt$ and the determination of its correct octant (if it turns out to be non-maximal) 
are also very vital tasks that need to be undertaken by the current and next generation neutrino oscillation experiments.  
These informations will provide crucial inputs to the theories of neutrino masses and 
mixings~\cite{Mohapatra:2006gs,Albright:2006cw,Albright:2010ap,King:2013eh}.
A number of excellent ideas, such as $\mu \leftrightarrow \tau$ symmetry~\cite{Fukuyama:1997ky,Mohapatra:1998ka,Lam:2001fb,Harrison:2002et,Kitabayashi:2002jd,Grimus:2003kq,Koide:2003rx,Mohapatra:2005yu}, 
$A_4$ flavor symmetry~\cite{Ma:2002ge,Ma:2001dn,Babu:2002dz,Grimus:2005mu,Ma:2005mw}, 
quark-lepton complementarity~\cite{Raidal:2004iw,Minakata:2004xt,Ferrandis:2004vp,Antusch:2005ca},
and neutrino mixing anarchy~\cite{Hall:1999sn,deGouvea:2012ac} have been proposed 
to explain the observed pattern of one small and two large mixing angles in the neutrino sector.
Future ultra-precise measurement of $\tmt$ will severely constrain these models leading to a better understanding of 
the theory of neutrino masses and mixing.

The outline of this review work is as follows. We start in Section~\ref{sec:oscillation} by revisiting the phenomenon of neutrino
oscillation in the three neutrino framework. Then, we discuss the importance of matter effect in neutrino oscillation in 
Section~\ref{sec:matter}. In Section~\ref{sec:status}, we take a look at our present 
understanding of neutrino oscillation parameters and we identify the fundamental missing links in the neutrino sector that
can be answered in the current and next generation long-baseline neutrino oscillation experiments. 
Section~\ref{sec:three-flavor} discusses in detail the three-flavor effects in long-baseline neutrino oscillation experiments
with the help of $\numu \rightarrow \nue$ oscillation probability, $P_{\mu e}$. Next, in Section~\ref{sec:current}, we study the
physics reach of current generation long-basline beam experiments, T2K and NO$\nu$A. In Section~\ref{sec:future}, we give an 
overview on the possible options for next generation accelerator-driven long-baselne oscillation facilities.
Finally, in Section~\ref{sec:conclusion}, we conclude with a summary of the main points.

\section{Three Neutrino Mixing and Oscillation Framework}
\label{sec:oscillation}

Bruno Pontecorvo was the pathfinder of neutrino oscillation~\cite{Bilenky:2006pd,Bilenky:2013re}. 
In 1957, he gave this concept~\cite{Pontecorvo:1957cp,Pontecorvo:1957qd} based on a two-level quantum system. 
Neutrino oscillation is a simple quantum mechanical phenomenon in which neutrino changes flavor as it travels. 
This phenomenon arises if neutrinos have non-degenerate masses and there is mixing.
First, we consider the fact that neutrinos $(\nue, \numu, \nutau)$ are produced or detected via weak interactions and therefore
they are referred to as weak-eigenstate neutrinos (denoted as $\nu_\alpha$). It means that they are the weak 
doublet-partners of $e^{-}, \mu^{-}$, and $\tau^{-}$ respectively. In such a case, if neutrinos are assumed to be massive,
then in general, it is not mandatory that the mass-matrix of neutrinos written in this weak (flavor) basis will have to be diagonal.
So, it follows that the mass eigenstate neutrinos $\nu_i, i = 1, 2, 3$ (the basis in which the neutrino mass matrix is diagonal) are
not identical to the weak or flavor basis\footnote{The charged lepton mass-matrix is diagonal in this basis.} and 
for three light active neutrinos we have
\beq
    |\nu_\alpha\rangle = \sum_{i=1}^{3} U^*_{\alpha i} |\nu_i\rangle,
\label{eq:flavor-mass}
\eeq
where $\alpha$ can be $e, \mu$ or $\tau$ and $U$ is the $3\times3$ unitary leptonic mixing matrix known as the 
Pontecorvo-Maki-Nakagawa-Sakata (PMNS) matrix~\cite{Maki:1962mu,Pontecorvo:1967fh}. 
This matrix is analogous to the CKM matrix in the quark sector. We use the standard Particle Data Group convention~\cite{Beringer:1900zz}
to parametrize the PMNS matrix in terms of the three mixing angles: $\tem$ (solar mixing angle), $\tmt$ (atmospheric mixing angle), 
$\tet$ (reactor mixing angle), and one Dirac-type CP phase $\dcp$ (ignoring Majorana phases). The mixing matrix $U$ can be parameterized as
\begin{equation}
U_{\mathrm{PMNS}}   =   \underbrace{ \left( 
\begin{array}{ccc} 
1 & 0 & 0 \\
0 & c_{23} & s_{23} \\
0 & -s_{23} & c_{23} 
\end{array} \right) }_{\mathrm{Atmospheric \, mixing}} \\
 \times  \underbrace{ \left( 
\begin{array}{ccc} 
c_{13} & 0 & s_{13} \, e^{-i \dcp} \\
0 & 1 & 0 \\
- s_{13} \, e^{i \dcp} & 0 & c_{13} 
\end{array} \right) }_{\mathrm{Reactor \, mixing}} \\
 \times   \underbrace{ \left( 
\begin{array}{ccc} 
c_{12} & s_{12} & 0 \\
-s_{12} & c_{12} & 0 \\
0 & 0 & 1 
\end{array} \right) }_{\mathrm{Solar \, mixing}} \, ,
\label{eq:euler_rotations}
\end{equation}
where $c_{ij}$ = $\cos\theta_{ij}$ and $s_{ij}$ = $\sin\theta_{ij}$. The neutrino mixing matrix finally takes the form
\beq
U_{\mathrm{PMNS}} = \left( 
\begin{array}{ccc}
c_{12} c_{13} & s_{12} c_{13} & s_{13} e^{-i \dcp} \\
-s_{12} c_{23} - c_{12} s_{23} s_{13} e^{i \dcp} & c_{12} c_{23} - s_{12} s_{23} s_{13} e^{i \dcp} & s_{23} c_{13} \\
s_{12} s_{23} - c_{12} c_{23} s_{13} e^{i \dcp} & -c_{12} s_{23} - s_{12} c_{23} s_{13} e^{i \dcp} & c_{23} c_{13}
\end{array} \right). 
\label{eq:upmns}
\eeq
It is quite interesting to note that three mixing angles are simply related to the flavor components of the three mass eigenstates as
\beq
~~~~~~~~~\frac{|U_{e2}|^2}{|U_{e1}|^2} = \tan^2\tem,~~~~~~~~~~~~~\frac{|U_{\mu3}|^2}{|U_{\tau3}|^2} = \tan^2\tmt,~~~~~~~~~~~~|U_{e3}|^2 = \sin^2\tet .
\label{eq:flavor_components}
\eeq
The transition probability that an initial $\nu_{\alpha}$ of energy $E$ gets converted to a $\nu_{\beta}$ after traveling a distance 
$L$ in vacuum is given by
\beq
P(\nu_{\alpha} \rightarrow \nu_{\beta}) = P_{\alpha\beta} = |\sum_jU_{\beta j}~e^{\frac{-im_j^2L}{2E}}~U^*_{\alpha j}|^{2}~,
\label{eq:mother}
\eeq
where the last factor arises from the decomposition of $\nu_{\alpha}$ into the mass eigenstates, the phase factor in the middle 
appears due to the propagation of each mass eigenstate over distance $L$, and the first factor emerges from their recomposition
into the flavor eigenstate $\nu_{\beta}$ at the end. Equation~\eqref{eq:mother} can also be written as
\beq
P_{\alpha\beta} \,\, = \,\, \delta_{\alpha\beta} \,\,
- \,\, \underbrace{4\sum_{i> j}^n\mbox{Re}[U^*_{\alpha i}U^*_{\beta j} U_{\beta i} U_{\alpha j}]\sin^2 X_{ij}}_{\mathrm{CP \, conserving}}
\,\, - \,\, \underbrace{2\sum_{i>j}^n\mbox{Im}[U^*_{\alpha i}U^*_{\beta j} U_{\beta i} U_{\alpha j}]\sin 2 X_{ij}}_{\mathrm{CP \, violating}} \,,
\label{eq:formula5}
\eeq
where
\begin{equation}
    X_{ij} = \frac{(m_i^2-m_j^2) L}{4 E} = 1.27 \,
    \frac{\Delta m^2_{ij}}{eV^2} \, \frac{L/E}{m/MeV} \,.
\label{eq:formula6}
\end{equation}
$\Delta m^2_{ij} = m_i^2 - m_j^2$ is known as the mass splitting and neutrino oscillations are only sensitive to this 
mass-squared difference but not to the absolute neutrino mass scale. The transition probability 
(given by equation~\eqref{eq:formula5}) has an oscillatory behaviour with oscillation lengths
\begin{equation}
L^{osc}_{ij}=\frac{4\pi E}{\Delta m^2_{ij}}\,\simeq \,2.48\;m\,\frac{E\,\mbox{(MeV)}}
{\Delta m^2_{ij}\, (\mbox{eV}^2)}=\,2.48\;km\,\frac{E\,\mbox{(GeV)}}
{\Delta m^2_{ij}\, (\mbox{eV}^2)}
\label{eq:formula7}
\end{equation}
and the amplitudes are proportional to the elements in the mixing matrix. Since neutrino oscillations can occur only if there is a 
mass difference between at least two neutrinos, an observation of this effect proves that at least one non-zero neutrino mass exists.
In a three-flavor framework, there are two independent mass-squared differences between the three neutrino masses:
$\Delta m^2_{21} = m_2^2-m_1^2$ (responsible for solar neutrino oscillations) and $\Delta m^2_{31} = m_3^2-m_1^2$ 
(responsible for atmospheric neutrino oscillations). The angle $\tet$ connects the solar sector with the atmospheric one and 
determines the impact of the three-flavor effects. The last term of equation~\eqref{eq:formula5} accommodates the CP violating 
part, proportional to $\sin\dcp$. This contribution can only be probed in a neutrino oscillation experiment measuring the
appearance probability of a new flavor, since for disappearance experiments ($\alpha = \beta$) 
the last term becomes zero identically. Also, the CP violating term changes sign in going from 
$P (\nu_{\alpha} \to \nu_{\beta})$ to $P (\bar\nu_{\alpha} \to \bar\nu_{\beta})$ (for anti-neutrino, we have to replace $U$ by $U^*$).
It also changes sign in going from $P (\nu_{\alpha} \to \nu_{\beta})$ to $P (\nu_{\beta} \to \nu_{\alpha})$, since the CPT
invariance ensures that $P (\bar\nu_{\alpha} \to \bar\nu_{\beta})$ = $P (\nu_{\beta} \to \nu_{\alpha})$.

\section{Neutrino Propagation through Matter}
\label{sec:matter}

Oscillation probability changes dramatically when neutrino passes through matter~\cite{Mikheev:1986gs,Mikheev:1986wj,Wolfenstein:1977ue}. 
During propagation through matter, the weak interaction couples the neutrinos to matter. Besides few hard scattering events, there is also coherent 
forward elastic scattering of neutrinos with matter particles they encounter along the path. The important fact is that the coherent forward elastic 
scattering amplitudes are not the same for all neutrino flavors. The ordinary matter consists of electrons,
protons and neutrons but it does not contain any muons or tau-leptons. Neutrinos of all three flavors ($\nu_e$, $\nu_\mu$ and $\nu_\tau$)
interact with the electrons, protons and neutrons of matter through flavor independent neutral current interaction mediated by $Z^0$ bosons.
These contributions are same for neutrinos of all three flavors, leading to an overall phase which can be subtracted.
Interestingly, the electron neutrinos have an additional contribution due to their charged current interactions with the ambient electrons of
the medium which are mediated by the $W^\pm$ exchange. This extra matter potential appears in the form
\begin{equation}
A = \pm 2\sqrt{2} G_F N_e E\,,
\label{eq:matter_potential} 
\end{equation}
where $G_F$ is the Fermi coupling constant, $N_e$ is the electron number density inside the Earth and $E$ is the neutrino energy.
The $+$ sign refers to neutrinos while the $-$ to anti-neutrinos. The connection between the electron density ($N_e$) and the matter
density ($\rho$) is given by
\begin{equation}
V_{CC} = \sqrt{2} G_F N_e \simeq 7.6 Y_e \frac{\rho}{10^{14}
\mbox{g}/\mbox{cm}^3} \mbox{eV}\,,
\label{eq:matter_density}
\end{equation}
where $Y_e = \frac{N_e}{N_p + N_n}$ is the relative electron number density. $N_p$, $N_n$ are the proton and neutron densities in Earth matter
respectively. In an electrically neutral, isoscalar medium, we have $N_e = N_p = N_n$ and $Y_e$ comes out to be 0.5.
If we compare the strength of $V_{CC}$ for the Earth with $\ma/2E$, then we can judge the importance of Earth's matter effect on 
neutrino oscillations. If we consider a neutrino of 5 GeV passing through the core of the Earth ($\rho \sim$ 10 g/cm$^3$) then 
$V_{CC}$ is comparable with $\ma/2E$ ($= 2.4 \times 10^{-13}$ eV if $\ma = 2.4 \times 10^{-3}$ eV$^2$).

In a two flavor formalism, the time evolution of the flavor eigenstates in matter is given by the following Schr\"{o}dinger equation
\beq
i\frac{d}{dt}\left( 
\begin{array}{c}
\nu_{\alpha} \\
\nu_{\beta}
\end{array}
\right)
= 
\frac{1}{2E}\left[U\left(
\begin{array}{cc}
m_{1}^{2} & 0 \\
0 & m_{2}^{2}
\end{array}
\right)
U^{\dagger}
+ \left(
\begin{array}{cc}
A(L) & 0  \\
0 & 0
\end{array}
\right)
\right]
\left( 
\begin{array}{c}
\nu_{\alpha} \\
\nu_{\beta}
\end{array}
\right).
\label{eq:evo_matter}
\eeq
In case of constant matter density, the problem boils down to a stationary one and a trivial diagonalization of the Hamiltonian
can provide the solution. In matter, the vacuum oscillation parameters are connected to the new 
parameters\footnote{The new parameters in matter carry a superscript $m$.} in the following way
\be
{{
{(\Delta m^2)^m} }} &=&
{{
\sqrt{(\Delta m^2 \cos 2 \theta - A)^2 +
(\Delta m^2 \sin 2 \theta)^2} }},\nonumber \\
{{\sin 2 \theta^m}}
&=& {{\sin 2 \theta ~\Delta m^2/
(\Delta m^2)^m }}.
\label{eq:matter_param}
\ee
The famous MSW-resonance~\cite{Mikheev:1986gs,Mikheev:1986wj,Wolfenstein:1977ue,Wolfenstein:1979ni} condition is satisfied at
\begin{equation}  
\Delta m^2 \cos 2 \theta = A.
\label{eq:resonance}
\end{equation}
At MSW-resonance, $\sin 2 \theta^m = 1$ (from equation~\eqref{eq:matter_param} and \ref{eq:resonance}) which immediately implies 
that independent of the value of the vacuum mixing angle $\theta$, the mixing in matter is maximal {\it i.e.} $\theta^m = \pi/4$.
This resonance occurs for neutrinos (anti-neutrinos) if $\Delta m^2$ is positive (negative). It suggests that the matter potential
modifies the oscillation probability differently depending on the sign of $\Delta m^2$. Following equation~\eqref{eq:resonance}, 
the resonance energy can be expressed as
\begin{equation} 
E_{res} = 10.83 \mbox{GeV} \left[\frac{\Delta m^{2}}{2.4 \times 10^{-3}~ 
\mbox{eV}^2}\right] \cdot \left[\frac{\cos 2 \theta}{0.96}\right] 
\cdot \left[\frac{2.8~\mbox{g}/\mbox{cm}^3}{\rho}\right].
\label{eq:resonance_energy}
\end{equation}
When neutrino travels through the upper part of the Earth mantle with $\rho = 2.8~\mbox{g}/\mbox{cm}^3$, the resonance occurs 
at roughly 10.8 GeV for positive $\Delta m^2$ of $2.4 \times 10^{-3}~\mbox{eV}^2$.
The MSW potential arises due to matter and not anti-matter and this fact is responsible for the observed asymmetry between neutrino 
and anti-neutrino oscillation probabilities even in the two neutrino case. In three-flavor scenario, besides the genuine
CP asymmetry caused by the CP phase $\dcp$, we also have fake CP asymmetry induced by matter which causes 
hindrances in extracting the information on $\dcp$.

\section{Global Status of Oscillation Parameters \& Missing Links}
\label{sec:status}

Oscillation data cannot predict the lowest neutrino mass. However, it can be probed in tritium beta decay~\cite{Osipowicz:2001sq} or
neutrinoless double beta decay~\cite{Avignone:2007fu} processes. We can also make an estimate of the lowest neutrino mass
from the contribution of neutrinos to the energy density of the universe~\cite{Lesgourgues:2012uu}.
Very recent measurements from the Planck experiment in combination with the WMAP polarization and baryon acoustic oscillation data
have set an upper bound over the sum of all the neutrino mass eigenvalues of $\sum m_i \leq 0.23$ eV at $95\%$ C.L.~\cite{Ade:2013zuv}. 
But, oscillations experiments are sensitive to the values of two 
independent mass-squared differences: $\ms$ and $\ma$. Recent global fit~\cite{GonzalezGarcia:2012sz} of all the available 
neutrino oscillation data in three-flavor framework gives as best-fit $\ms = 7.5 \times 10^{-5}~\mbox{eV}^2$ and
$|\ma| = 2.4 \times 10^{-3}~\mbox{eV}^2$ with the relative $1\sigma$ precision of  2.4\% and 2.8\% respectively.
The atmospheric mass splitting is 32 times larger than the solar mass splitting, showing the smallness of the ratio 
$\alpha=\ms/\ma \simeq 0.03$. At present, the $3\sigma$ allowed range for $\ms$ is 
$7.0 \times 10^{-5}~\mbox{eV}^2 \to 8.1 \times 10^{-5}~\mbox{eV}^2$ and the same for $|\ma|$ is
$2.2 \times 10^{-3}~\mbox{eV}^2 \to 2.7 \times 10^{-5}~\mbox{eV}^2$.
$\ms$ is required to be positive to explain the observed energy dependence of the electron
neutrino survival probability in solar neutrino experiments but $\ma$ is allowed to be either positive or negative by the present 
oscillation data. Hence, two patterns of neutrino masses are possible: $m_3 > m_2 > m_1$, called NH where $\ma$ is positive 
and $m_2 > m_1 >  m_3$, called IH where $\ma$ is negative. Determining the sign of $\ma$ is one of the prime goals of the current 
and next generation long-baseline neutrino oscillation experiments.

\begin{table}[tp]
\begin{center}
\scalebox{0.9}{
\begin{tabular}{|l||c|c|c|}
\hline
Reference
& Forero et.al.~\cite{Tortola:2012te}
& Fogli et.al.~\cite{Fogli:2012ua}
& Gonzalez-Garcia et.al.~\cite{GonzalezGarcia:2012sz}
\\
\hline\hline
$\sin^2\theta_{23}$ (NH)
& $0.427^{+0.034}_{-0.027}\oplus 0.613^{+0.022}_{-0.040}$
& $0.386^{+0.024}_{-0.021}$
& $0.41^{+0.037}_{-0.025}\oplus 0.59^{+0.021}_{-0.022}$
\\
$3\sigma$ range
& $0.36\rightarrow 0.68$
& $0.331\rightarrow 0.637$
& $0.34\rightarrow 0.67$
\\
\cline{1-3}
$\sin^2\theta_{23}$ (IH)
& $0.600^{+0.026}_{-0.031}$
& $0.392^{+0.039}_{-0.022}$
&
\\
$3\sigma$ range
& $0.37\rightarrow 0.67$
& $0.335\rightarrow 0.663$
&
\\
\hline\hline
\end{tabular}
}
\caption{\footnotesize{$1\sigma$ bounds on $\sin^2\theta_{23}$ from the global fits performed in 
references~\cite{Tortola:2012te}, \cite{Fogli:2012ua}, and \cite{GonzalezGarcia:2012sz}. The numbers cited from 
reference~\cite{GonzalezGarcia:2012sz} have been obtained by keeping the reactor fluxes free in the fit and also 
including the short-baseline reactor data with $L \lesssim 100$ m, with the mass hierarchy marginalized.
This table has been taken from reference~\cite{Agarwalla:2013ju}.}}
\label{sin2theta23}
\end{center}
\end{table}

As far as the mixing angles are concerned, the solar neutrino mixing angle $\tem$ is now pretty well determined with a best-fit 
value of $\sss = 0.3$ and the relative $1\sigma$ precision on this parameter is 4\%. The $3\sigma$ allowed range for this parameter
is $0.27 \to 0.35$. The smallest lepton mixing angle $\tet$ has been discovered very recently with a moderately large best-fit value of
$\sch =  0.023$. The relative $1\sigma$ precision achieved on this parameter is also quite remarkable which is around 10\%.
The error in the measurement of $\sch$ lies in the range of $0.016 \to 0.029$ at $3\sigma$ confidence level. 
The uncertainty associated to the choice of reactor $\anue$ fluxes~\cite{Mueller:2011nm,Mention:2011rk,Huber:2011wv} and 
its impact on the determination of $\tet$ has been discussed in detail in Section 3 of reference~\cite{GonzalezGarcia:2012sz}. 
Here, we would like to emphasize on that fact that the values of $\tet$ measured by the recent reactor and accelerator experiments 
are consistent with each other within errors, providing an important verification of the framework of three neutrino mixing.
These recent onset of data from reactor and accelerator experiments will also enable us to explore the long expected 
complementarity between these two independent measurements~\cite{Minakata:2002jv,Huber:2003pm,Roy:2012ch,Pascoli:2013wca}.
Our understanding of the 2-3 mixing angle $\tmt$ has also been refined a lot in recent years. The best-fit values and ranges of 
$\tmt$ obtained from the three recent global fits~\cite{Tortola:2012te}, \cite{Fogli:2012ua}, and \cite{GonzalezGarcia:2012sz} 
are listed in Table~\ref{sin2theta23}. A common feature that has emerged from all the three global fits is that we now have 
hint for non-maximal $\theta_{23}$, giving two degenerate solutions: either $\tmt$ belongs to the LO ($\sin^2\theta_{23} \approx 0.4$) 
or it lies in the HO ($\sin^2\theta_{23}\approx0.6$). This octant ambiguity of $\tmt$, in principal, can be resolved with the help of
$\nu_{\mu} \leftrightarrow \nu_{e}$ oscillation data. The preferred value would depend on the choice of the neutrino mass ordering.
However, as can be seen from Table~\ref{sin2theta23}, the fits of reference~\cite{Tortola:2012te} do not agree on which value
should be preferred, even when the mass ordering is fixed to be NH. LO is preferred over HO for both NH and IH in 
reference~\cite{Fogli:2012ua}. Reference~\cite{GonzalezGarcia:2012sz} marginalizes over the mass ordering, 
so the degeneracy remains. The global best-fits in references~\cite{Tortola:2012te,GonzalezGarcia:2012sz} do not see 
any sensitivity to the octant of $\tmt$ unless they add the information from the atmospheric neutrinos. But, in~\cite{Fogli:2012ua}, 
they do find a preference for LO even without adding the atmospheric neutrino data. At present, the relative $1\sigma$ precision
on $\sa$ is around 11\%. Further improvement in the measurement of $\tmt$ and settling the issue of its octant 
(if it turns out to be non-maximal) are also the crucial issues that need to be addressed in current and next generation long-baseline
experiments. Leptonic CP violation can be established if CP violating phase $\dcp$ is shown to differ from 0 and $180^\circ$.
We have not seen any signal for CP violation in the data so far. Thus, $\dcp$ can have any value in the range 
$[-180^\circ,180^\circ]$. Measuring the value of $\dcp$ and establishing the CP violation in the neutral lepton sector 
would be the top most priorities for the present and future long-baseline experiments. 

Due to the fact that both $\alpha$ and $\tet$ are small, so far, it was possible to analyze the the data from each neutrino 
oscillation experiment adopting an appropriate, effective two flavor oscillation approach. 
This method has been quite successful in measuring the solar and atmospheric neutrino parameters. 
The next step must involve probing the full three-flavor effects, including the sub-leading ones which are proportional to 
$\alpha$. These are the key requirements to discover neutrino mass hierarchy, CP violation, and octant of $\tmt$ 
in long-baseline experiments~\cite{Feldman:2012qt,Diwan:2013eha}.

\section{Three-Flavor Effects in $\numu \rightarrow \nue$ Oscillation Channel}
\label{sec:three-flavor}

To illustrate the impact of three-flavor effects, the most relevant oscillation channels are $\numu \rightarrow \nue$ and
$\anumu \rightarrow \anue$. A study of these oscillation channels at long-baseline superbeam experiments is capable 
of addressing all the three major issues discussed in the previous Section. In particular, the use of an appearance channel 
in which the neutrino changes flavor between production and detection is mandatory to explore CP violation in neutrino
oscillations. Earth matter effects are also going to play a significant role in probing these fundamental unknowns.
The exact expressions of the three-flavor oscillation probabilities including matter effects are very complicated. 
Therefore, to demonstrate the nature of neutrino oscillations as a function of baseline and/or neutrino energy,
it is quite useful to have an approximate analytic expression for $P_{\mu e}$ (the T-conjugate of  $P_{e \mu}$) in
matter~\cite{Wolfenstein:1977ue,Mikheev:1986gs,Barger:1980tf}, keeping terms only up to second order in
the small quantities $\theta_{13}$ and $\alpha$~\cite{Cervera:2000kp,Freund:2001ui,Akhmedov:2004ny}:
\be
P_{\mu e} &\simeq&
{\underbrace{\sin^2\theta_{23} \sin^22\theta_{13}
    \frac{\sin^2[(1-\hat{A})\Delta]}{(1-\hat{A})^2}}_{C_0}}
    + {\underbrace{\alpha^2 \cos^2\theta_{23} \sin^22\theta_{12}
    \frac{\sin^2(\hat{A}\Delta)}{{\hat{A}}^2}}_{C_1}} \nonumber \\
&\mp& {\underbrace{\alpha \sin2\theta_{13}\cos\theta_{13} \sin2\theta_{12}
    \sin2\theta_{23} \sin(\Delta) \frac{\sin(\hat{A}\Delta)}{\hat{A}}
    \frac{\sin[(1-\hat{A})\Delta]}{(1-\hat{A})}}_{C_-}} \sin\dcp \nonumber \\
&+& {\underbrace{\alpha \sin2\theta_{13}\cos\theta_{13} \sin2\theta_{12}
    \sin2\theta_{23} \cos(\Delta) \frac{\sin(\hat{A}\Delta)}{\hat{A}}
    \frac{\sin[(1-\hat{A})\Delta]}{(1-\hat{A})}}_{C_+}} \cos\dcp ,
\label{eq:pmue}
\ee
where
\be
\Delta\equiv \frac{\ma L}{4E},~~~~~~\hat{A} \equiv \frac{A}{\ma}.
\label{eq:matt}
\ee
Equation~\eqref{eq:pmue} has been derived under the constant matter density approximation. The matter effect is expressed by the
dimensionless quantity $\hat{A}$. The `$-$' sign which precedes the term $C_{-}$ refers to neutrinos whereas the `$+$' refers to to anti-neutrinos.
In equation~\eqref{eq:pmue}, $\alpha$, $\Delta$, and $\hat{A}$ are sensitive to the sign of $\ma$ i.e., the type of the neutrino mass ordering.
Note that the sign of $\hat{A}$ changes with the sign of $\ma$ as well as in going from neutrino to the corresponding anti-neutrino 
mode. The former suggests that the matter effect can be utilized to determine the sign of $\ma$, while the latter implies that it can
mimic a CP violating effect and hence complicate the extraction of $\dcp$ by comparing neutrino and anti-neutrino data.
For large $\tet$, the first term of Eq.~\eqref{eq:pmue} ($C_0$) dominates and it contains the largest Earth matter effect which can 
therefore be used to measure the sign of $\ma$. This term also depends on $\sa$ and therefore is sensitive to the octant of $\tmt$.
The sub-dominant terms $C_{-}$ and $C_{+}$ are suppressed by $\alpha$ and provide information on $\dcp$.
The term $C_{-}$ is the CP-violating part. The term $C_{+}$, although $\delta_{CP}$-dependent, is CP-conserving. 
The term $C_1$ is independent of both $\theta_{13}$ and $\delta_{CP}$ and depends mainly on the 
solar parameters, $\ms$ and $\tem$.

\subsection{Hierarchy-$\dcp$ Degeneracy}
\label{subsec:hierarchy-dcp}

Since the hierarchy and $\dcp$ are both unknown, the interplay of the terms $C_0$, $C_{-}$, and $C_{+}$ 
in equation~\eqref{eq:pmue} gives rise to hierarchy-$\dcp$ degeneracy~\cite{Minakata:2001qm}. 
This degeneracy can be broken completely using the large Earth matter effects provided by the baselines 
which are $>$ 1000 km~\cite{Agarwalla:2011hh,Agarwalla:2013txa,Agarwalla:2013hma}. 
For these long baselines, we can also observe both the first and second oscillation maxima quite efficiently using the detectors like
Liquid Argon Time Projection Chamber (LArTPC)~\cite{Amerio:2004ze}. It helps to evade the problem of 
($sgn(\ma),\dcp$) degeneracy~\cite{Minakata:2001qm} and ($\tet,\dcp$) intrinsic degeneracy~\cite{BurguetCastell:2001ez}
which can cause the $\pi$-transit~\cite{Huber:2002mx} effect even for large values of $\stch$.
Adding data from two different experiments with different baselines can also be very useful to resolve this 
degeneracy~\cite{Minakata:2001qm,Barger:2002rr,BurguetCastell:2002qx,Minakata:2002qi,Barger:2002xk,Huber:2002rs}.
Another elegant way to tackle these degeneracies is to kill the spurious clone solutions at the 
``magic'' baseline~\cite{Barger:2001yr,Huber:2003ak,Smirnov:2006sm,Agarwalla:2009xc}.
When $\sin(\hat{A}\Delta)=0$, the last three terms in equation~\eqref{eq:pmue} drop out and the $\dcp$ dependence 
disappears from the $P_{\mu e}$ channel, which provides a clean ground for $\theta_{13}$ and $sgn(\ma)$ measurements.
Since $\hat{A}\Delta = \pm  (2 \sqrt{2} G_F n_e L)/4$ by definition, the first non-trivial solution for the condition,
$\sin(\hat{A}\Delta)=0$ reduces to $\rho L = \sqrt{2}\pi/G_F Y_e$. This gives $\frac{\rho}{[{\rm g/cc}]}\frac{L}{[km]} \simeq 32725$,
which for the PREM density profile of the Earth is satisfied for the ``magic baseline'', $L_{\rm magic} \simeq 7690~{\rm km}$.
The CERN to India-based Neutrino Observatory (INO)~\cite{INO} distance corresponds to $L=7360$ km, which is tantalizingly close 
to this ``magic'' baseline. Performing a long-baseline experiment at ``Bimagic'' baseline can be also very promising to suppress the 
effect of these degeneracies~\cite{Raut:2009jj, Dighe:2010js}.
Note that the low order expansion of the probability $P_{\mu e}$ given by equation~\eqref{eq:pmue} is valid only for values of $E$ 
and Earth matter density $\rho$ (and hence $L$) where flavor oscillations are far from resonance, i.e., $\hat{A} \ll 1$. 
In the limit $\hat{A} \sim 1$, one can check that even though the analytic expression for $P_{\mu e}$ given by equation~\eqref{eq:pmue} 
remains finite, the resultant probability obtained is incorrect~\cite{Takamura:2005df,Agarwalla:2013tza}. 
While we will use this analytical formula to explain our results in some cases, all the simulations presented in this review article are based 
on the full three-flavor neutrino oscillation probabilities in matter, using the Preliminary Reference Earth Model (PREM)~\cite{Dziewonski:1981xy}.

\begin{figure}[tp]

        \begin{tabular}{lr}
        
                \hspace*{-0.85in} \includegraphics[width=0.72\textwidth]{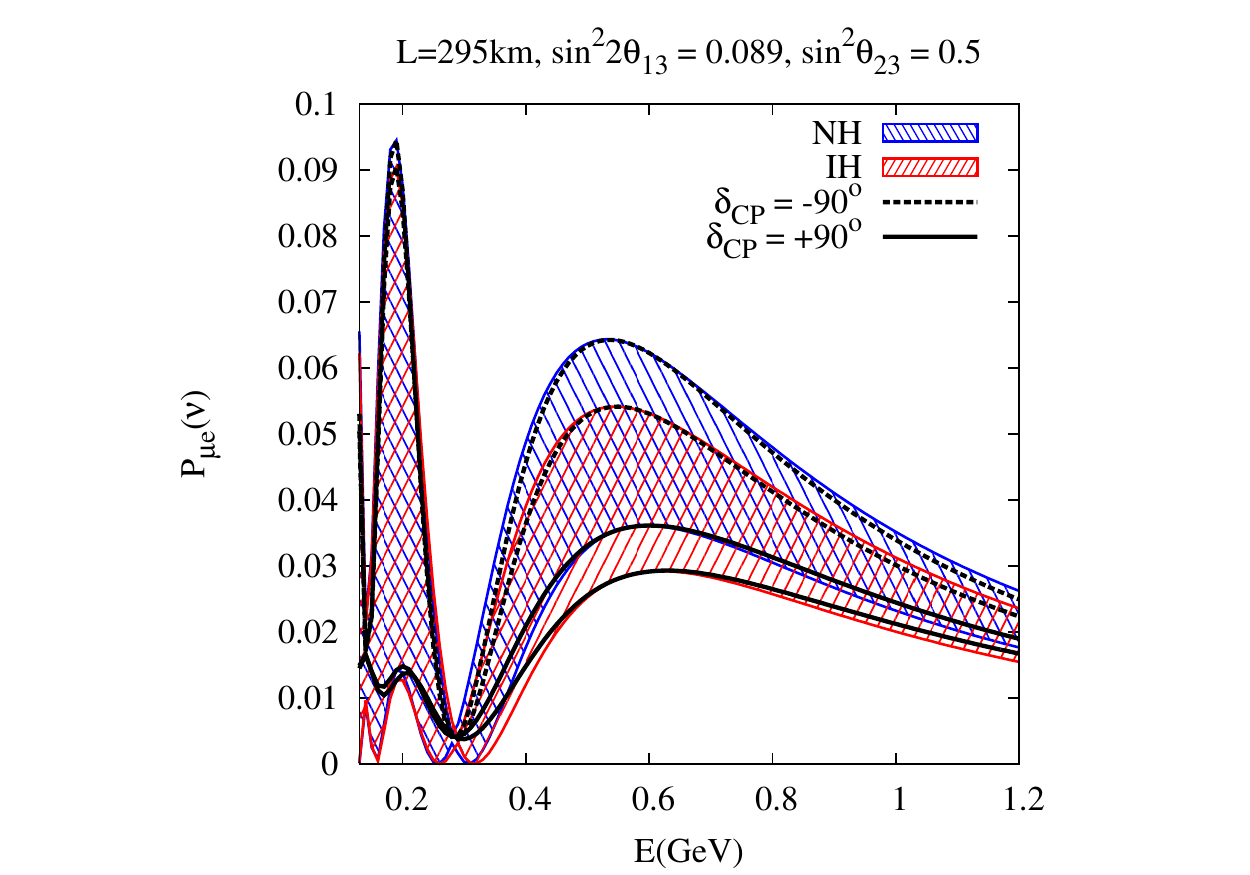}
                &
                \hspace*{-1.7in} \includegraphics[width=0.72\textwidth]{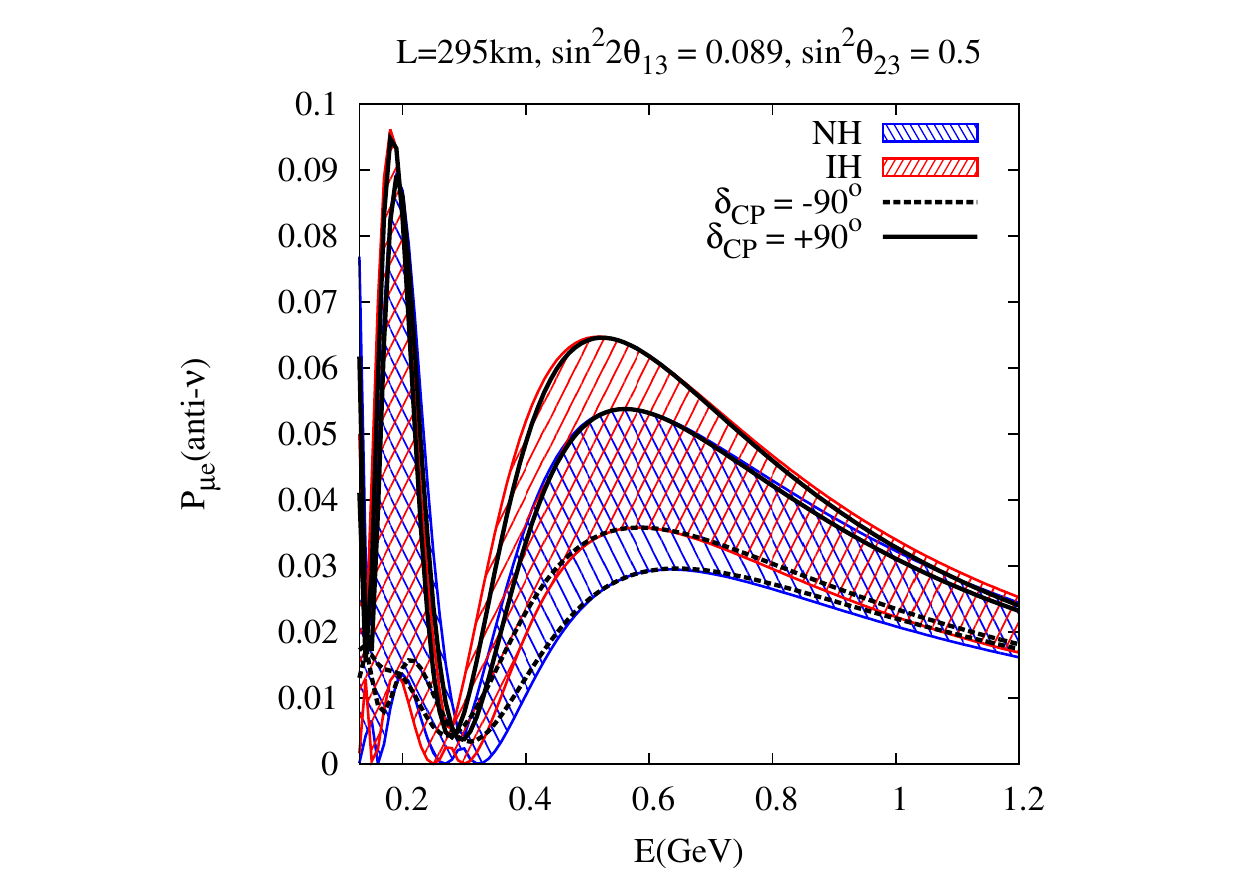}
        
         \end{tabular}
         
        \begin{tabular}{lr}
                 
                \hspace*{-0.85in} \includegraphics[width=0.72\textwidth]{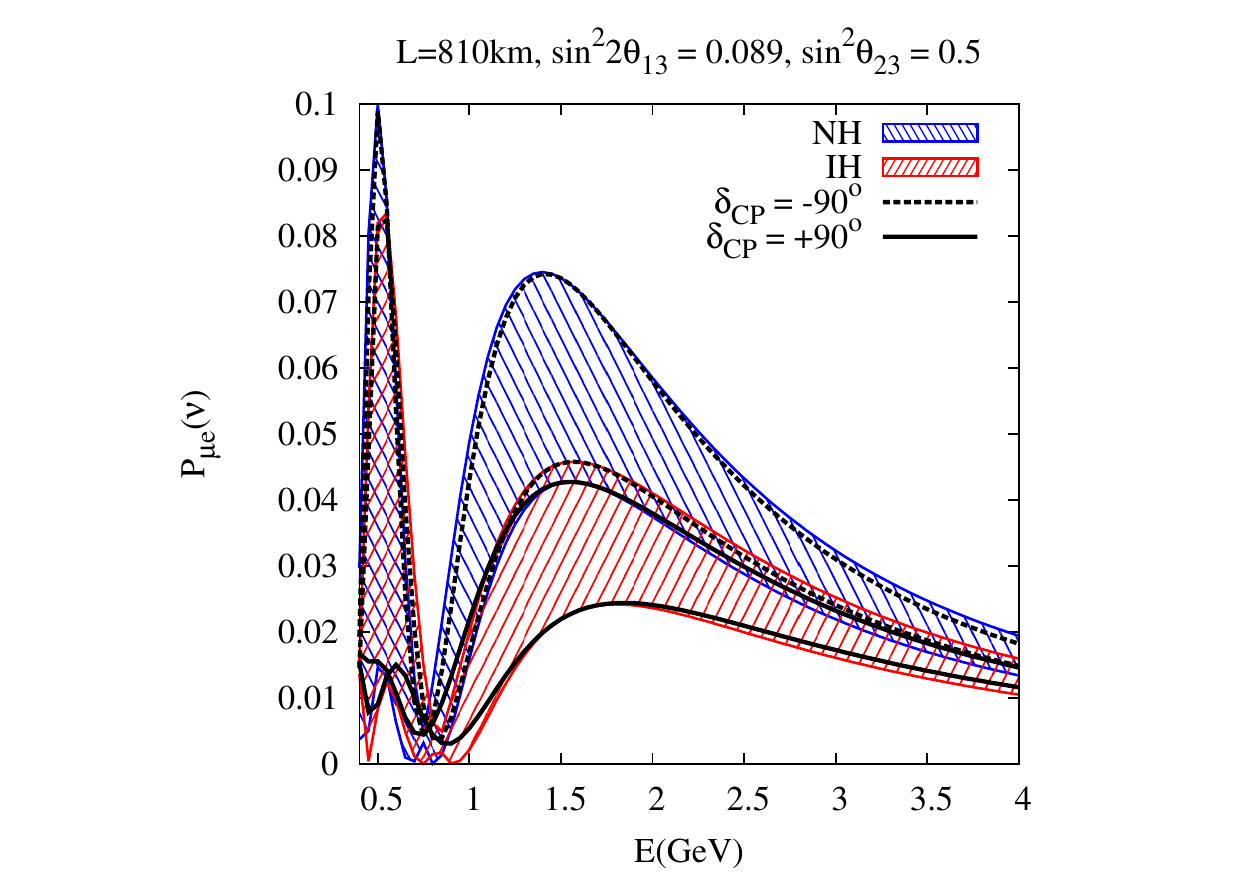}
                &
                \hspace*{-1.7in} \includegraphics[width=0.72\textwidth]{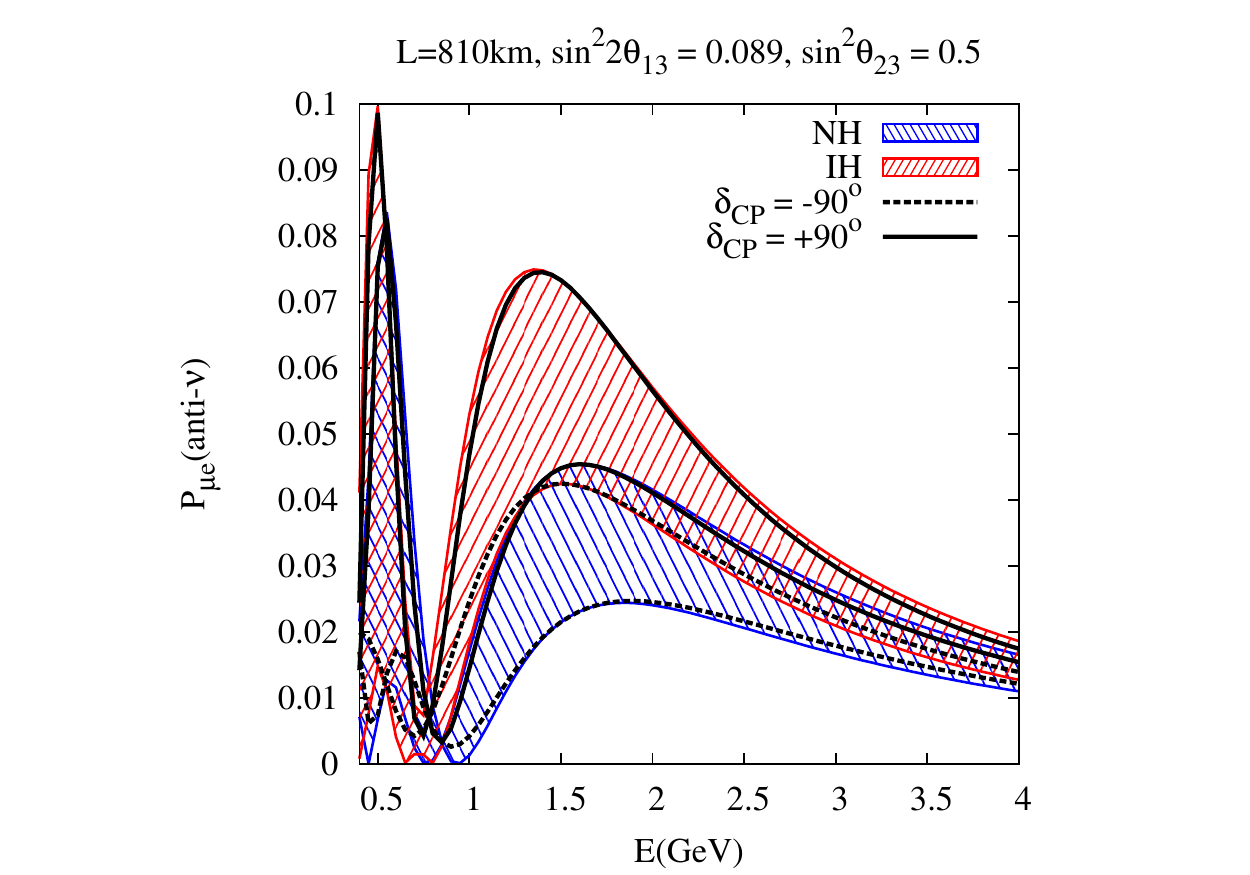}
                
        \end{tabular}

\caption{\footnotesize{The transition probability $P_{\mu e}$ as a function of neutrino energy. The band reflects the effect of 
unknown $\dcp$. Inside each band, the probability for $\dcp = 90^\circ$ ($\dcp = -90^\circ$) case is shown by the solid (dashed) line. 
The blue (red) band is for NH (IH).
The left panel (right panel) is for $\nu$ ($\bar\nu$). The upper panels
are drawn for the T2K baseline of 295 km. The lower panels are for the NO$\nu$A baseline of 810 km. Here, we take
$\stch = 0.089$ and $\sa = 0.5$.}}

\label{hierarchy-dcp-nu-antinu-T2K-NOvA}
\end{figure}

\begin{figure}[tp]

        \begin{tabular}{lr}
        
                \hspace*{-0.85in} \includegraphics[width=0.72\textwidth]{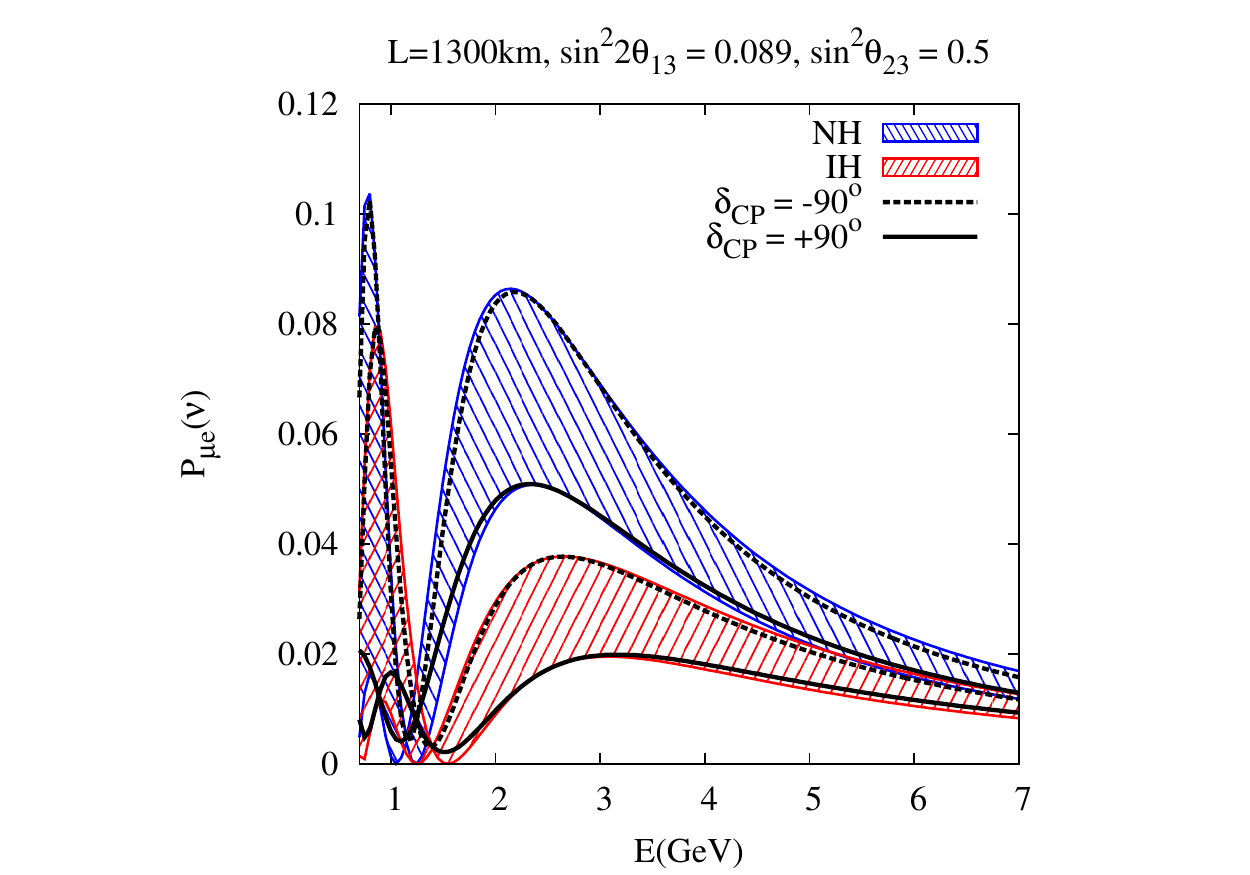}
                &
                \hspace*{-1.7in} \includegraphics[width=0.72\textwidth]{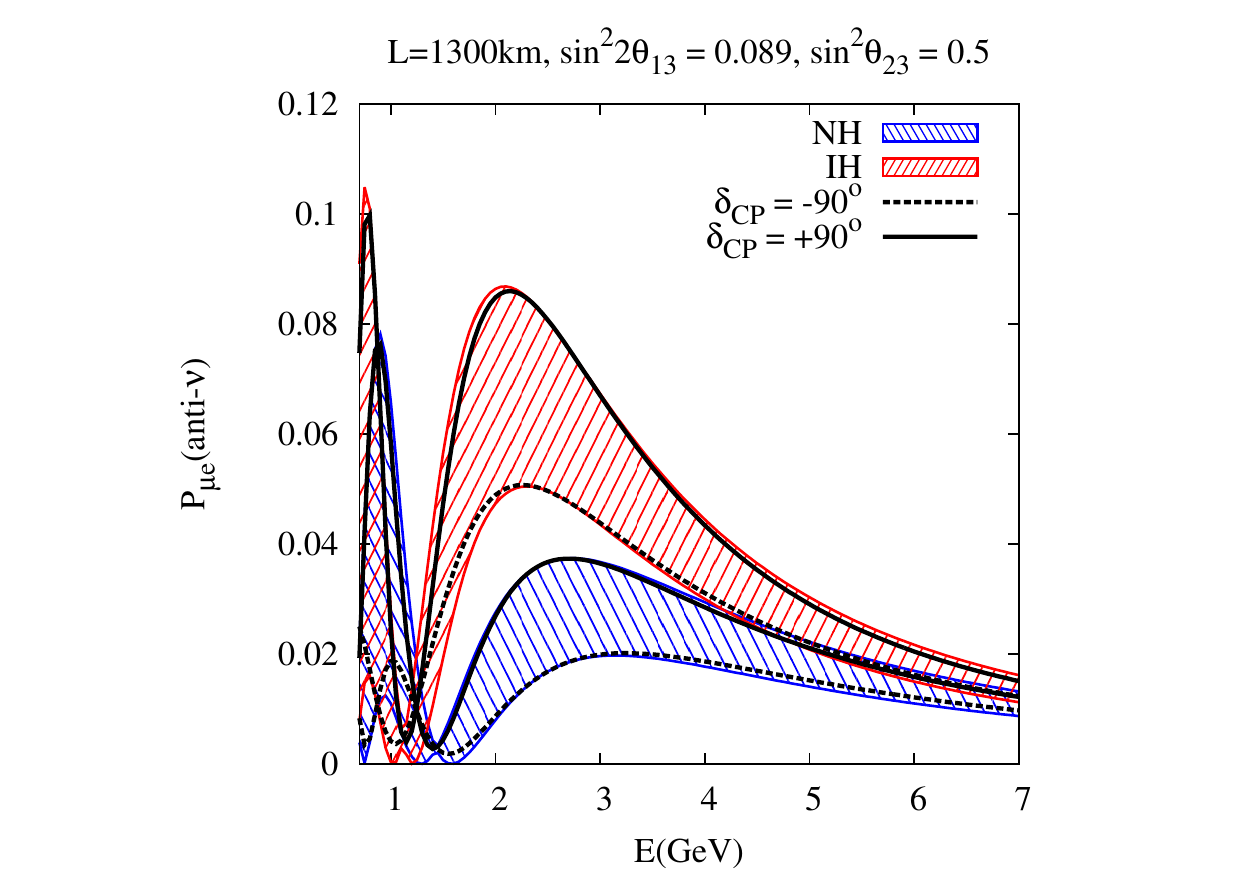}
        
         \end{tabular}
         
        \begin{tabular}{lr}
                 
                \hspace*{-0.85in} \includegraphics[width=0.72\textwidth]{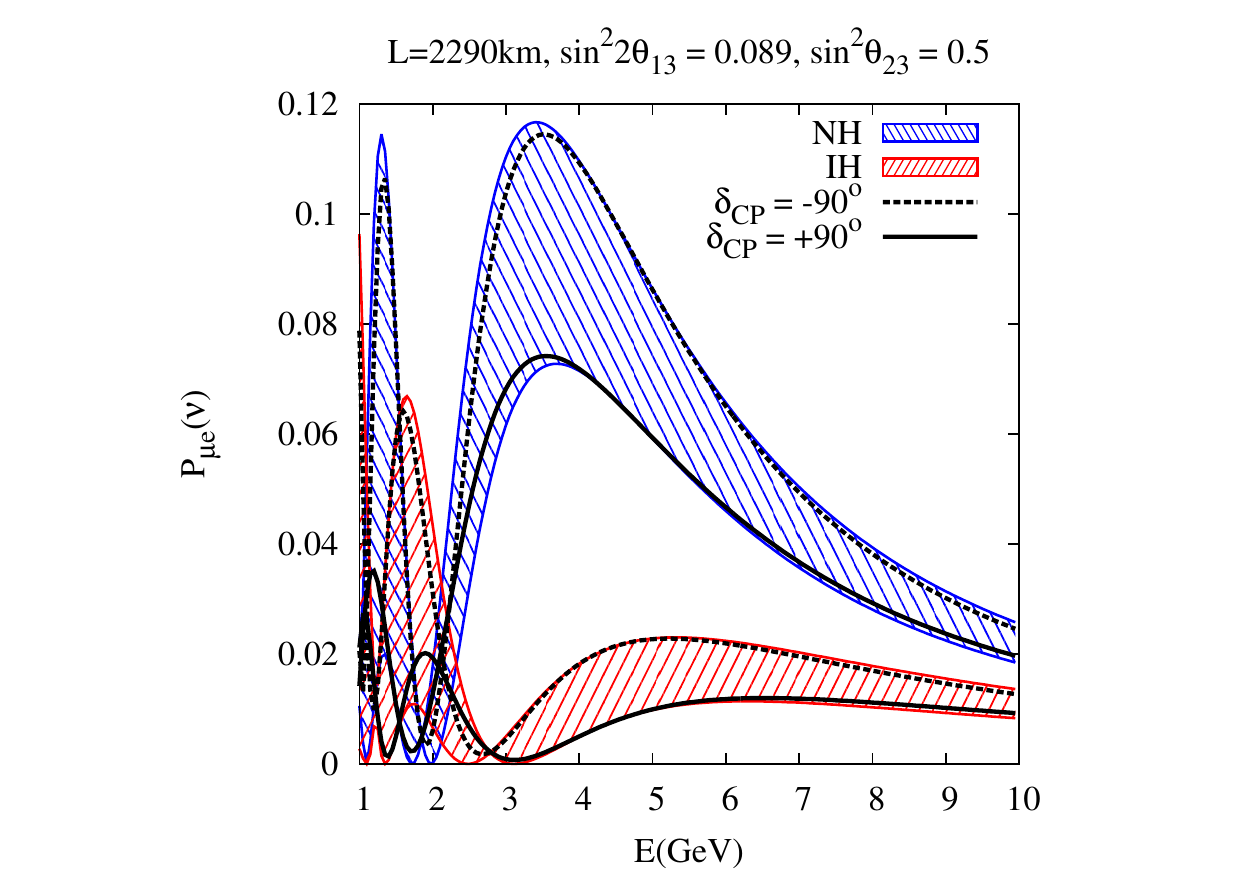}
                &
                \hspace*{-1.7in} \includegraphics[width=0.72\textwidth]{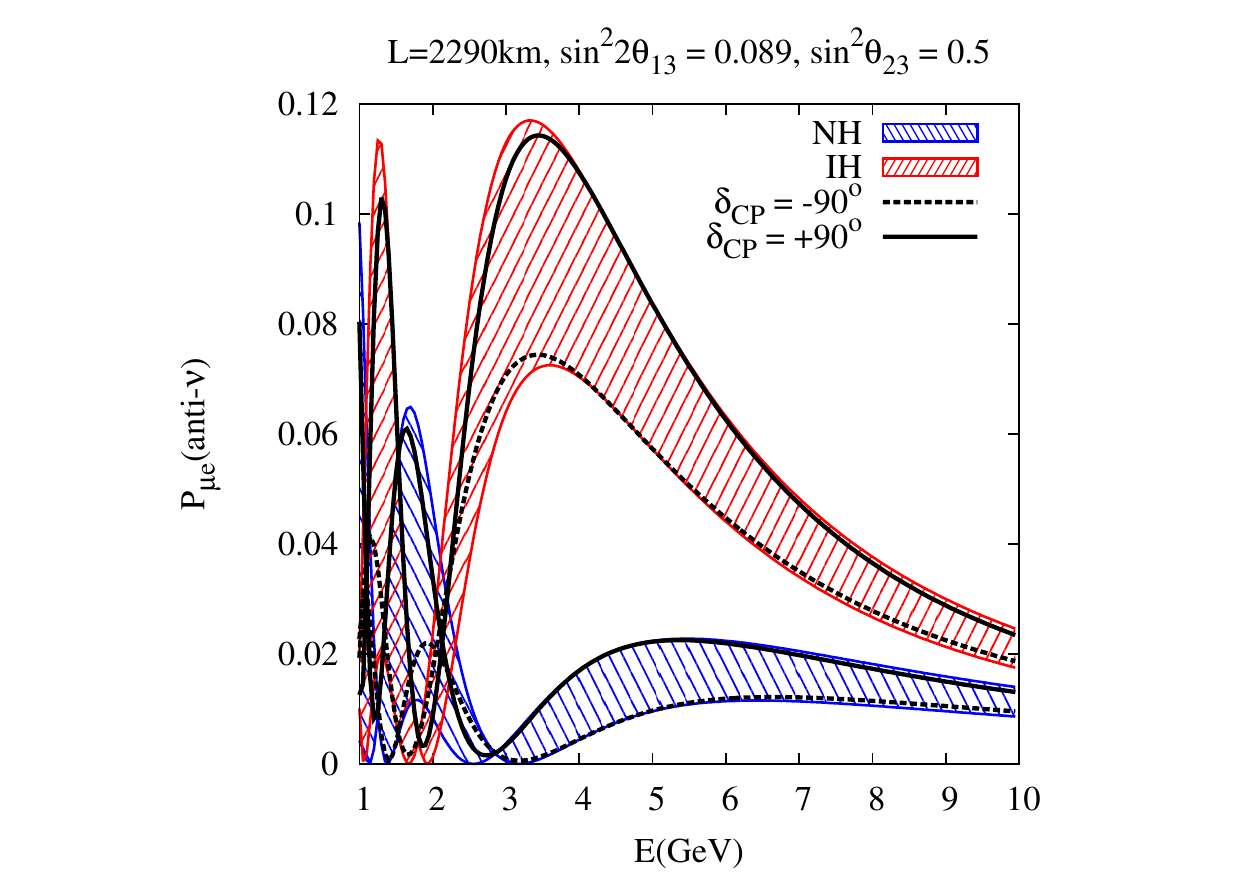}
                
        \end{tabular}

\caption{\footnotesize{$P_{\mu e}$ as a function of neutrino energy. Here, the bands correspond to different
values of $\dcp$ from $-180^\circ$ to $180^\circ$. 
Inside each band, the probability for $\dcp = 90^\circ$ ($\dcp = -90^\circ$) case is shown by the solid (dashed) line.
The blue (red) band is for NH (IH).
The left panel (right panel) is for $\nu$ ($\bar\nu$). The upper panels are drawn for the LBNE baseline of 1300 km. 
The lower panels are for the LBNO baseline of 2290 km. Here, we take $\stch = 0.089$ and $\sa = 0.5$.}}
 
\label{hierarchy-dcp-nu-antinu-LBNE-LBNO}
\end{figure}

The transition probability $P_{\mu e}$ as a function of the neutrino energy is shown in Figure~\ref{hierarchy-dcp-nu-antinu-T2K-NOvA}.
We allow $\dcp$ to vary within the range $-180^\circ$ to $180^\circ$ and the resultant probability is shown as a band, with the thickness 
of the band reflecting the effect of $\dcp$ on $P_{\mu e}$. Inside each band, the probability for $\dcp = 90^\circ$ ($\dcp = -90^\circ$) case 
is shown explicitly by the solid (dashed) line. In each panel, the blue (red) band is for NH (IH). Left panel (right panel) 
depicts the probability for neutrino (anti-neutrino). In upper panels, we take the baseline of 295 km which matches with the distance of 
Tokai-to-Kamioka (T2K) experiment~\cite{Itow:2001ee,Abe:2011ks} in Japan. In lower panels, we consider the baseline of 810 km
which is the distance between Fermilab and Ash River, chosen for the NuMI Off-axis Neutrino Appearance (NO$\nu$A)
experiment~\cite{Ayres:2002ws,Ayres:2004js,Ayres:2007tu} in the United States. 
Matter effect increases $P(\numu \to \nue)$ for NH and decreases it for IH and vice verse for $P(\anumu \to \anue)$. 
For $\dcp$ in the lower half-plane (LHP, $-180^\circ \leq \dcp \leq 0$), $P(\numu \to \nue)$ is larger and for $\dcp$ in the upper half-plane
(UHP, $0 \leq \dcp \leq 180^\circ$), $P(\numu \to \nue)$ is smaller. Hence, for the combination (NH, LHP), the values of $P(\numu \to \nue)$  
are much higher than those for IH (and  $P(\anumu \to \anue)$ values are much lower). Similarly, for the combination (IH, UHP), the values 
of $P(\numu \to \nue)$ are much lower than those of NH (and  $P(\anumu \to \anue)$ values are much higher). Thus, LHP is the favorable 
half-plane for NH and UHP is for IH~\cite{Prakash:2012az,Agarwalla:2012bv}. For T2K baseline, the matter effect is very small and 
therefore the $\dcp$ bands drawn for NH and IH overlap for almost entire rage of $\dcp$ 
(except for the most favorable combinations like: NH, $\dcp = -90^\circ$ and IH, $\dcp = 90^\circ$) for almost all the choices of $E$.
NO$\nu$A has better chances to discriminate between NH and IH compared T2K because of its larger baseline causing larger matter
effect. But, for the unfavorable combinations like: NH, $\dcp = 90^\circ$ and IH, $\dcp = -90^\circ$, the NH and IH bands still overlap
with each other. In the upper panels of Figure~\ref{hierarchy-dcp-nu-antinu-LBNE-LBNO}, we study the same for the 
Long-Baseline Neutrino Experiment (LBNE)~\cite{Diwan:2003bp,Barger:2007yw,Huber:2010dx,Akiri:2011dv} baseline of 1300 km 
which is the distance between the Fermilab and the Homestake mine in South Dakota in the United States. The lower panels of 
Figure~\ref{hierarchy-dcp-nu-antinu-LBNE-LBNO} depict the hierarchy-$\dcp$ degeneracy pattern for the Long-Baseline Neutrino 
Oscillation Experiment (LBNO)~\cite{Autiero:2007zj,Rubbia:2010fm,Angus:2010sz,Rubbia:2010zz,Stahl:2012exa} baseline of 2290 
km which is the distance between the CERN and the Pyh\"asalmi mine in Finland. For both the LBNE and LBNO baselines, the matter
effects are substantial and they break the hierarchy-$\dcp$ degeneracy completely.

\subsection{Octant-$\dcp$ Degeneracy}
\label{subsec:octant-dcp}

\begin{figure}[tp]

        \begin{tabular}{lr}
        
                \hspace*{-0.85in} \includegraphics[width=0.72\textwidth]{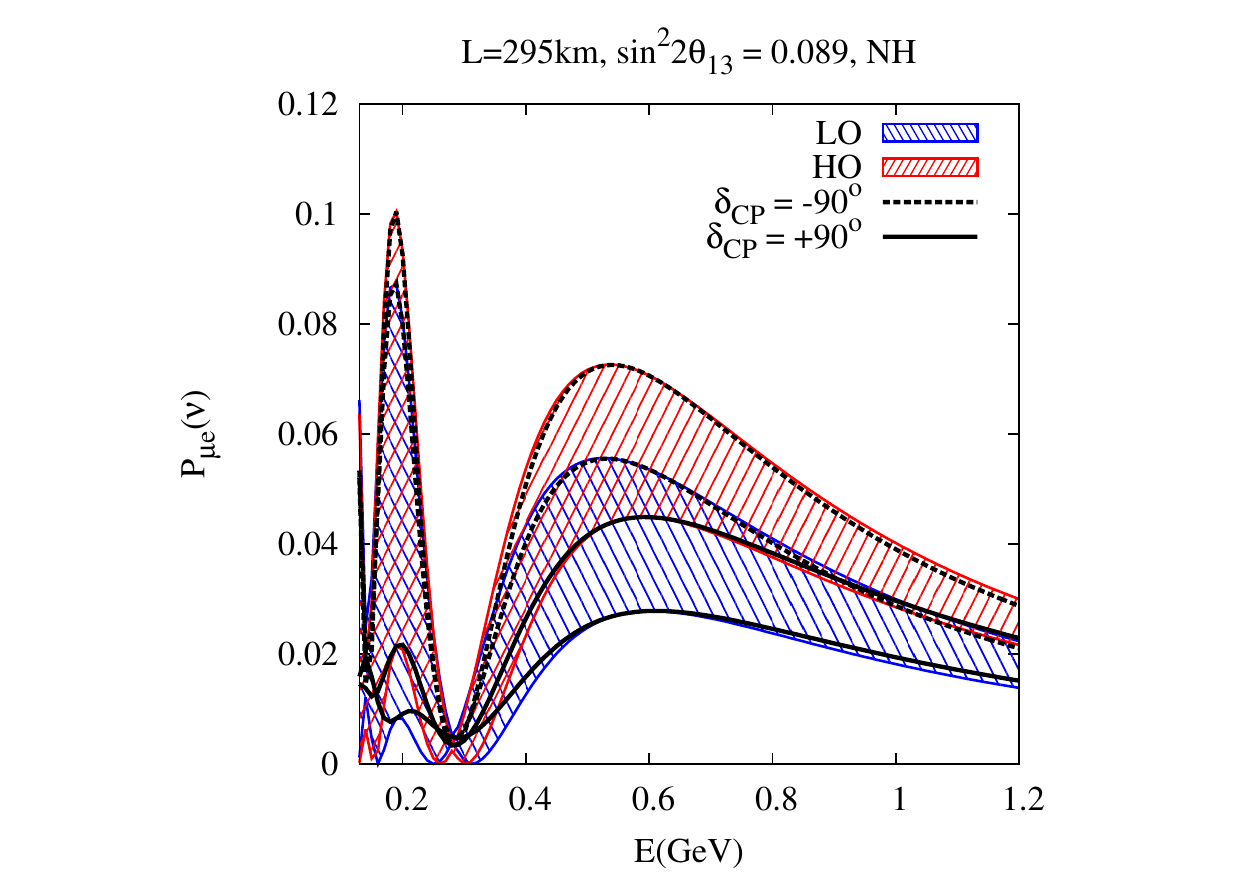}
                &
                \hspace*{-1.7in} \includegraphics[width=0.72\textwidth]{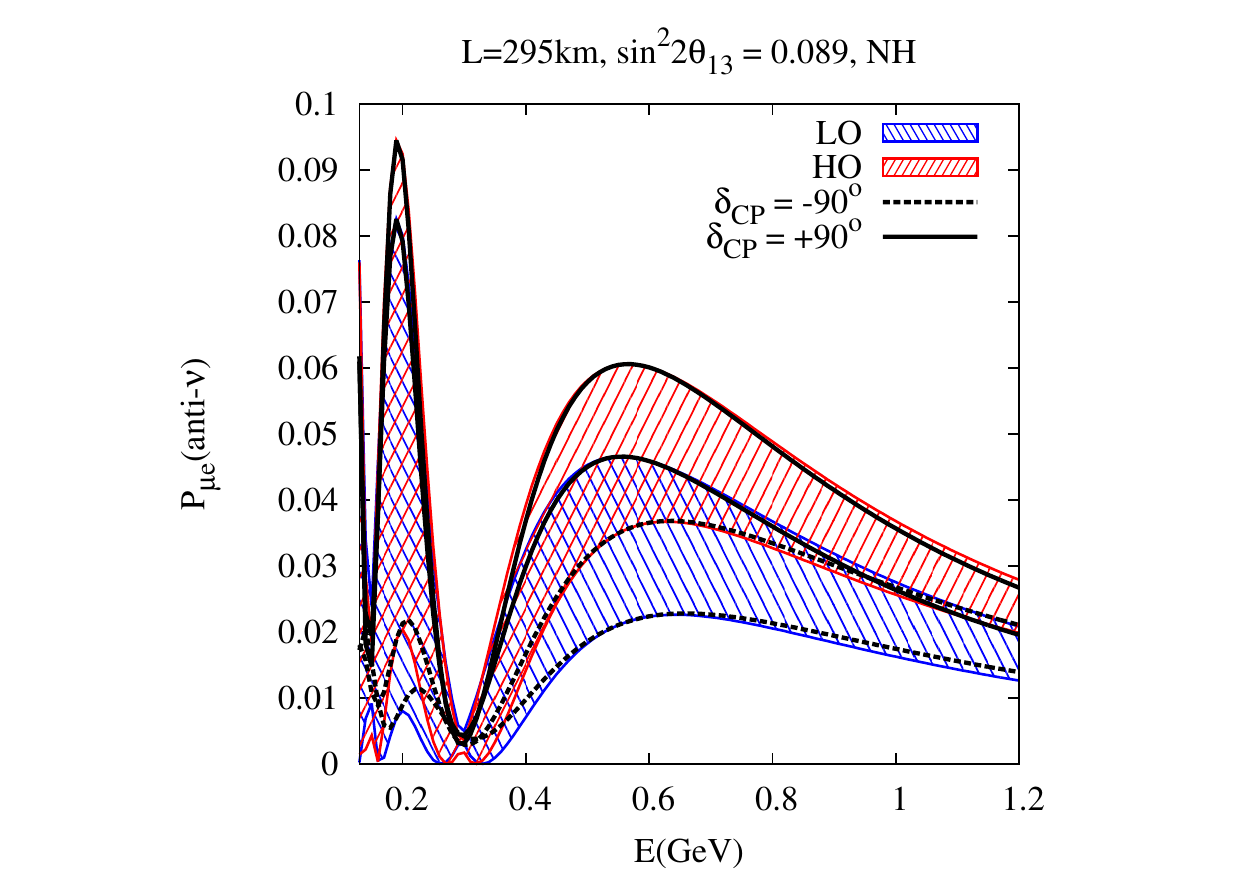}
        
         \end{tabular}
         
        \begin{tabular}{lr}
                 
                \hspace*{-0.85in} \includegraphics[width=0.72\textwidth]{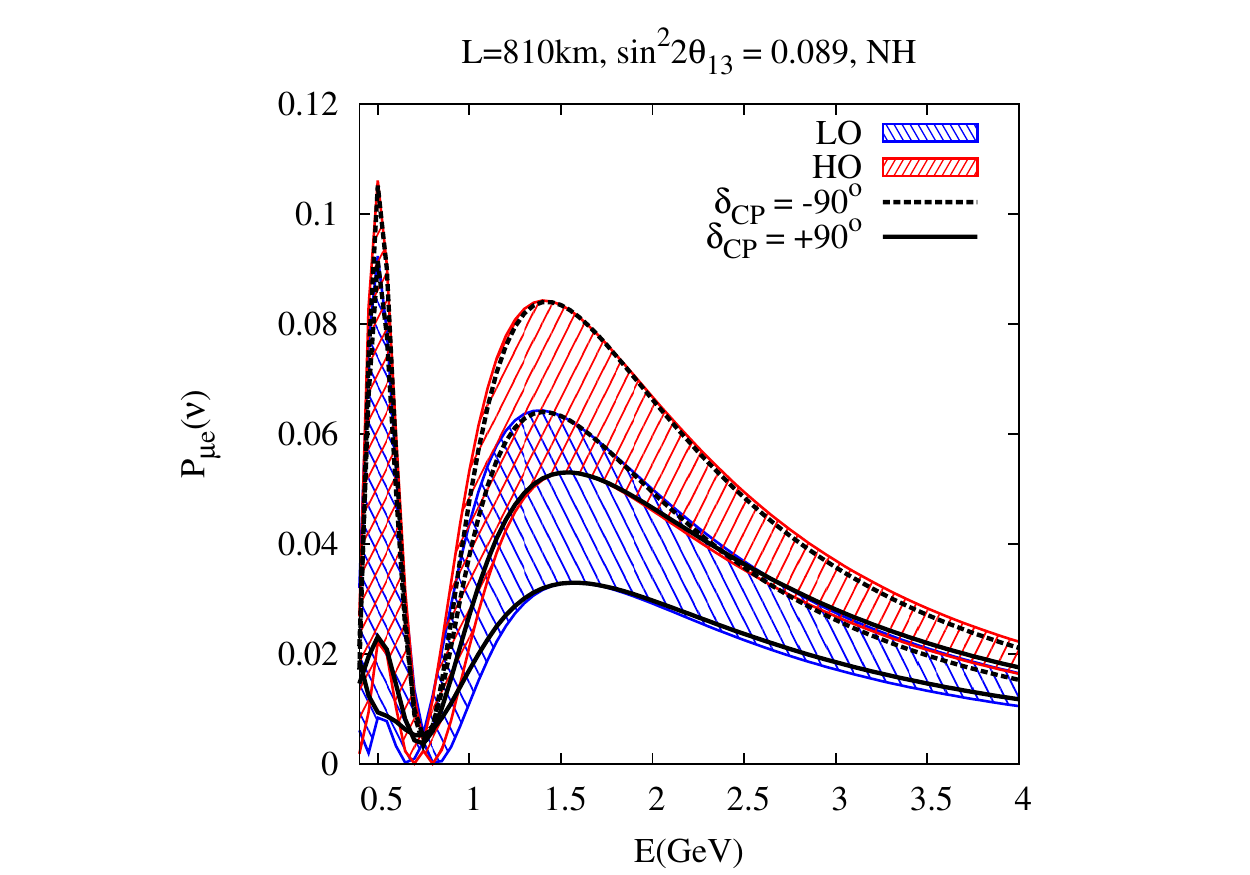}
                &
                \hspace*{-1.7in} \includegraphics[width=0.72\textwidth]{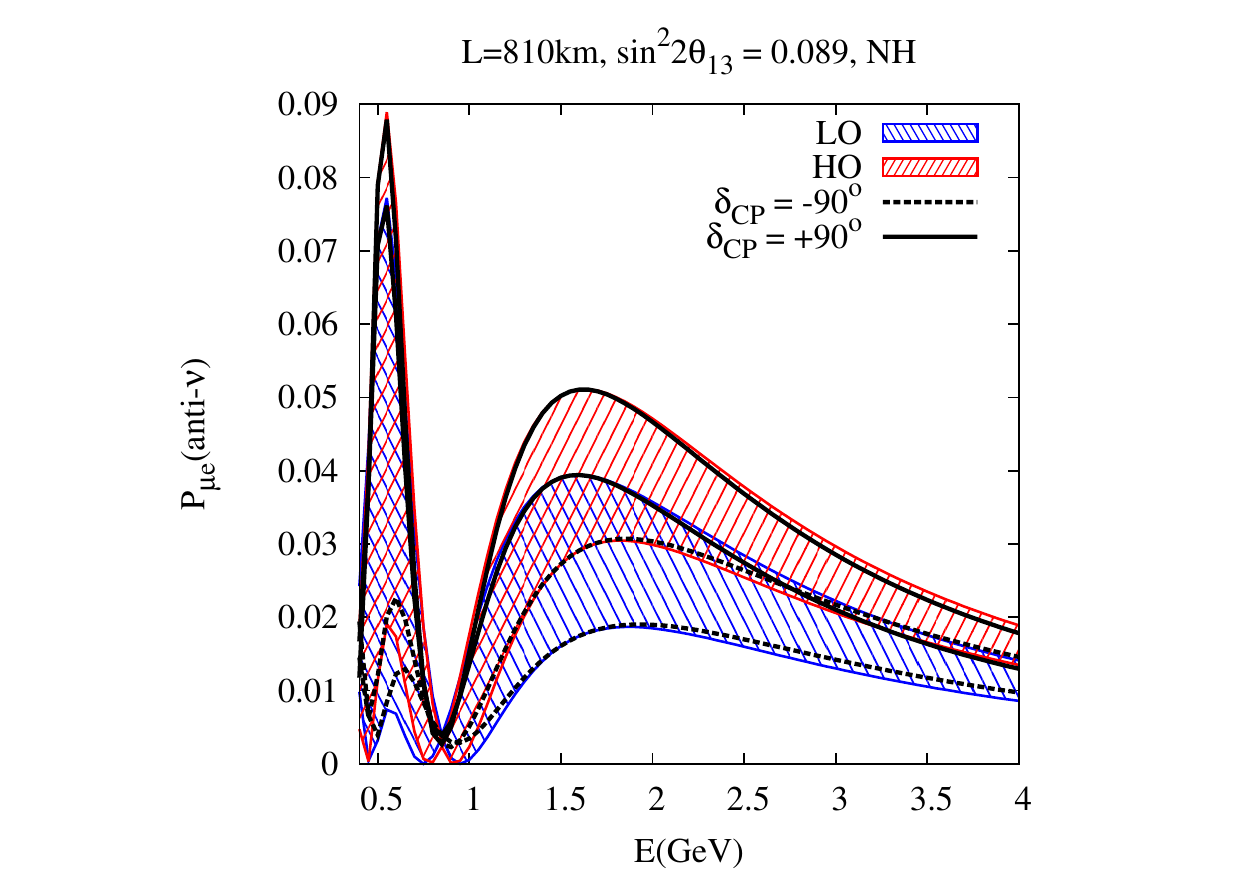}
                
        \end{tabular}

\caption{\footnotesize{$P_{\mu e}$ as a function of neutrino energy. Here, the bands correspond to different
values of $\dcp$ ranging from $-180^\circ$ to $180^\circ$.
Inside each band, the probability for $\dcp = 90^\circ$ ($\dcp = -90^\circ$) case is shown by the solid (dashed) line.
The red (blue) band is for HO with $\sa = 0.59$ (LO with $\sa = 0.41$).
The left panel (right panel) is for $\nu$ ($\bar\nu$). The upper panels are drawn for the T2K baseline of 295 km. 
The lower panels are for the NO$\nu$A baseline of 810 km. Here, we consider $\stch = 0.089$ and NH.}}

\label{octant-dcp-nu-antinu-T2K-NOvA}
\end{figure}

\begin{figure}[tp]

        \begin{tabular}{lr}
        
                \hspace*{-0.85in} \includegraphics[width=0.72\textwidth]{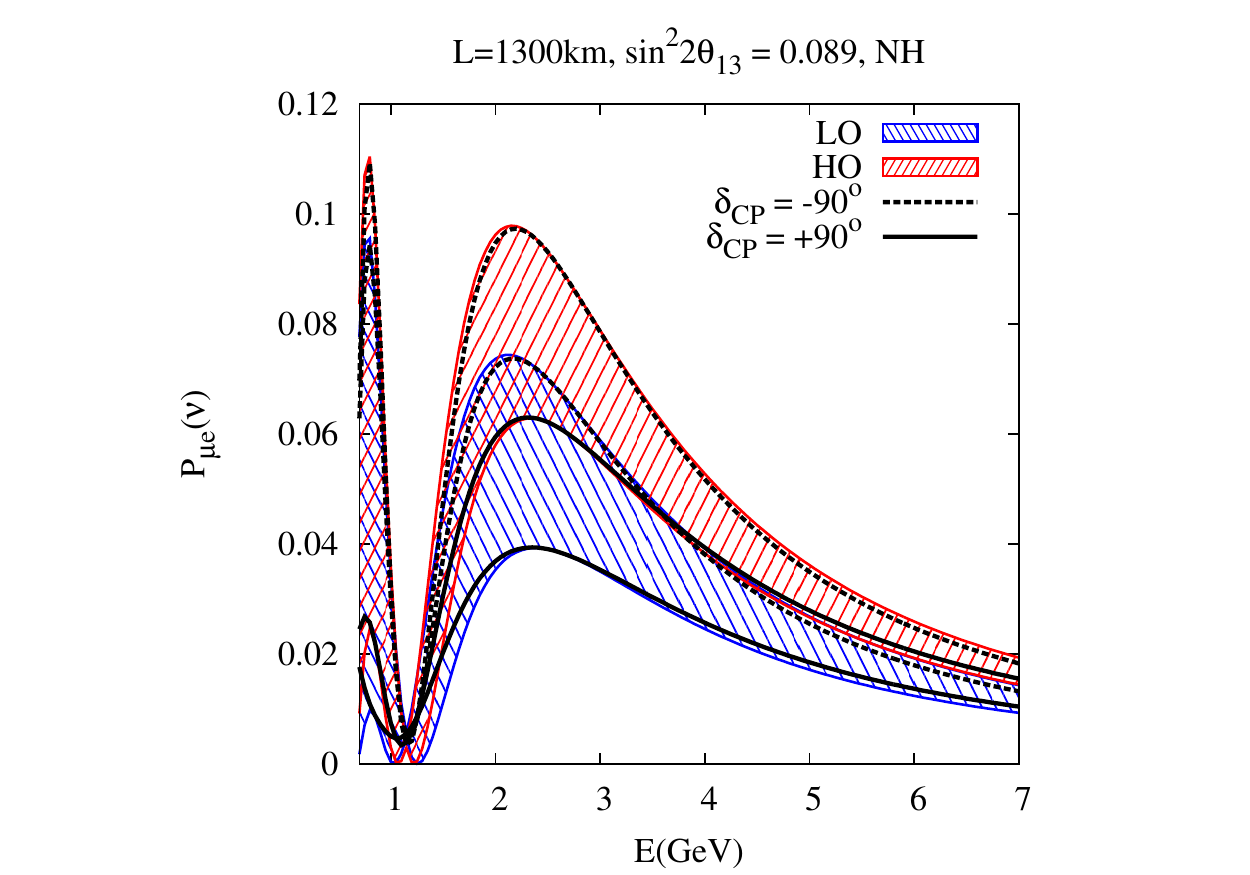}
                &
                \hspace*{-1.7in} \includegraphics[width=0.72\textwidth]{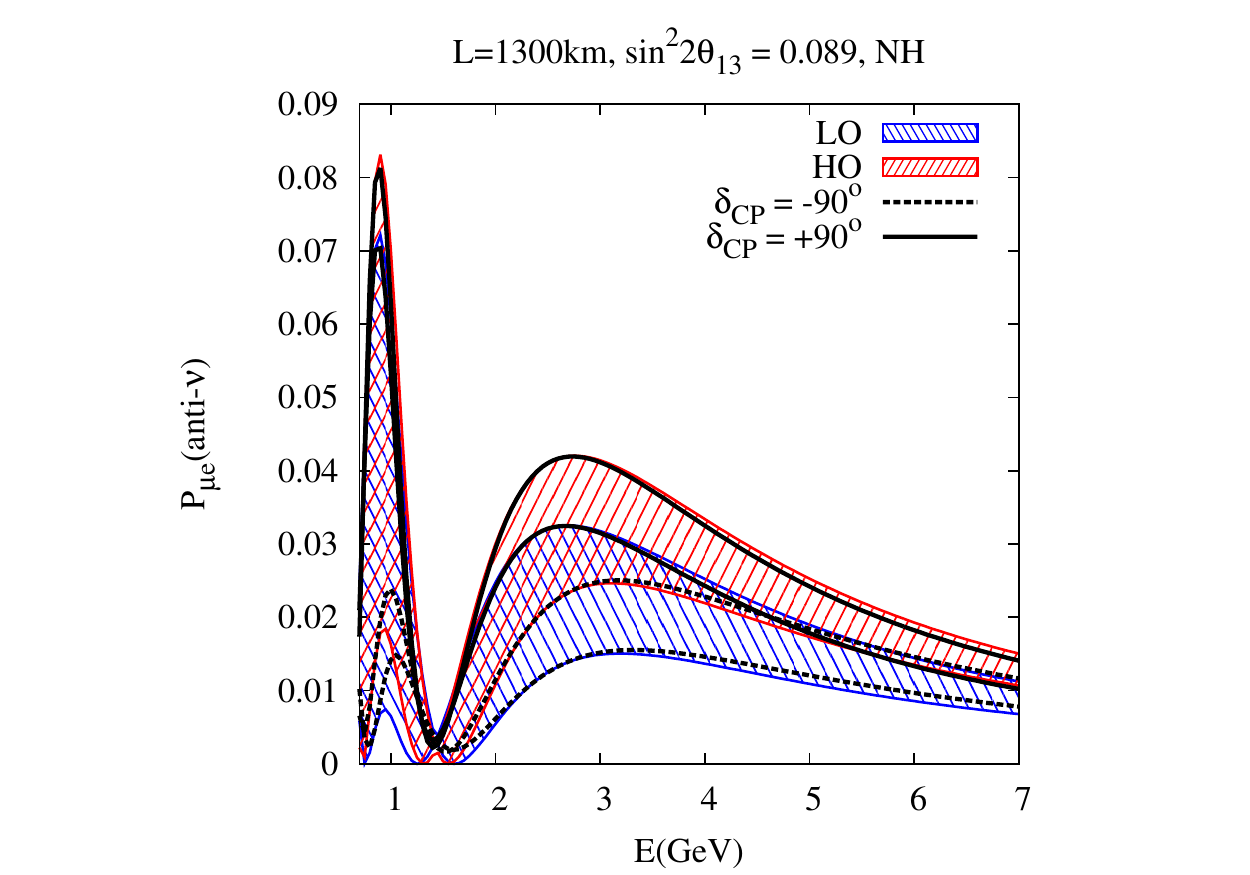}
        
         \end{tabular}
         
        \begin{tabular}{lr}
                 
                \hspace*{-0.85in} \includegraphics[width=0.72\textwidth]{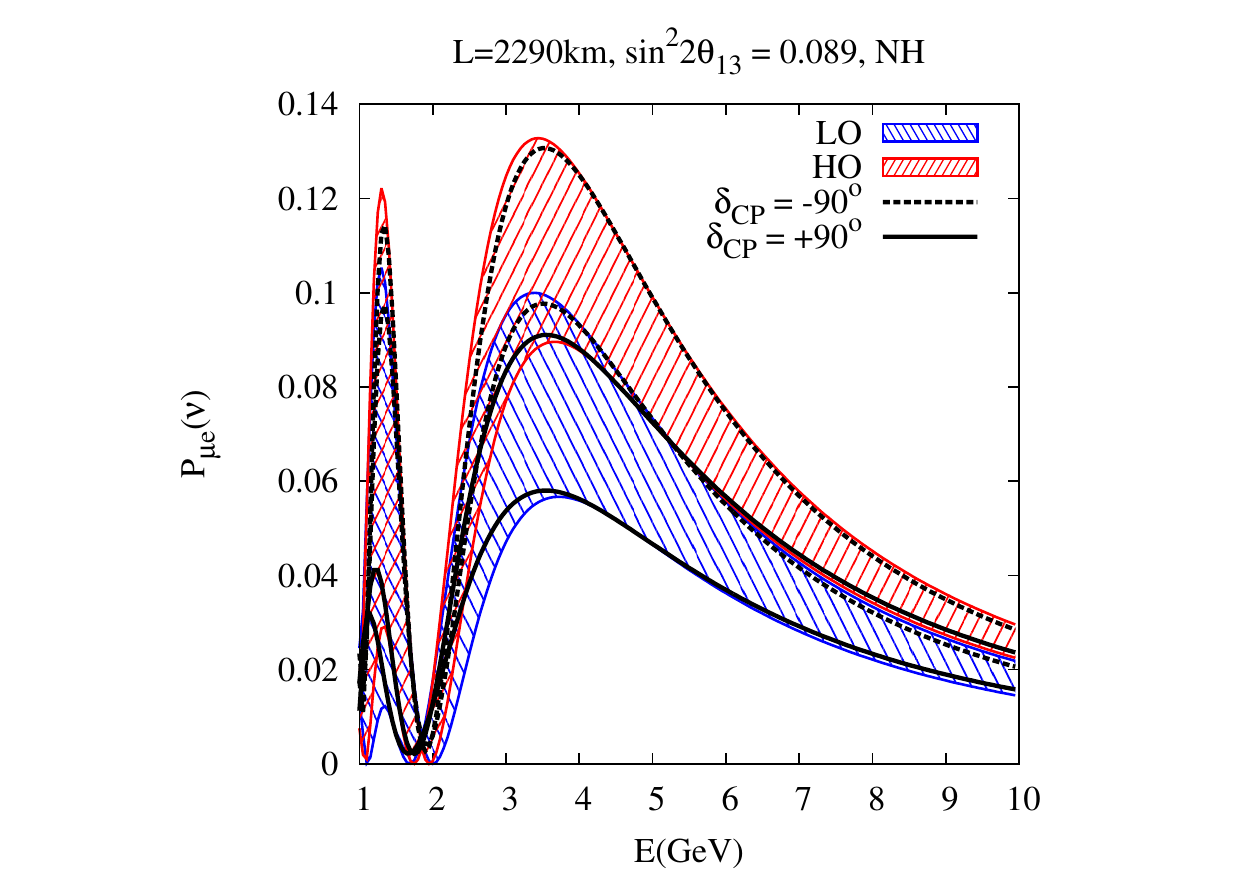}
                &
                \hspace*{-1.7in} \includegraphics[width=0.72\textwidth]{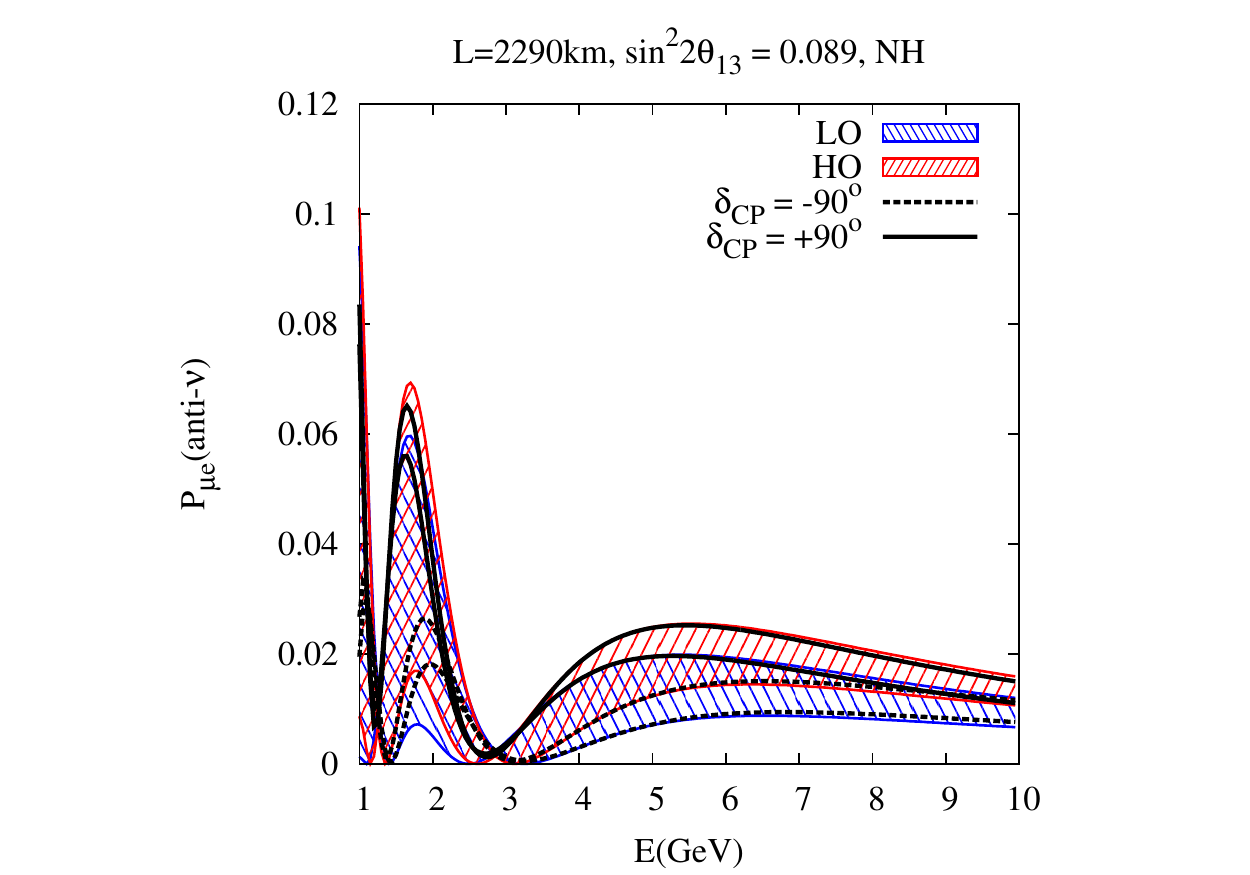}
                
        \end{tabular}

\caption{\footnotesize{$P_{\mu e}$ as a function of neutrino energy. Here, the bands correspond to different
values of $\dcp$ ranging from $-180^\circ$ to $180^\circ$. 
Inside each band, the probability for $\dcp = 90^\circ$ ($\dcp = -90^\circ$) case is shown by the solid (dashed) line.
The red (blue) band is for HO with $\sa = 0.59$ (LO with $\sa = 0.41$).
The left panel (right panel) is for $\nu$ ($\bar\nu$). The upper panels are drawn for the LBNE baseline of 1300 km. 
The lower panels are for the LBNO baseline of 2290 km. Here, we consider $\stch = 0.089$ and NH.}}

\label{octant-dcp-nu-antinu-LBNE-LBNO}
\end{figure}

There is a similar octant-$\dcp$ degeneracy also in the $P_{\mu e}$ channel, which limits our ability to determine the correct 
octant of $\tmt$~\cite{Fogli:1996pv}. The upper left (right) panel of Figure~\ref{octant-dcp-nu-antinu-T2K-NOvA} shows
$P_{\mu e}$ vs. $\textrm{E}_{\nu}$ ($P_{\bar{\mu}\bar{e}}$ vs. $\textrm{E}_{\bar{\nu}}$) for all possible values of $\dcp$ and 
for the two different values of $\sa$, assuming NH to be the true hierarchy. These plots are drawn for the T2K experiment.
The lower panels show the same for the NO$\nu$A baseline.
As can be seen from the upper and lower left panels of Figure~\ref{octant-dcp-nu-antinu-T2K-NOvA}, for neutrino data,
the two octant bands overlap for some values of $\dcp$ and are distinct for other values. The combinations of octant and $\dcp$
which lie farthest from overlap will be favorable combinations for octant determination. For example, LO and $\dcp$ of $90^\circ$
and HO and $\dcp$ of $-90^\circ$ form the favorable combinations. For the combinations with overlap, HO and $\dcp$ of $90^\circ$ 
and LO and $\dcp$ of $-90^\circ$, it is impossible to determine octant using neutrino data alone. However, as we see from the 
upper and lower right panels, these unfavorable combinations for neutrino case are the favorable ones for the anti-neutrino case.
Thus, a combination of neutrino and anti-neutrino data will have a better capability to determine octant compared to neutrino data
alone. This is in contrast to the hierarchy-$\dcp$ degeneracy, where for a given hierarchy, the favorable $\dcp$ region is the same for 
both neutrino and anti-neutrino. Thus, we draw the conclusion that a balanced neutrino and anti-neutrino data is imperative for resolving 
the octant ambiguity of $\tmt$ for all values of $\dcp$~\cite{Agarwalla:2013ju}.
The octant-$\dcp$ degeneracy pattern for the LBNE (LBNO) experiment can be seen from the upper (lower) panels of
Figure~\ref{octant-dcp-nu-antinu-LBNE-LBNO}. 

\section{Present Generation Beam Experiments: T2K \& NO$\nu$A}
\label{sec:current}

With the aim to unravel the $\tet$-driven $\numu \to \nue$ appearance oscillation, the T2K experiment~\cite{Itow:2001ee,Abe:2011ks}
started its journey in 2010 and the NO$\nu$A experiment~\cite{Ayres:2002ws,Ayres:2004js,Ayres:2007tu} in the United States is now
under construction and will start taking data near the end of this year. The detection of electron neutrino appearance in a $\numu$ beam
is the prime goal of these experiments and their experimental setups are optimized to achieve this target. Both the T2K and NO$\nu$A
experiments use the classic off-axis beam technique~\cite{Para:2001cu} that delivers a narrow peak in the energy spectrum, tuned to be
at the expected oscillation maximum. Furthermore, this off-axis technology helps to reduce the background coming from the intrinsic
$\nue$ contamination in the beam and a smaller fraction of high energy tails reduces the background coming from neutral current events.
As a result, it improves the signal-to-background ratio a lot. With the recent discovery of a moderately large value of $\tet$, these
current generation experiments are now poised to probe the impact of full three-flavor effects to discover neutrino mass hierarchy,
CP violation and octant of $\tmt$. But, to achieve these goals, they need to have very high proton beam powers of order 1 MW and 
detectors with huge fiducial masses (of order 10 kilotons) and therefore, these experiments are knows as `superbeam' experiments. 
Next, we briefly describe the main features of the T2K and NO$\nu$A experiments and then we present the physics reach of these
experiments in light of the recently discovered moderately large value of $\tet$.

\subsection{T2K}
\label{subsec:t2k}

T2K uses the 50 $\mathrm{kilotons}$ Super-Kamiokande water Cherenkov detector (fiducial volume 22.5 $\mathrm{kilotons}$) 
as the far detector for the neutrino beam from J-PARC. The detector is at a distance of 295 $\mathrm{km}$ from the source at an 
off-axis angle of $2.5^\circ$~\cite{Itow:2001ee}. The neutrino flux is peaked sharply at the first oscillation maximum of 
$0.6$ $\mathrm{GeV}$. The experiment is scheduled to run for 5 years in the neutrino mode with a power of $0.75$ MW. 
Because of the low energy of the peak flux, the neutral current backgrounds are small and they can be rejected based on energy cut. 
The signal efficiency is $87\%$. To estimate the physics sensitivity, the background information and other details are taken 
from~\cite{fechnerthesis,Huber:2009cw}.

\subsection{NO$\nu$A}
\label{subsec:nova}

NO$\nu$A is a 14 $\mathrm{kilotons}$ totally active scintillator detector (TASD) placed at a distance of 810 $\mathrm{km}$ from Fermilab, 
at a location which is $0.8^\circ$ off-axis from the NuMI beam. Because of the off-axis location, the flux of the neutrinos is reduced 
but is sharply peaked around 2 $\mathrm{GeV}$, again close to the first oscillation maximum energy of 1.7 GeV in $P(\numu \to \nue)$.
The most problematic background in NO$\nu$A experiment is neutral current interactions which mostly consist of the single $\pi^0$ production.
However, the measured energy of this background is shifted to values of energy below the region where the flux is significant. 
Hence this background can be rejected using a simple kinematic cut. The experiment is scheduled to have three years run in neutrino 
mode first and then later, three years run in anti-neutrino mode as well with a NuMI beam power of $0.7$ MW, corresponding to 
$6\times 10^{20}$ protons on target per year. The details of the experiment are given in~\cite{Ayres:2007tu}. In light of the recent measurement 
of large $\tet$, NO$\nu$A has reoptimized its event selection criteria. Relaxing the cuts, they now allow more events in both signal and background. 
Additional neutral current backgrounds are reconstructed at lower energies and can be managed by a kinematical cut.
In our calculations, we use these reoptimized values of signal and background, the details of which are given 
in~\cite{Patterson:2012zs,Agarwalla:2012bv}.

\subsection{Mass Ordering and CP Violation Discovery}
\label{subsec:mh-cpv}

In this section, we describe the capabilities of the T2K and NO$\nu$A experiments for the determination of mass hierarchy and CP violation.
We use GLoBES~\cite{Huber:2004ka,Huber:2007ji} software to simulate the data for these experiments. For the atmospheric/accelerator 
neutrino parameters, we take the following central (true) values:
\beq
|\Delta m^2_{\mathrm{eff}}| = 2.4 \cdot 10^{-3}\,\mathrm{eV}^2, \quad \sin^2 2 \theta_{23} = 1.0,
\label{eq:atmospheric-parameters}
\eeq
where $\Delta m^2_{\mathrm{eff}}$ is the effective mass-squared difference measured by the accelerator experiments in 
$\numu \rightarrow \nu_\mu$ disappearance channel~\cite{Nichol:2013caa,Adamson:2013whj}. It is related to the 
$\ma$ (larger) and $\ms$ (smaller) mass-squared differences through the expression~\cite{Nunokawa:2005nx,deGouvea:2005hk}
\beq
\Delta m^2_{\mathrm{eff}} = \Delta m^2_{31} - \Delta m^2_{21} (\cos^2 \theta_{12} - \cos \dcp \sin \theta_{13} \sin 2 \theta_{12} \tan \theta_{23}).
\label{effective-atmospheric-splitting}
\eeq
The value of $\ma$ is calculated separately for NH and for IH using this equation where $\Delta m^2_{\mathrm{eff}}$ is taken to be +ve for NH 
and -ve for IH. For $\tmt$, we take the maximal mixing as still favored by the Super-Kamiokande atmospheric 
data~\cite{Wendell:2013kxa,Itow:2012,Wendell:2010md}. For $\theta_{13}$, we take the best-fit value of $\sch = 0.026$.
The uncertainties in the above parameters are taken to be $\sigma(\sch) = 13\%$~\cite{Tortola:2012te}, 
$\sigma(|\Delta m^2_{\mathrm{eff}}|) = 4\%$, and $\sigma(\sin^2 2 \theta_{23}) = 2\%$~\cite{Itow:2001ee}. In the calculation, 
these informations are included in the form of priors. In our $\chi^2$ fit, we marginalize over {\it all} oscillation parameters within 
their $\pm~3\sigma$ ranges, as well as the mass hierarchy, by allowing these parameters to vary in the fit and picking the smallest 
value of the $\chi^2$ function. We take the solar parameters to be
\beq 
\ms = 7.62\cdot10^{-5}\,\mathrm{eV}^2,  \quad \sss = 0.32.
\label{eq:solar-parameters}
\eeq
We keep the solar parameters to be fixed throughout the calculation because varying them will have negligible effect.
We also take the Earth matter density to be a constant $2.8$ g/cm$^3$ because the variations and the uncertainties in 
density can be neglected for the T2K and NO$\nu$A baselines. While calculating the sensitivity for T2K, we include a 2\% 
systematic error on appearance signal events and a (uncorrelated) 5\% systematic error on backgrounds. 
For NO$\nu$A, we have assumed 5\% systematic error on appearance signal events and a (uncorrelated) 10\% systematic 
error on background events. These informations on the systematic errors are included in the $\chi^2$ function using the
pull method as described in {\it e.g.} references~\cite{Huber:2002mx,Fogli:2002au}. In our definition of the $\chi^2$ function, 
we have assumed that the neutrino and anti-neutrino channels are completely uncorrelated, all the energy bins for a given channel 
are fully correlated, and the systematic errors on signal and background are fully uncorrelated. 
We perform the usual $\chi^2$ analysis using a Poissonian likelihood function adding the information coming from
$\nue$ appearance and $\numu$ disappearance channel.

For long-baseline experiments, the measurement of the mass hierarchy is easier than a measurement of $\dcp$ because matter effects 
enhance the separation between the oscillation spectra, and therefore the event rates, of a NH and an IH. Additionally, this measurement 
is one that is `discrete' as we only need to differentiate between two possibilities. A `discovery' of the mass hierarchy is defined as the ability 
to exclude any degenerate solution for the wrong (fit) hierarchy at a given confidence level. A `discovery' of CP violation, if it exists, means 
being able to exclude the CP-conserving values of $0^{\circ},\,180^{\circ}$ at a given confidence level.
Clearly, this measurement becomes very difficult for the $\dcp$ values which are closer to $0^{\circ},\,180^{\circ}$.
Therefore, whilst it is possible to discover the mass hierarchy for \emph{all} possible values of $\dcp$, the same is not true for CP violation.

\begin{figure}[tp]
\centering
\includegraphics[width=0.49\textwidth]{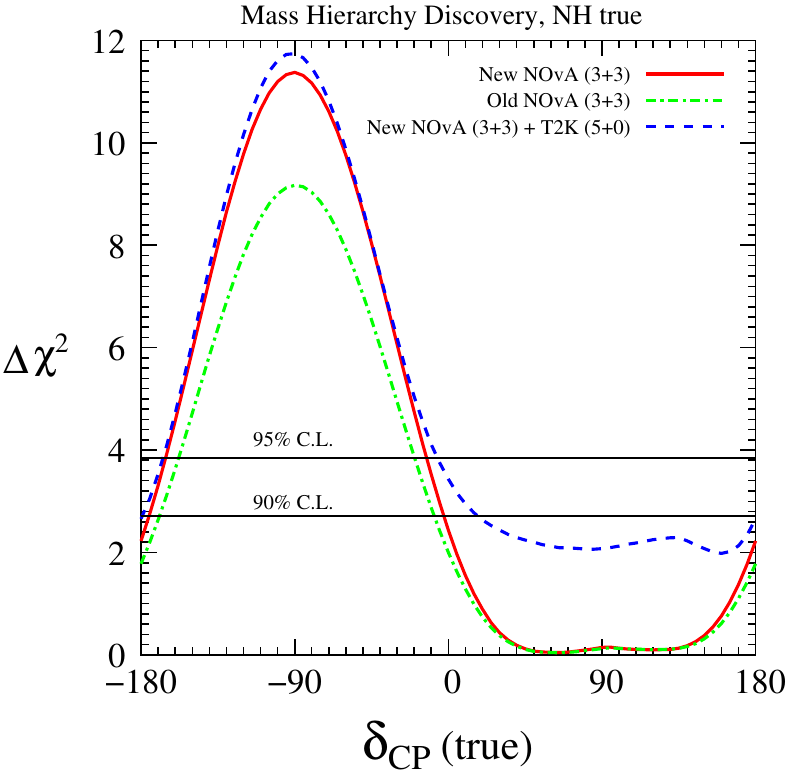}
\includegraphics[width=0.49\textwidth]{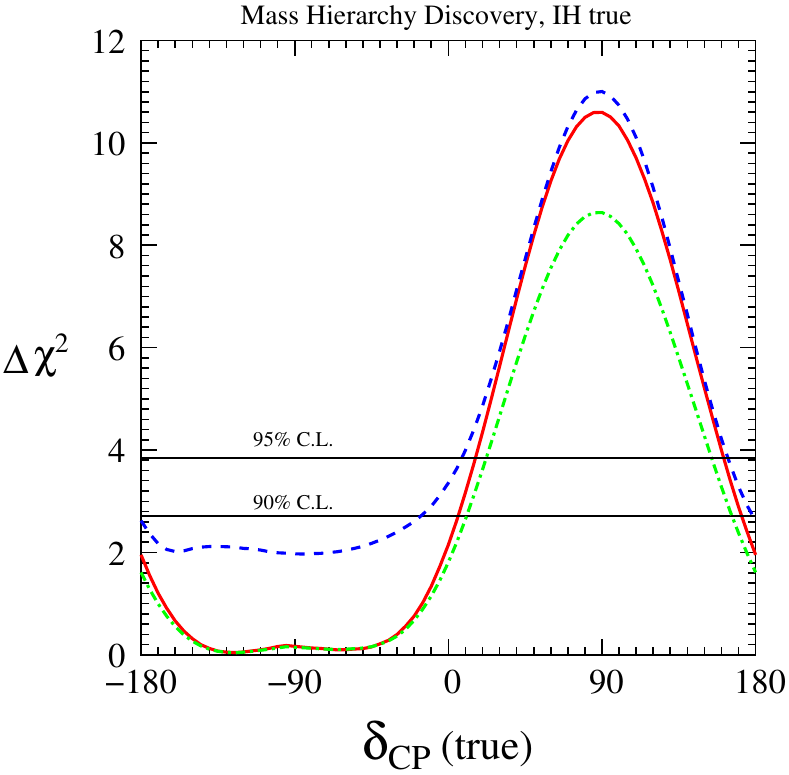}
\caption{\footnotesize{Left panel (right panel) shows the $\Delta\chi^2$ for the mass hierarchy discovery as a function of true 
value of $\dcp$ assuming NH (IH) as true hierarchy. This Figure has been taken from reference~\cite{Agarwalla:2012bv}.}}
\label{fig:mh-discovery}
\end{figure}

In Figure~\ref{fig:mh-discovery}, we plot the hierarchy discrimination sensitivity of the old NO$\nu$A, the new NO$\nu$A 
(with reoptimized event selection criteria for large $\tet$), and the combined sensitivity of new NO$\nu$A and T2K, as a function 
of the true value of $\dcp$. In the left (right) panel, we have assumed NH (IH) to be the true hierarchy. We see that the wrong hierarchy 
can be ruled out very effectively for $\dcp$ in the favorable half-plane, which is LHP (UHP) for NH (IH).
The new event selection criteria of NO$\nu$A make the experiment even more effective in ruling out the wrong hierarchy for $\dcp$ in
the favorable half-plane. In the unfavorable half-plane, both the old and the new criteria are equally ineffective. However, the addition
of T2K data improves the situation significantly and $\Delta\chi^2$ increases from 0 to $\geq 2$ for all the true values of $\dcp$, thus 
making it possible to get a $90\%$ C.L. hint of hierarchy with some additional data. We have checked that a further increment in the
exposure of T2K or addition of anti-neutrino data from T2K does not improve the hierarchy sensitivity much.

The prospects of determining the neutrino mass hierarchy with the combined data from T2K, NO$\nu$A, Double Chooz, RENO, 
Daya Bay, and the atmospheric neutrino experiment ICAL@INO~\cite{INO} have been studied in detail in reference~\cite{Ghosh:2012px}. 
With 10 years of atmospheric ICAL@INO data collected by 50 kilotons magnetized iron calorimeter detector combined with T2K, NOvA, 
and reactor data, a $2.3\sigma$ -- $5.7\sigma$ discovery of the neutrino mass hierarchy could be achieved depending on the true 
values of $\sa$ [0.4 -- 0.6], $\stch$ [0.08 -- 0.12], and $\dcp$ [0 -- 2$\pi$]~\cite{Ghosh:2012px}.

\begin{figure}[tp]
\centering
\includegraphics[width=0.49\textwidth]{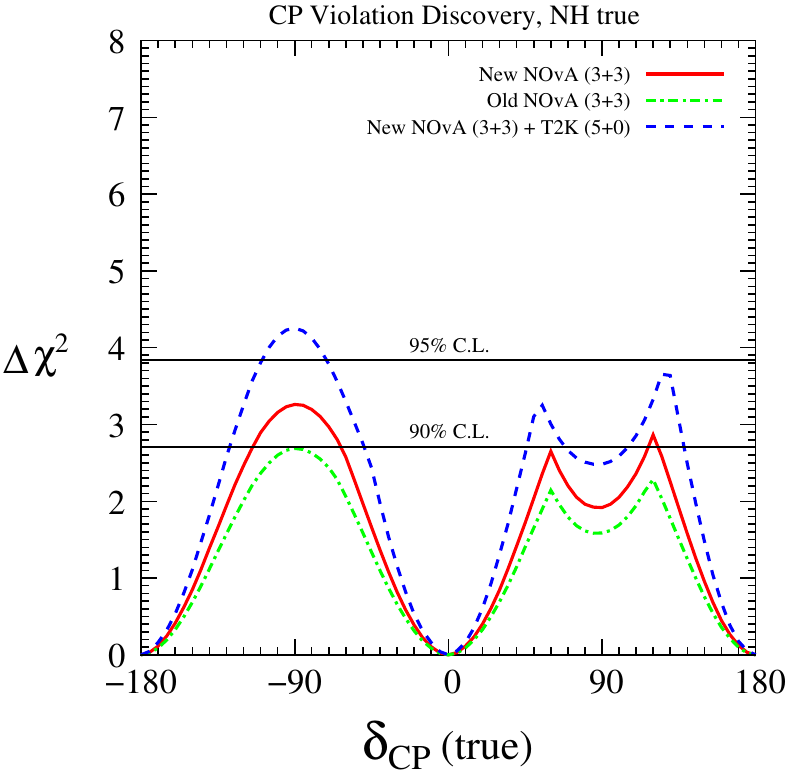}
\includegraphics[width=0.49\textwidth]{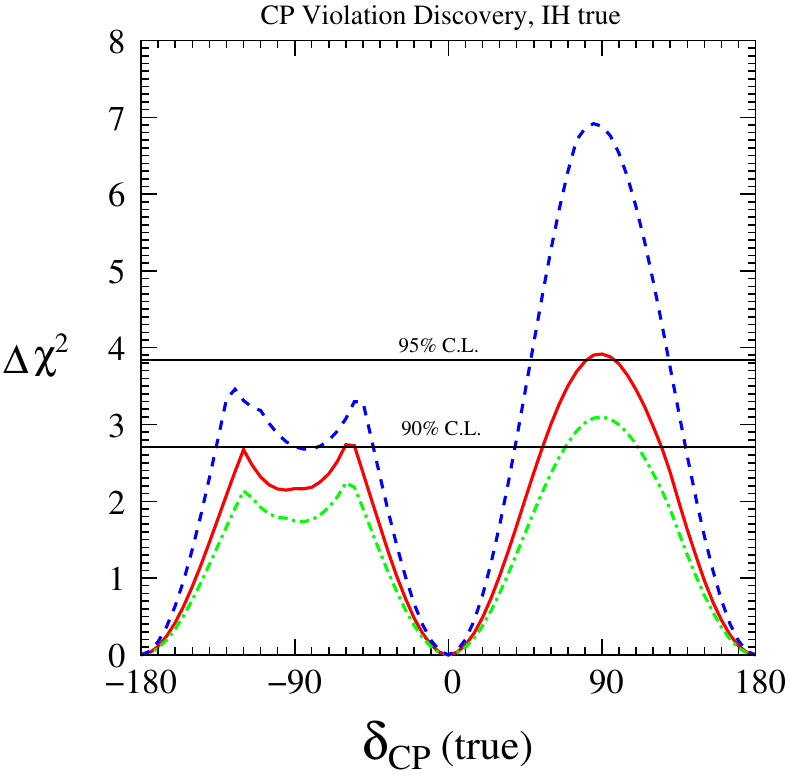}
\caption{\footnotesize{Left panel (right panel) shows the $\Delta\chi^2$ for the CP violation discovery as a function of true 
value of $\dcp$ assuming NH (IH) as true hierarchy. This Figure has been taken from reference~\cite{Agarwalla:2012bv}.}}
\label{fig:cp-discovery}
\end{figure}

Reoptimization of the event selection criteria of NO$\nu$A has the most dramatic effect on the CP violation discovery potential
of the experiment. In Figure~\ref{fig:cp-discovery}, we plot the sensitivity to rule out the CP conserving scenarios, as a function 
of true $\dcp$ in the left (right) panel for NH (IH) being the true hierarchy. We notice that, while in the case of old NO$\nu$A there
is no CP violation sensitivity at all at 90\% C.L., there is such a sensitivity in new NO$\nu$A, for about one third fraction of the 
favorable half-plane. Addition of T2K data leads to CP violation sensitivity for about half the region in both favorable half planes 
at 90\% confidence level. It can be shown that, T2K by itself, has no CP violation sensitivity. But, the synergistic combination of 
NO$\nu$A and T2K leads to much better CP violation sensitivity compared to the individual capabilities. Here, we would like to 
mention that a large value of $\tet$ always does not help for CP violation discovery. As $\tet$ becomes large, the number of
electron appearance event increases, reducing the statistical error. However, the large atmospheric term acts as a background
in the measurement of CP phase. In fact, the CP asymmetry term is proportional to $1/\sin2\tet$~\cite{Dick:1999ed,Donini:1999jc}.
These two contradictory issues make the measurement of CP phase quite complicated. The systematic uncertainties are also going 
to play a crucial role for CP violation discovery in light of large $\tet$~\cite{Coloma:2012ji}.

A summary of our results is given in Table~\ref{tab:compare1} in terms of the fraction of $\dcp$ values for which mass hierarchy can be 
determined/CP violation can be detected. Please note that in deriving the results given in Table~\ref{tab:compare1}, we have considered
the best-fit value of $\sch = 0.026$ and maximal mixing for $\tmt$. In Table~\ref{tab:compare2}, we present the same for the rather
conservative choices of neutrino mixing angles: $\sch = 0.023$ (the best-fit value suggested by the Daya Bay experiment) and 
$\sa = 0.413$ (the LO value of $\sa$ as indicated by the recent MINOS accelerator data). 

\begin{table}[tp]
\begin{center}
\begin{tabular}{|c|c|c||c|c|} \hline\hline
\multirow{4}{*}{Setups} & \multicolumn{4}{c|}{{\rule[0mm]{0mm}{6mm}Fraction of $\dcp$(true)}}
\cr\cline{2-5}
& \multicolumn{2}{c||}{{\rule[0mm]{0mm}{4mm}{\bf Mass Hierarchy}}} & \multicolumn{2}{c|}{\rule[0mm]{0mm}{4mm}{{\bf CP violation}}}
\cr
\cline{2-5}
& NH true & IH true & NH true & IH true \cr
\hline
 NO$\nu$A (3+3) & 0.48 (0.43) & 0.46 (0.41) & 0.16 (0) & 0.21(0.04) \cr
\hline
 NO$\nu$A (3+3) + T2K (5+0) & 0.55 (0.45) & 0.54 (0.43) & 0.38 (0.11) & 0.49 (0.23) \cr
\hline\hline
\end{tabular}
\caption{\footnotesize{Fractions of true values of $\dcp$ for which a discovery is possible for mass hierarchy and CP violation.
The numbers without (with) parentheses correspond to 90\% (95\%) C.L. Here we take the central values: $\sch = 0.026$ and
$\sa = 0.5$. The results are shown for both NH and IH as true hierarchy.
This table has been taken from reference~\cite{Agarwalla:2012bv}.}}
\label{tab:compare1}
\end{center}
\end{table}

\begin{table}[tp]
\begin{center}
\begin{tabular}{|c|c|c||c|c|} \hline\hline
\multirow{4}{*}{Setups} & \multicolumn{4}{c|}{{\rule[0mm]{0mm}{6mm}Fraction of $\dcp$(true)}}
\cr\cline{2-5}
& \multicolumn{2}{c||}{{\rule[0mm]{0mm}{4mm}{\bf MH}}} & \multicolumn{2}{c|}{\rule[0mm]{0mm}{4mm}{{\bf CPV}}}
\cr
\cline{2-5}
& NH true & IH true & NH true & IH true \cr
\hline
 NO$\nu$A (3+3) & 0.39 (0.33) & 0.37 (0.31) & 0.2 (0.1) & 0.22 (0.13) \cr
\hline
 NO$\nu$A (3+3) + T2K (5+0) & 0.41 (0.34) & 0.39 (0.31) & 0.28 (0.22) & 0.3 (0.25) \cr
\hline\hline
\end{tabular}
\caption{\footnotesize{Fractions of true values of $\dcp$ for which a discovery is possible for mass hierarchy and CP violation.
The numbers without (with) parentheses correspond to 90\% (95\%) C.L. Here we take the central value for $\sch$ to be 0.023
as predicted by Daya Bay. For $\sa$, the best-fit value that we consider is 0.413.
The results are shown for both NH and IH as true hierarchy.
This table has been taken from reference~\cite{Agarwalla:2012bv}.}}
\label{tab:compare2}
\end{center}
\end{table}

In~\cite{Agarwalla:2012bv}, we further explore the improvement in the hierarchy and CP violation sensitivities for T2K and NO$\nu$A 
due to the addition of a $10\,\mathrm{kilotons}$ LArTPC placed close to NO$\nu$A site and exposed to the NuMI beam during NO$\nu$A running. 
It is expected, of course, that such a detector will come on line much later than NO$\nu$A. The capabilities of such a detector are equivalent 
to those of NO$\nu$A in all respects. We find that combined data from $10\,\mathrm{kilotons}$ LArTPC (3 years of $\nu$ + 3 years of $\bar\nu$ run),
NO$\nu$A (6 years of $\nu$ + 6 years of $\bar\nu$ run) and T2K (5 years of $\nu$ run) can give a close to $2\sigma$ hint of hierarchy discovery 
for all values of $\dcp$. With this combined data, we can achieve CP violation discovery at 95\% C.L. for roughly 60\% values of $\dcp$.
A similar proposal considering 6 kilotons LArTPC has been studied recently in detail in reference~\cite{Adamson:2013jsa}.

\subsection{Resolving the Octant ambiguity of $\tmt$}
\label{subsec:octant}

In this section, we study whether the expected appearance data from the ongoing T2K experiment and the upcoming NO$\nu$A
experiment can resolve the octant ambiguity of $\tmt$ or not?

\begin{figure}[tp]

        \begin{tabular}{lr}
                \hspace*{-0.85in} \includegraphics[width=0.72\textwidth]{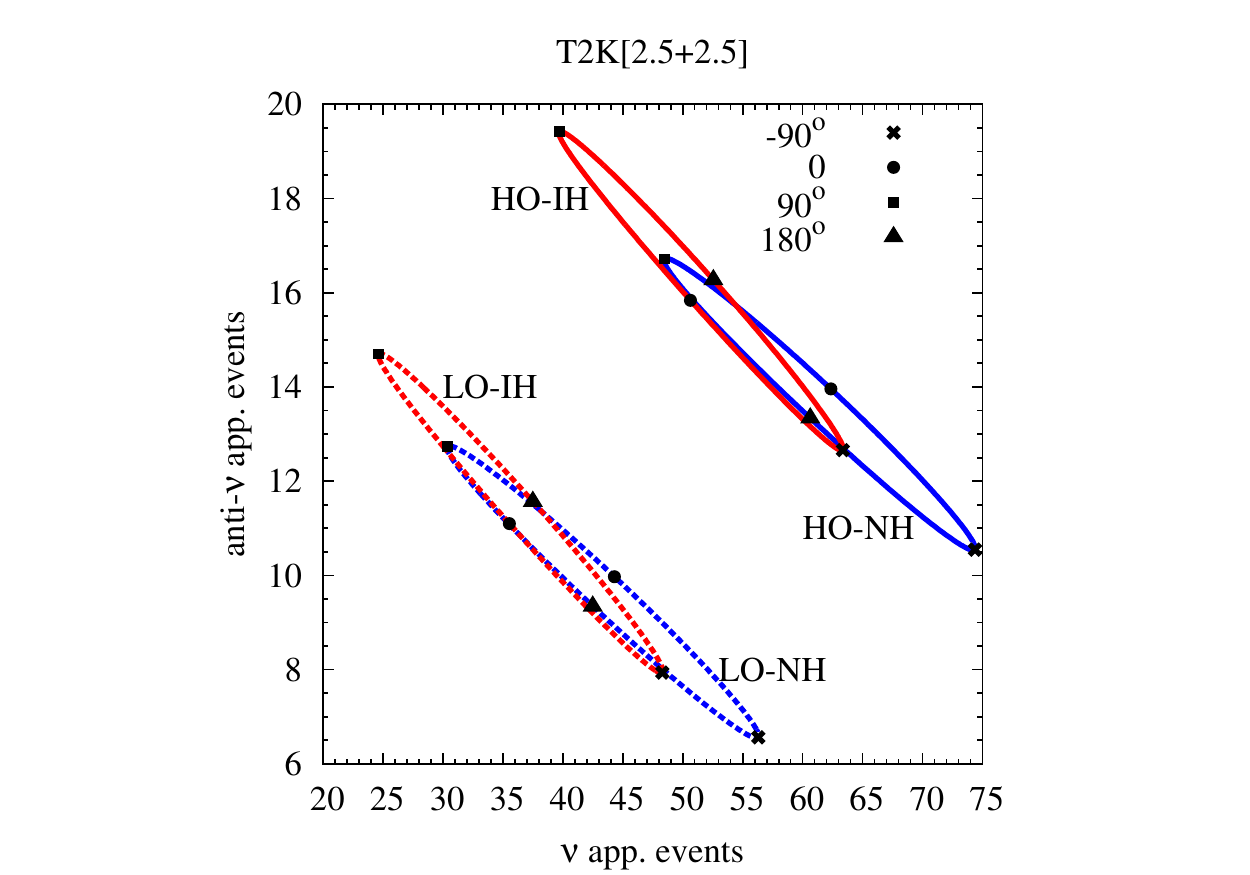}
                &
                \hspace*{-1.7in} \includegraphics[width=0.72\textwidth]{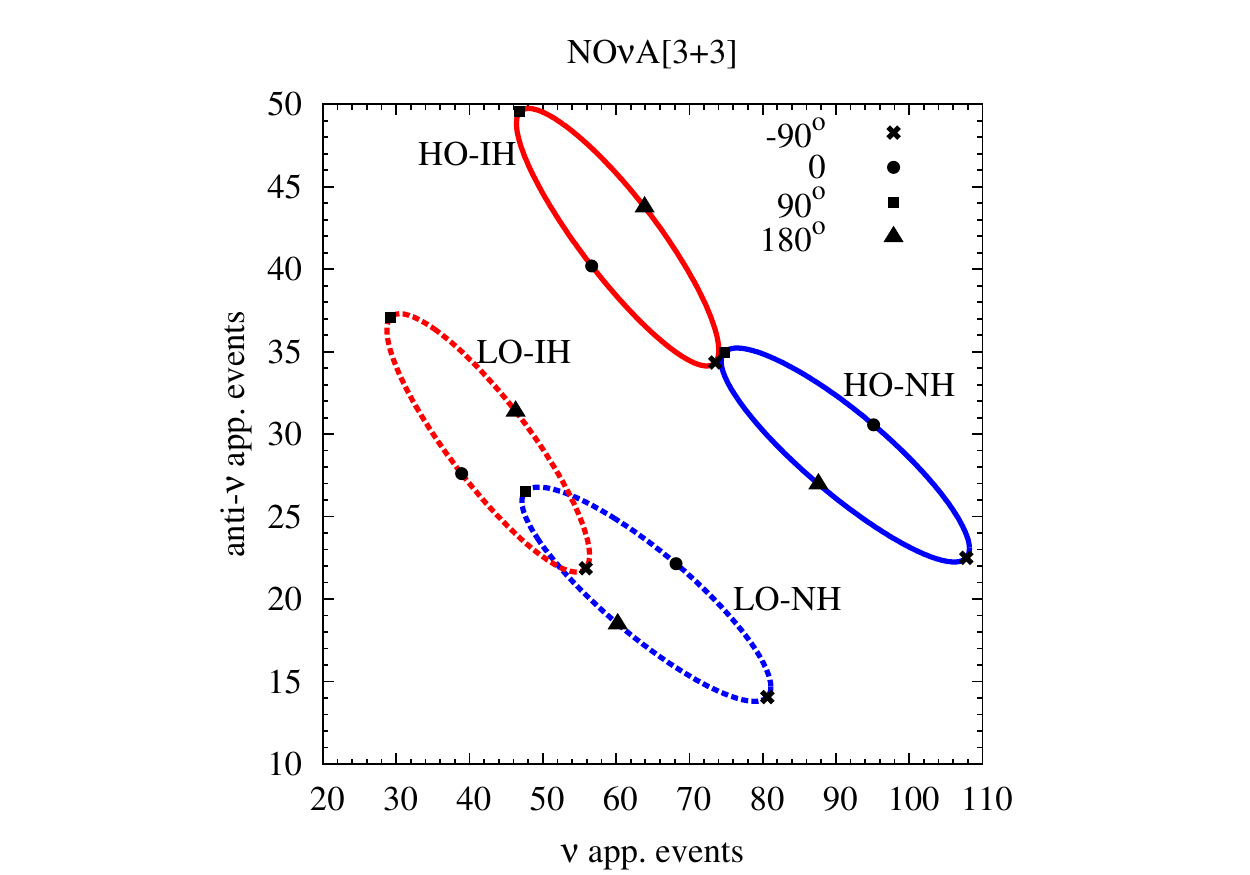}
        \end{tabular}

\caption{\footnotesize{Neutrino and anti-neutrino appearance events for all possible combinations of hierarchy, octant and $\dcp$. 
The left (right) panel is for T2K (NO$\nu$A). Here $\sin^22\tet = 0.089$. For LO (HO), $\sin^2\tmt = 0.41~(0.59)$. Note that for T2K,
equal $\nu$ and $\bar\nu$ runs of 2.5 years each has been assumed. This Figure has been taken from reference~\cite{Agarwalla:2013ju}.}}
\label{iso-contour-event}
\end{figure}

In Figure~\ref{iso-contour-event}, we show neutrino events vs. anti-neutrino events for various octant-hierarchy combinations. 
In each case, with varying values of $\dcp$, the plot becomes an ellipse. The left panel depicts these ellipses for T2K whereas the right 
panel shows the same for NO$\nu$A. Here, we assume that T2K will have equal $\nu$ and $\bar\nu$ runs of 2.5 years each. 
In the right panel, we see that the ellipses for the two mass orderings overlap whereas the ellipses of LO are well separated 
from those of HO. Hence, we can expect that NO$\nu$A. will have better octant resolution capability than hierarchy discrimination. 
This situation is even more dramatic in the left panel where there is large overlap between the two hierarchies but clear separation
between the octants. Thus, it is very likely that anti-neutrino data from T2K may play an important role in the determination of octant.

\begin{figure}[tp]

        \begin{tabular}{lr}
                \hspace*{-0.85in} \includegraphics[width=0.72\textwidth]
                {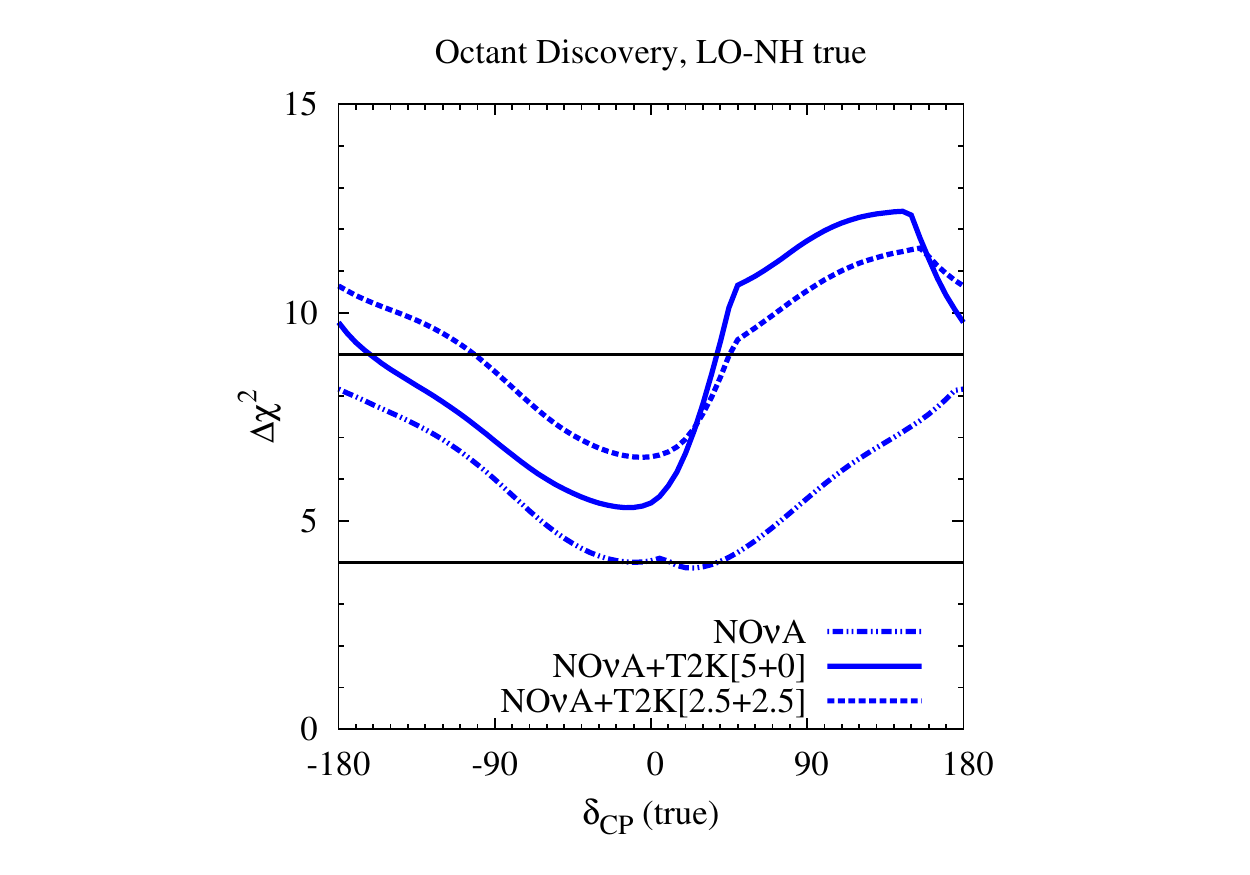}
                &
                \hspace*{-1.7in} \includegraphics[width=0.72\textwidth]
                {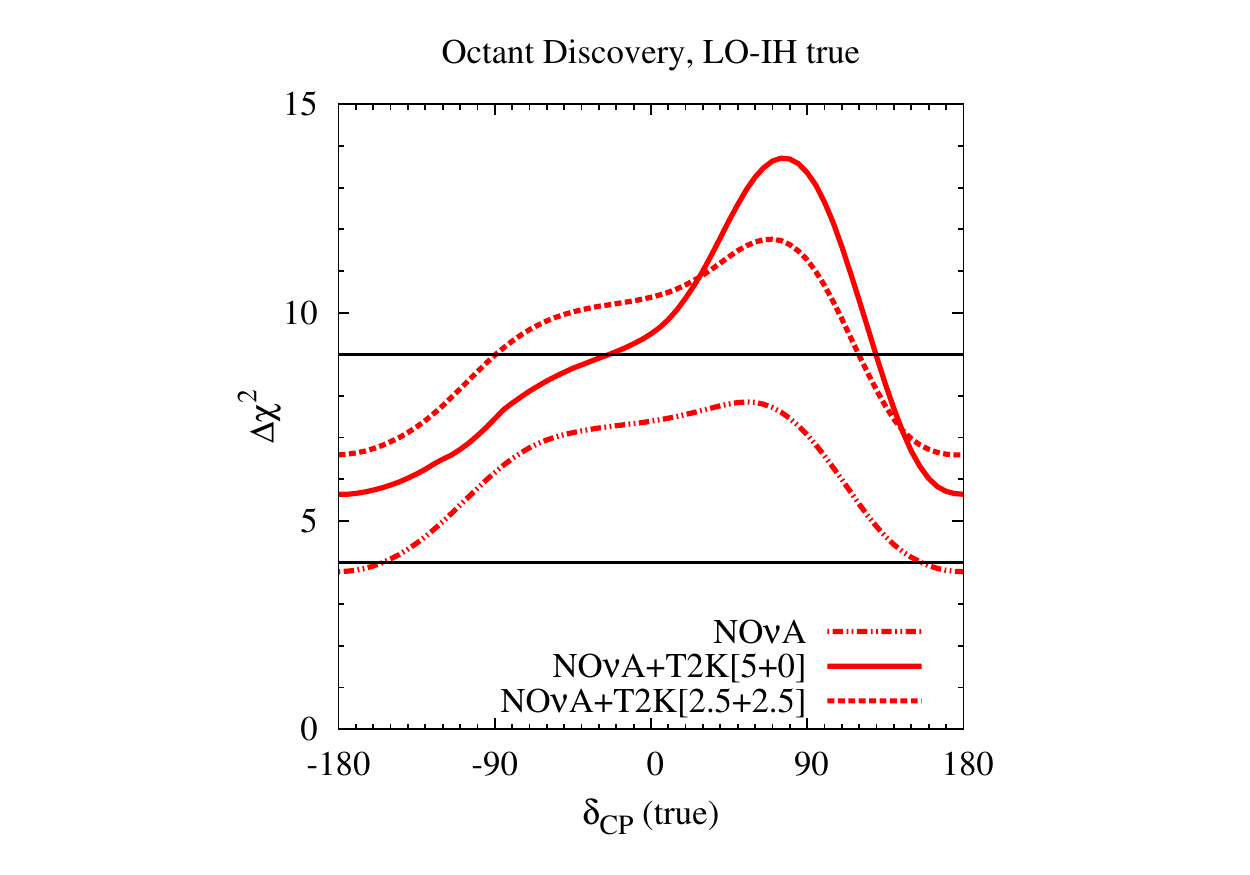}
        \end{tabular}

\caption{\footnotesize{Octant resolving capability as a function of true $\dcp$ for various set-ups . In these plots, LO is assumed to be
the true octant. The left (right) panel corresponds to NH (IH) being the true hierarchy. This Figure has been taken from reference~\cite{Agarwalla:2013ju}.}}

\label{fig:dchsqLO}
\end{figure}

\begin{figure}[tp]

        \begin{tabular}{lr}
                \hspace*{-0.85in} \includegraphics[width=0.72\textwidth]
                {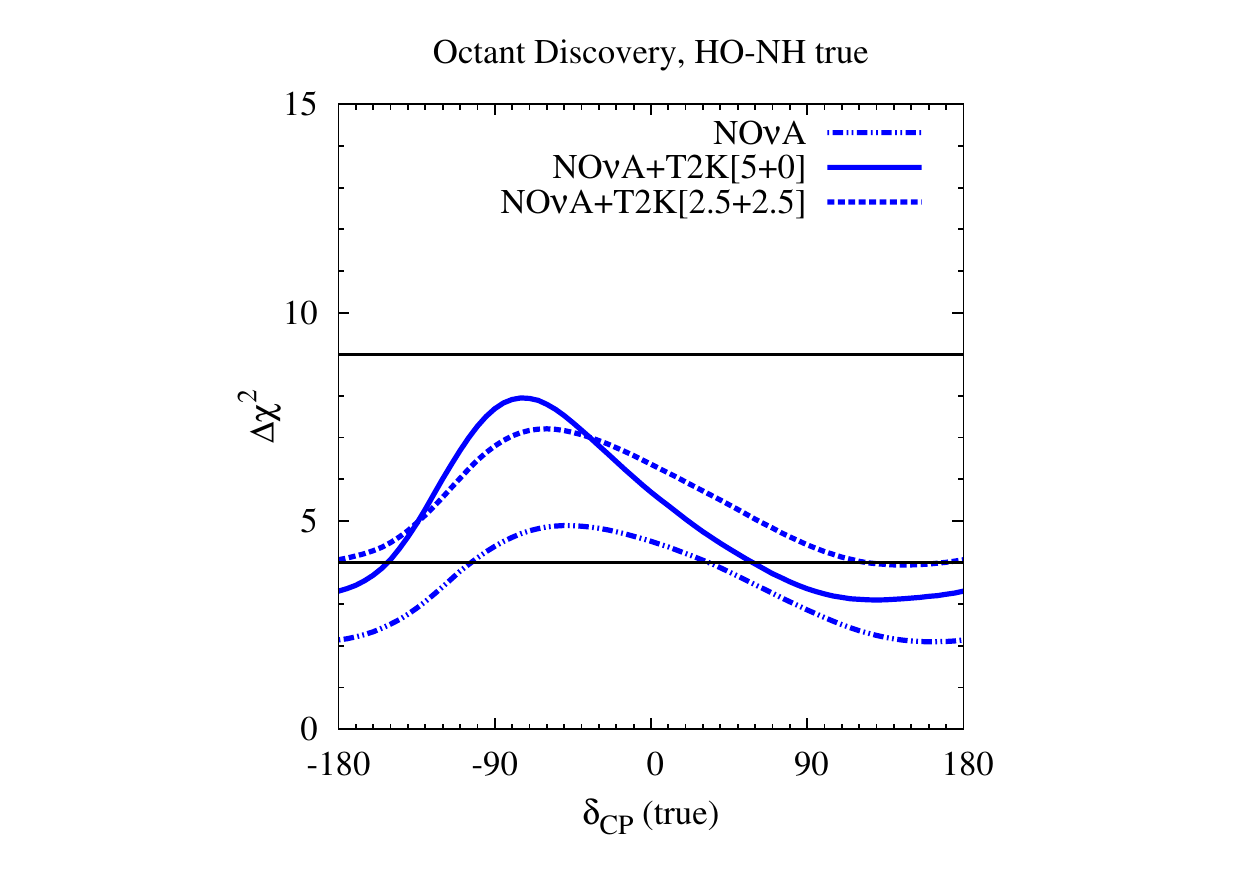}
                &
                \hspace*{-1.7in} \includegraphics[width=0.72\textwidth]
                {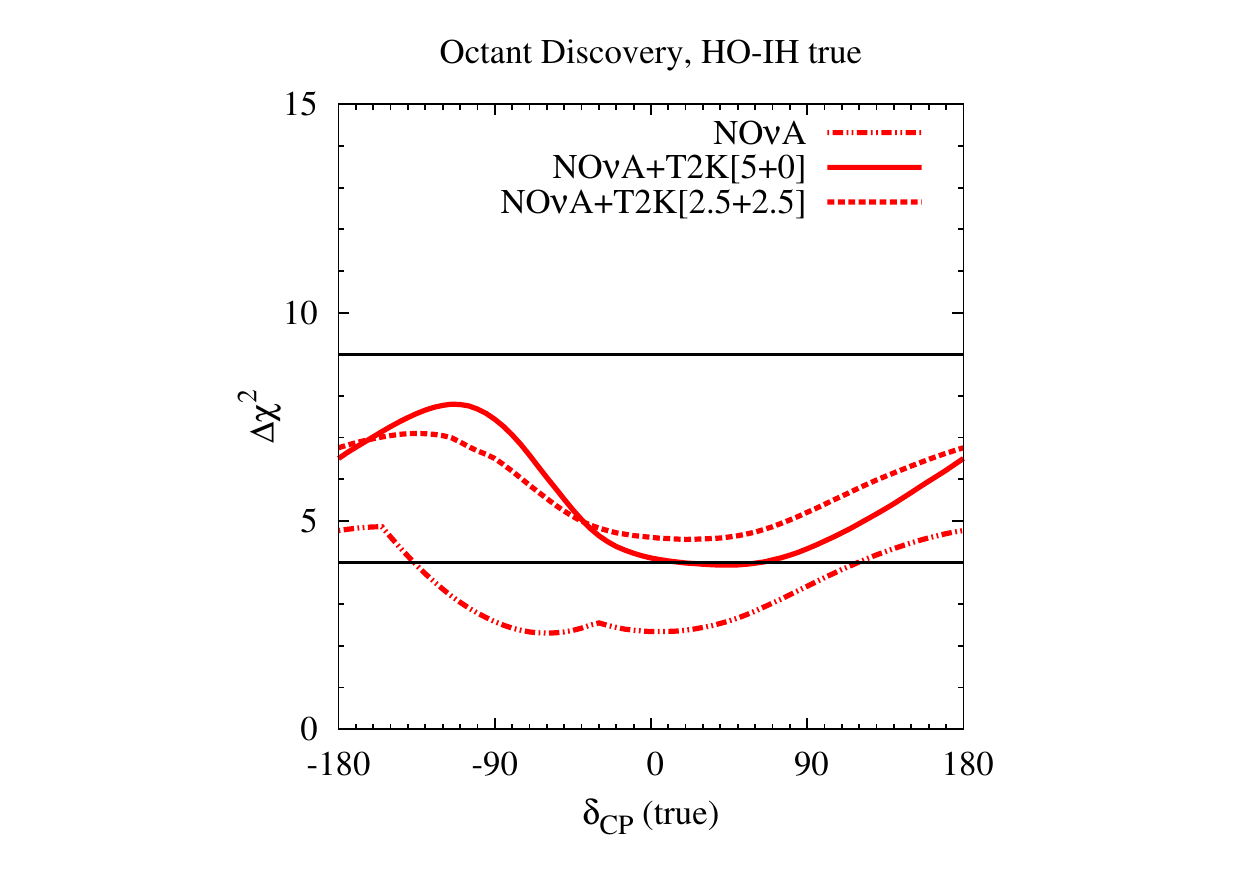}
        \end{tabular}

\caption{\footnotesize{Octant resolving capability as a function of true $\dcp$ for various set-ups . In these plots, HO is assumed to be
the true octant. The left (right) panel corresponds to NH (IH) being the true hierarchy. This Figure has been taken from reference~\cite{Agarwalla:2013ju}.}}

\label{fig:dchsqHO}
\end{figure}

In Figures~\ref{fig:dchsqLO} and \ref{fig:dchsqHO}, we study the behavior of $\Delta\chi^2$ between the true and the wrong octants as a function 
of true $\dcp$. Here, the $\Delta\chi^2$ is estimated in the following way. First, we fix the true value of $\dcp$. We take $\sin^2\tmt$ to be its best-fit 
value in the true octant: 0.41 for LO and 0.59 for HO. If the LO (HO) is the true octant, the test values of $\sin^2\tmt$ in the HO (LO) are varied within
the range $[0.5,0.63]$ ($[0.36,0.5]$), where 0.63 (0.36) is the $2\sigma$ upper (lower) limit of the allowed range of $\sin^2\tmt$.
The $\Delta\chi^2$ is computed between the spectra with the best-fit $\sin^2\tmt$ of the true octant and that with various test values in the wrong octant 
and is marginalized over other neutrino parameters, especially the hierarchy, $\sin^22\tet$ and $\dcp$.
Figures~\ref{fig:dchsqLO} and \ref{fig:dchsqHO} portray the minimum of this $\Delta\chi^2$ vs. the true value of $\dcp$.

From Figure~\ref{fig:dchsqLO}, we observe that the NO$\nu$A data by itself can almost rule out the wrong octant at $2\sigma$, if LO is the true octant. 
If HO is the true octant, then NO$\nu$A data is not sufficient to rule out the wrong octant as can be seen from Figure~\ref{fig:dchsqHO}.
In fact, the wrong octant can be ruled out only for about half of the true $\dcp$ values. As illustrated in Figures~\ref{fig:dchsqLO} and \ref{fig:dchsqHO}, 
addition of T2K data improves the octant determination ability significantly. From Figure~\ref{fig:dchsqLO}, we see that the combined data from NO$\nu$A 
and T2K (5 years $\nu$ run) give a 2$\sigma$ octant resolution for all values of true $\dcp$ if LO is the true octant. From Figure~\ref{fig:dchsqHO},
we see that this combined data can rule out the wrong octant at $2\sigma$ for HO-IH, but not for HO-NH. The problem of HO-NH can be solved 
if the T2K has equal $\nu$ and $\bar\nu$ runs of 2.5 years each. This change improves the octant determination for the unfavorable values of 
true $\dcp$ (where $\Delta\chi^2$ is minimum) for all four combinations of hierarchy and octant. In particular, for the case of HO-NH, it leads to a complete 
ruling out of the wrong octant at $2\sigma$ for all values of true $\dcp$. Thus, balanced runs of T2K in $\nu-\bar\nu$ mode is preferred over a pure $\nu$ 
run because of better octant determination capability. We observe that this feature of LO being more favorable compared to HO is a consequence of marginalization over the oscillation parameters (mainly $\dcp$) and the systematic uncertainties. We checked that in the absence of any kind of 
marginalization $\Delta\chi^2_{\textrm{HO}}$ is consistently larger than $\Delta\chi^2_{\textrm{LO}}$.

\begin{figure}[tp]

        \begin{tabular}{lr}
                \hspace*{-0.85in} \includegraphics[width=0.72\textwidth]
                {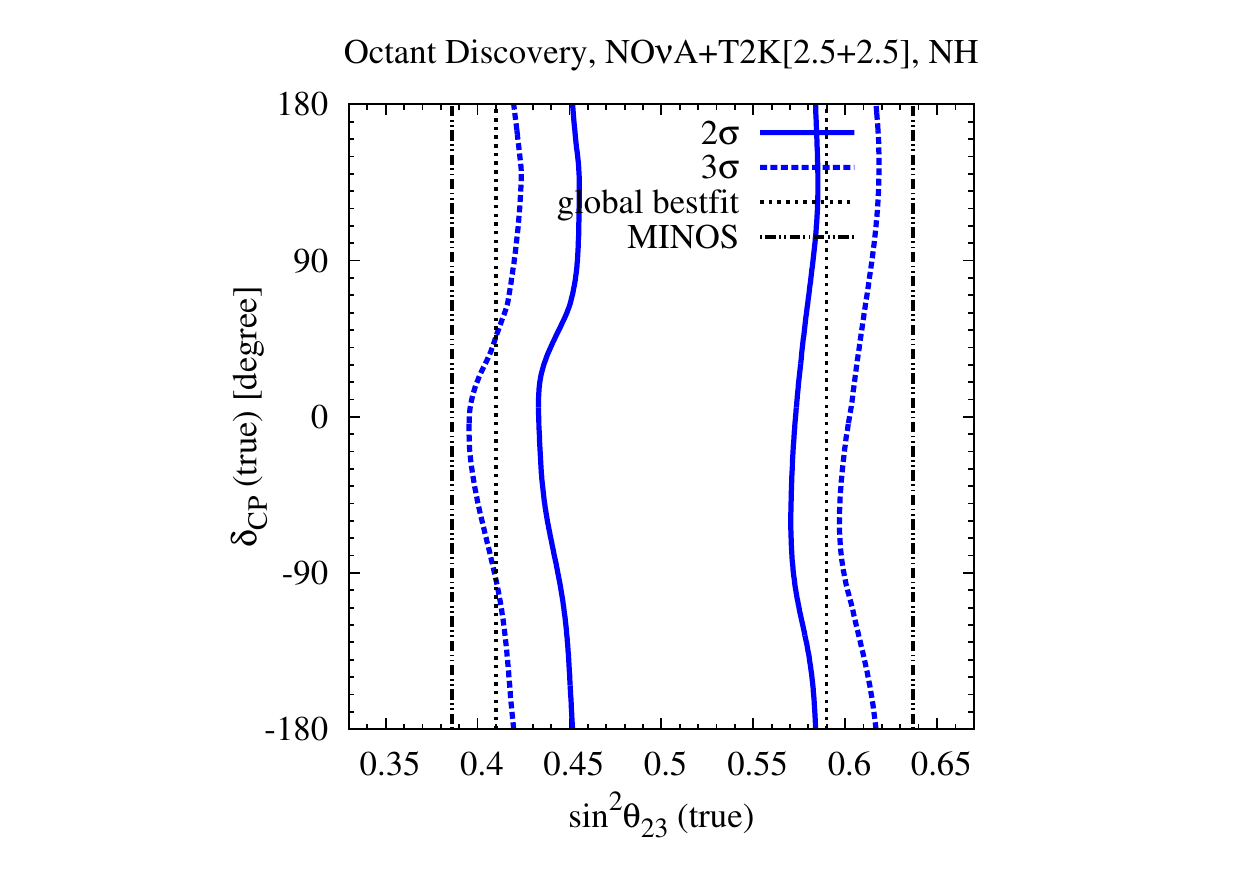}
                &
                \hspace*{-1.7in} \includegraphics[width=0.72\textwidth]
                {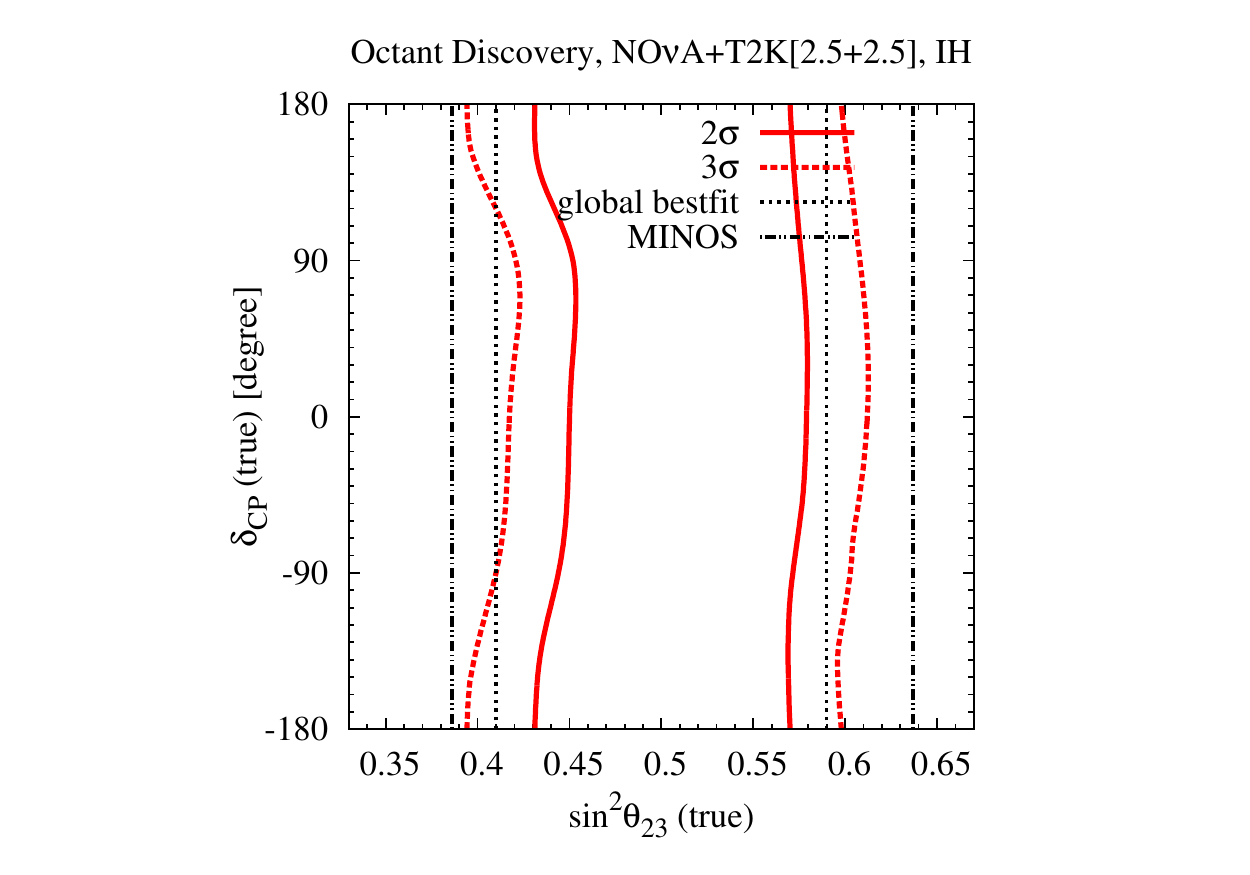}
        \end{tabular}

\caption{\footnotesize{Octant resolving capability in the true $\sin^2\tmt$ - true $\dcp$ plane for the combined $3\nu+3\bar\nu$ runs of
NO$\nu$A and $2.5\nu+2.5\bar\nu$ runs of T2K. Both $2\sigma$ and $3\sigma$ C.L. contours are plotted. The vertical lines correspond 
to the best-fit values of global data and those of MINOS accelerator data. The left (right) panel corresponds to NH (IH) being the true hierarchy.
This Figure has been taken from reference~\cite{Agarwalla:2013ju}.}}

\label{fig:true-theta23-true-dcp}
\end{figure}

In the discussion so far, we have assumed the true values of $\sin^2\tmt$ to be 0.41 for LO and 0.59 for HO. These are, of course, the best-fit points 
from the global analyses. But, we must consider the octant resolution capability for values of true $\sin^2\tmt$ in the full allowed range (0.34 to 0.67). 
In Figure~\ref{fig:true-theta23-true-dcp}, we plot the $2\sigma$ and $3\sigma$ octant resolution contours in true $\sin^2\tmt$ - true $\dcp$ plane. 
Octant resolution is possible only for points lying outside the contours. These Figures clearly show that octant resolution is possible at $2\sigma$ for 
global best-fit points and at $3\sigma$ for MINOS best-fit points. The results for the two hierarchies are quite similar. 
From Figure~\ref{fig:true-theta23-true-dcp}, we can see that if T2K experiment would have equal neutrino and anti-neutrino runs of 2.5 
years each, a $2\sigma$ resolution of the octant becomes possible provided $\sin^2\tmt \leq 0.43~\textrm{or}~ \geq0.58$ for any value of $\dcp$.
In reference~\cite{Chatterjee:2013qus}, the possibility of determining the octant of $\tmt$ in the long-baseline experiments T2K and NO$\nu$A 
in conjunction with future atmospheric neutrino detectors has been studied. The combined data from T2K and NO$\nu$A can provide a 
$\sim$ 2\% precision on $\sa$ at $2\sigma$ using the information coming from the disappearance channel~\cite{Agarwalla:2013ju}. 

In this section, we have discussed in detail the physics reach of current generation long-baseline beam experiments: T2K and NO$\nu$A
to unravel the neutrino mass hierarchy, CP violation and octant of $\tmt$. Given their relatively short baselines, narrow band beams and 
limited statistics, these experiments suffer a lot from the hierarchy-$\dcp$ and octant-$\dcp$ degeneracies. They can provide a hint for
these unknown issues only for favorable ranges of parameters at limited confidence level.
Hence, new long-baseline experiments with intense neutrino beam sources and advanced detector technologies are 
mandatory~\cite{Bandyopadhyay:2007kx,Choubey:2011zzq,Agarwalla:2011hh,Coloma:2012wq} to fathom the hitherto uncharted parameter 
space of the neutrino mixing matrix well beyond the capabilities of T2K and NO$\nu$A.

\section{Next Generation Long-baseline Beam Experiments}
\label{sec:future}

In this section, we briefly review the possible options for future high precision long-baseline beam experiments with a special emphasis 
on high power superbeam facilities using liquid argon and water Cherenkov detectors. As discussed in the literature 
(see e.g., references~\cite{Bandyopadhyay:2007kx,Mezzetto:2010zi}), the ability of future long-baseline neutrino experiments
to discover mass hierarchy, octant of $\tmt$, and CP violation depends on the achievable event statistics and hence strongly on the value 
of $\tet$. The fact that $\tet$ is large will have a significant impact on the realization of future long-baseline neutrino oscillation
experiments, for which planning till 2011 was focused on a staged approach to achieve sensitivity to increasingly smaller 
values of $\tet$. This approach was exemplified in the optimization of the neutrino factory~\cite{Huber:2006wb,Agarwalla:2010hk} and 
beta-beam~\cite{Agarwalla:2006vf,Agarwalla:2008gf} experiments for which it was possible to discover a value of $\stch$ as small as $10^{-4}$.
However, following the recent discovery of a moderately large value of $\tet$, the focus of future optimizations will be on the possibility to explore 
mass hierarchy, octant of $\tmt$, and CP violation for a {\it given} value of $\tet$. A relatively large value of $\tet$ also allows us to pursue an 
incremental program, staged in terms of the size of the experiment~\cite{Agarwalla:2011hh}, producing significant new results at each stage.
Out of the three major unknowns, the discovery of CP violation is the most toughest goal to achieve. Hence, the determination of mass hierarchy 
and octant should be considered as the first step towards the discovery of leptonic CP violation. Future facilities must be developed with the 
requirements that they should have the capability to determine hierarchy and octant at $3\sigma$ confidence level or better for any possible 
value of $\dcp$ during the first stage and discover CP violation and measure $\dcp$ during the second stage.
An incremental approach is also justified in view of the challenges (some of them are unknown) involved in operating very high power 
superbeams and in building giant underground neutrino detectors, which makes such an approach effectively safer and possibly more 
cost-effective. Both the proposed LBNE and LBNO facilities have adopted this staged approach in light of large $\tet$.
First we present a comparative study of the physics reach that can be achieved during the first phase of these two proposed experiments.
Then, we talk about the proposals of J-PARC to Hyper-Kamiokande (T2HK) long-baseline superbeam experiment~\cite{Abe:2011ts}
and CERN to MEMPHYS (at Fr{\'e}jus) SPL superbeam experiment~\cite{Campagne:2006yx,Agostino:2012fd,Baussan:2012wf,Edgecock:2013lga}.
We also mention about the possibility to explore leptonic CP violation based on the European Spallation Source (ESS) proton linac which can deliver
very intense, cost effective and high performance neutrino beam in parallel with the production of spallation 
neutron~\cite{Baussan:2012cw,Baussan:2013zcy}. Megaton-size water Cherenkov detector is one of the key components of these three experimental setups.
Next, we discuss the physics prospects of Low-Energy Neutrino Factory (LENF)~\cite{Geer:2007kn,Bross:2007ts,FernandezMartinez:2010zza,Ballett:2012rz} 
in conjunction with magnetized iron detector (MIND)~\cite{Cervera:2010rz,Agarwalla:2010hk,Bayes:2012ex} which seems to be a very 
promising setup to explore leptonic CP violation for large $\tet$. Finally, we mention about few proposals based on mono-flavor 
beta-beam concept~\cite{Zucchelli:2002sa,Mezzetto:2003ub,Agarwalla:2006vf,Agarwalla:2008gf,Benedikt:2011za}.  

\subsection{Discovery Reach of LBNE and LBNO}
\label{subsec:lbne-lbno}

\underline{\bf {LBNE:}}
The Long-Baseline Neutrino Experiment (LBNE)~\cite{Akiri:2011dv} is one of the major components of Fermilab's intensity frontier program.
In its first phase (LBNE10), it will have a new, high intensity, on-axis neutrino beam directed towards a $10\,\mathrm{kilotons}$ LArTPC
located at Homestake with a baseline of $1300\,\mathrm{km}$. This facility is designed for initial operation at a proton beam power of 
$708\,\mathrm{kW}$, with proton energy of $120\,\mathrm{GeV}$ that will deliver $6 \times 10^{20}$ protons on target in 230 days per 
calendar year. In our simulation, we have used the latest fluxes being considered by the collaboration, which have been estimated 
assuming the smaller decay pipe and the lower horn current compared to the previous studies~\cite{mbishai}.
We have assumed five years of neutrino run and five years of anti-neutrino run. The detector characteristics have been taken 
from Table 1 of ~\cite{Agarwalla:2011hh}. To have the LArTPC cross-sections, we have scaled the inclusive charged current
cross sections of water by 1.06 (0.94) for the $\nu$ ($\bar{\nu}$) case~\cite{zeller,petti-zeller}.

\underline{\bf {LBNO:}}
The long baseline neutrino oscillation experiment (LBNO)~\cite{Stahl:2012exa} plans to use an experimental setup where neutrinos produced 
in a conventional wide-band beam facility at CERN would be observed in a proposed $20\,\mathrm{kilotons}$ (in its first phase) LArTPC 
housed at the Pyh\"asalmi mine in Finland, at a distance of $2290\,\mathrm{km}$. The fluxes have been computed~\cite{poster} assuming an 
exposure of $1.5 \times 10^{20}$ protons on target in 200 days per calendar year from the SPS accelerator at $400\,\mathrm{GeV}$ with a 
beam power of $750\,\mathrm{kW}$. For LBNO also, we consider five years of neutrino run and five years of anti-neutrino run. 
We assume the same detector properties as that of LBNE. 

\begin{figure}[tp]
\centering
\includegraphics[width=0.49\textwidth]{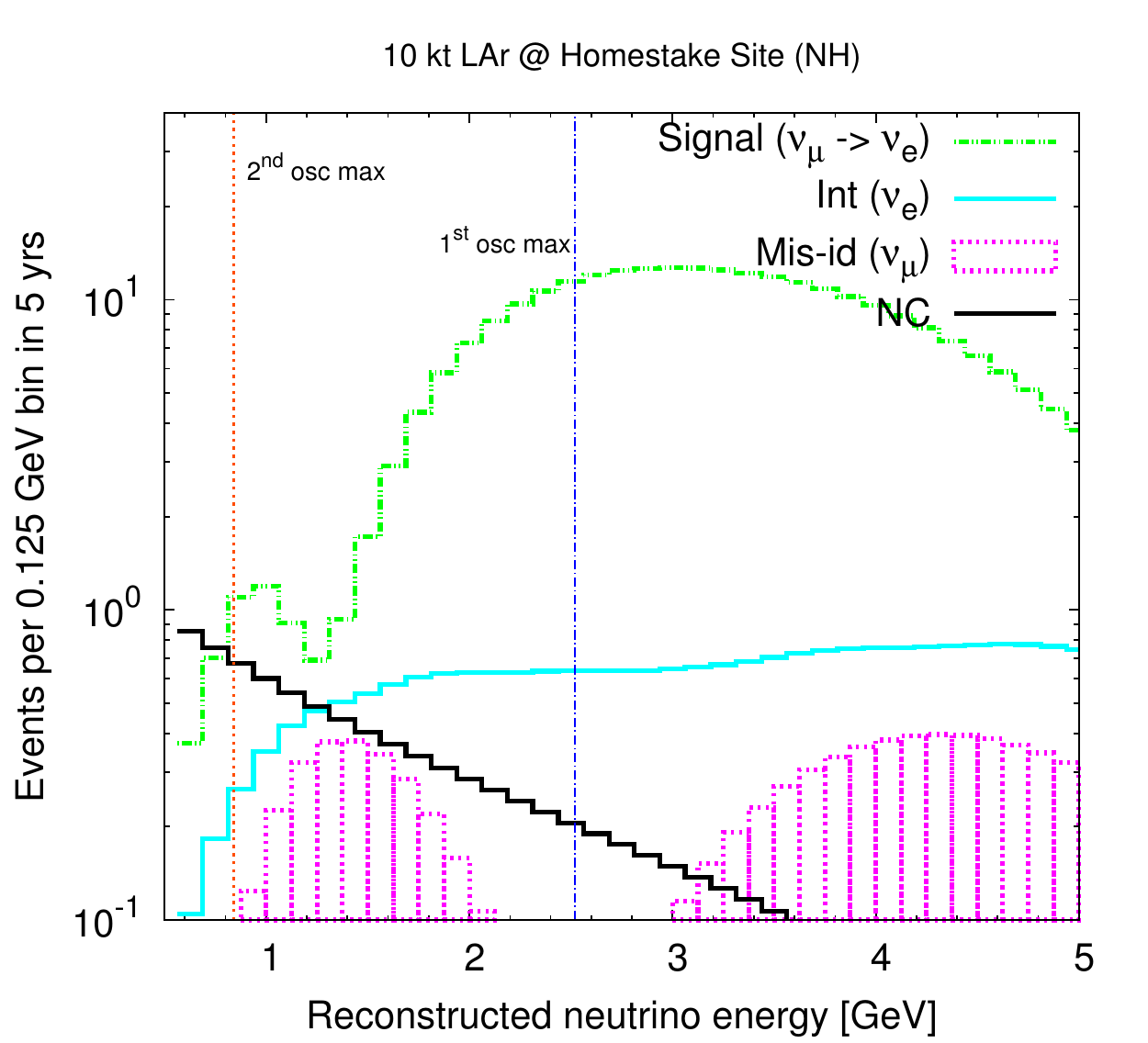}
\includegraphics[width=0.49\textwidth]{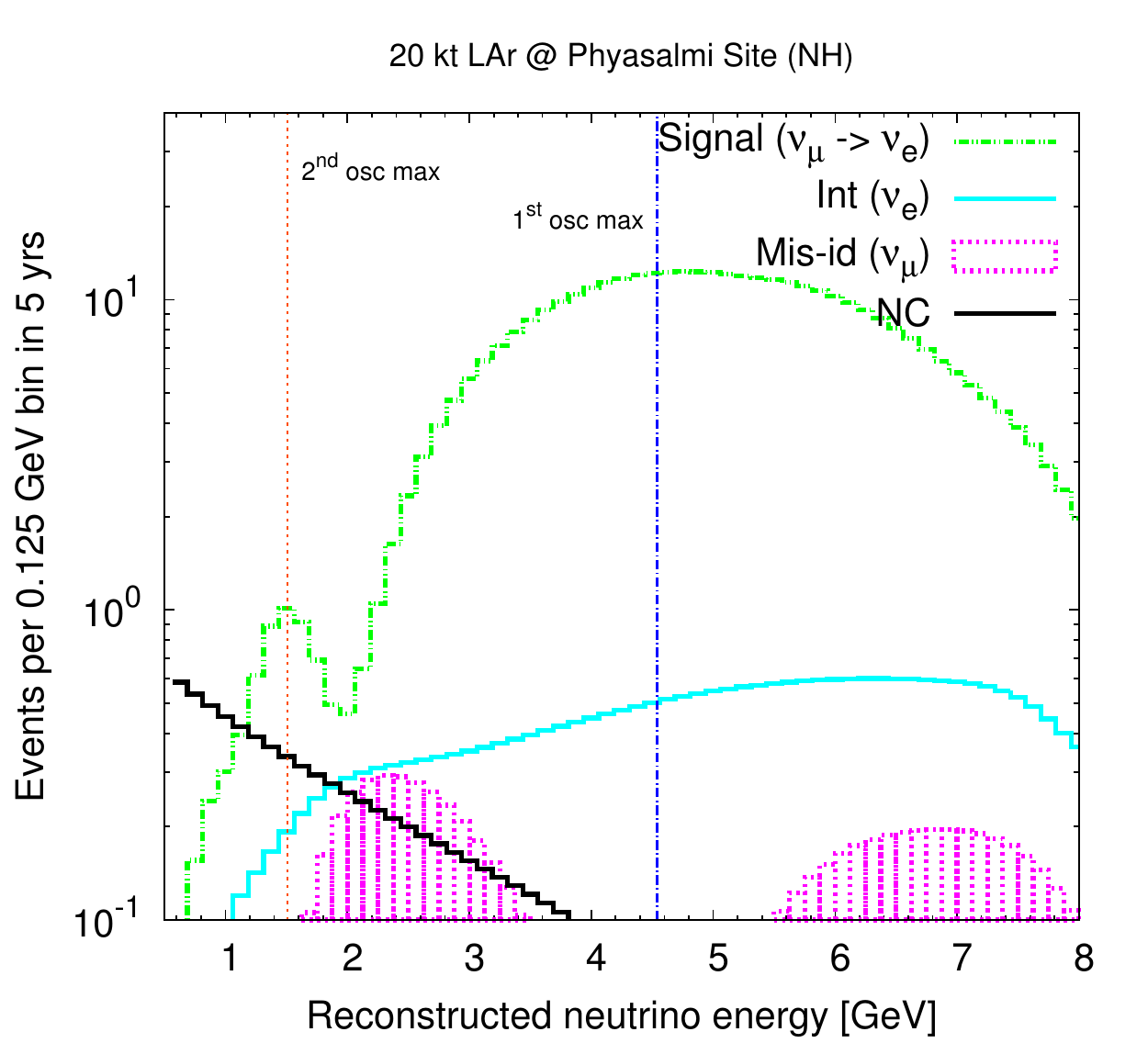}
\caption{\footnotesize{Expected signal and background event rates in the $\nue$ appearance channel as a function of the reconstructed 
neutrino energy including the efficiency and background rejection capabilities. Here, we consider $\stch$ = 0.0975 and $\dcp$ = 0$^{\circ}$.
Left panel (right panel) is for LBNE10 (LBNO). A normal hierarchy has been assumed. In both the panels, the blue dot-dashed and the orange 
dotted vertical lines display the locations of the first and second oscillation maxima.}}
\label{fig:event-spectrum-LBNE10-LBNO}
\end{figure}

\underline{\bf {Event Spectrum at LBNE10 and LBNO:}}
Figure~\ref{fig:event-spectrum-LBNE10-LBNO} portrays the expected signal and background event spectra in the $\nue$ appearance channel 
as a function of reconstructed neutrino energy including the efficiency and background rejection capabilities for LBNE10 (left panel) and 
LBNO (right panel) setups. In both the panels of Figure~\ref{fig:event-spectrum-LBNE10-LBNO}, one can clearly see a systematic downward bias 
in the reconstructed energy for neutral current background events due to the final state neutrino included using the migration matrices. 
The blue dot-dashed and the orange dotted vertical lines display the locations of the first and second oscillation maxima. 
The green double-dotted-dashed histogram shows the signal event rate. Although, we have some statistics around the second oscillation 
maximum for both the baselines, its impact is limited due to the fact that the event samples are highly contaminated with neutral current and 
other backgrounds at lower energies.

\begin{figure}[tp]
\centering
\includegraphics[width=12cm, height=8.0cm]{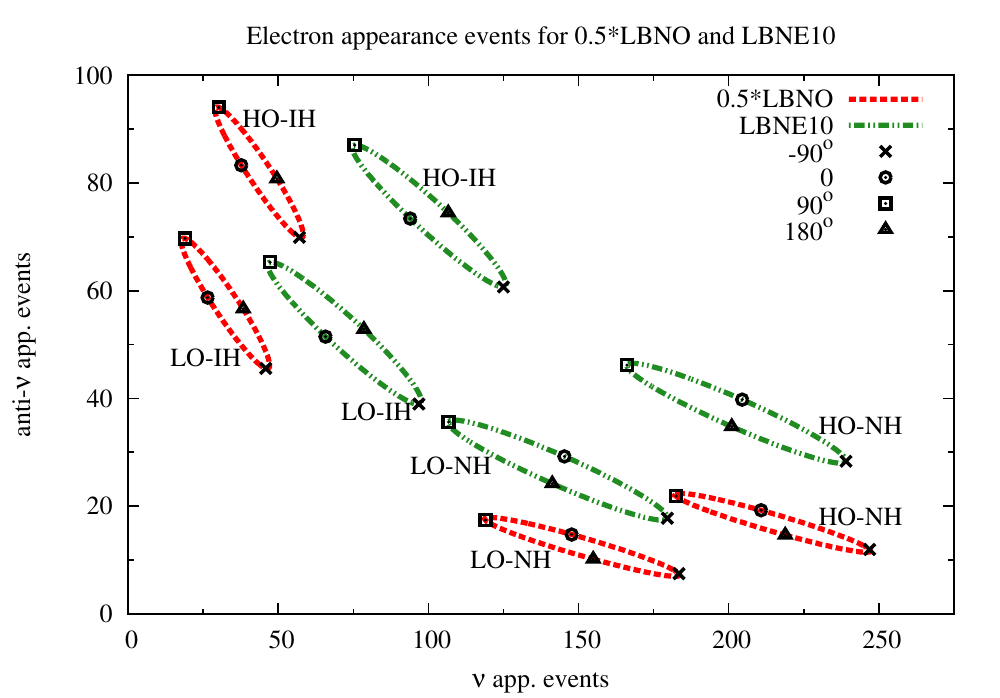}
\caption{\footnotesize{Bi-events ($\nu_e$ and $\bar\nu_e$ appearance) plots for the four octant-hierarchy combinations and all possible
$\dcp$ values. The experiments considered are LBNE10 and 0.5*LBNO. Here $\sin^22\tet = 0.089$. For LO (HO), $\sin^2\tmt = 0.41(0.59)$.
This Figure has been taken from reference~\cite{Agarwalla:2013hma}.}}
\label{bi-events}
\end{figure}

\underline{\bf{Physics with bi-events plots:}}
In Figure~\ref{bi-events}, we have plotted $\nu_e$ vs. $\bar{\nu}_e$ appearance events, for LBNE10 and 0.5*LBNO for the four possible combinations 
of hierarchy and octant. Since $\dcp$ is unknown, events are generated for $[-180^\circ, 180^\circ]$, leading to the ellipses. Here, we take $\stch = 0.089$.
For the lower octant (LO) of $\theta_{23}$, the value $\sin^2 \theta_{23} = 0.41$ is chosen and for higher octant (HO), it is taken to be $0.59$. 
Note that, in Figure~\ref{bi-events} we have plotted the total number of events,whereas the actual analysis will be done based on the
spectral information. Nevertheless, the contours in Figure~\ref{bi-events} contain very important information regarding the physics capabilities of the experiments. An experiment can determine both the hierarchy and the octant, if every point on a given ellipse is well separated from every point
on each of the other three ellipses. The larger the separation, the better is the confidence with which the above parameters can be determined.

For 0.5*LBNO, the two (LO/HO)-IH ellipses are well separated from the two (LO/HO)-NH ellipses, in their number of neutrino events.
Hence, 0.5*LBNO has excellent hierarchy determination capability with just neutrino data. However, only neutrino data alone will not be sufficient
to determine the octant in case of IH, because various points on (LO/HO)-IH ellipses have the same number of neutrino events. Likewise, only
anti-neutrino data cannot determine the octant in case of NH. Therefore, balanced neutrino and anti-neutrino data is mandatory to make an effective 
distinction between (LO/HO)-IH ellipses and also between (LO/HO)-NH ellipses.

For LBNE10, $\nu$ data alone can not determine hierarchy because various points on LO-NH and HO-IH ellipses
have the same number of $\nu_e$ events. Thus, $\bar\nu$ data is also needed. Even with $\bar\nu$ data, 
hierarchy determination can be difficult to achieve, if nature chooses LO and one of the two worst case combinations of hierarchy
and $\dcp$ which are (NH, $90^\circ$) or (IH, $-90^\circ$). In such a situation, the $\nu_e$ and $\bar\nu_e$ events are rather close to each other
and it will be very difficult for LBNE10 to reject the wrong combination. Regarding octant determination, the capability of LBNE10 is very similar to that of 0.5*LBNO because the separations between the ellipses, belonging to LO and HO are very similar for the two experiments.

\underline{\bf{Hierarchy and Octant discovery with LBNE10 and LBNO:}}
Measurement of hierarchy and octant should be considered as a prerequisite for the discovery of leptonic CP violation.
To present the results for mass hierarchy and octant discovery, we consider three experimental set-ups: LBNE10, LBNO and a possible 
LBNO configuration with a detector half the mass (10 kilotons), which we denote as 0.5*LBNO. Scaling down the detector size of LBNO 
by half makes the exposures of these two experiments very similar. Therefore, considering 0.5*LBNO enables us to make a direct comparison 
between the inherent properties of the two baselines involved. Both LBNE and LBNO will operate at multi-GeV energies with very long-baselines. 
This will lead to large enough matter effect to break the hierarchy-$\dcp$ degeneracy completely. They are also scheduled to have equal neutrino 
and anti-neutrino runs, tackling the octant-$\dcp$ degeneracy. These experiments are planning to use LArTPCs ~\cite{Amerio:2004ze,Rubbia:2009md} 
which have excellent kinematic reconstruction capability for all the observed particles. This feature helps in rejecting quite a large fraction of 
neutral current background.

We study the hierarchy discovery potential for two true values of $\sin^2\tmt$: 0.41 (LO) and 0.5 (MM). If HO is true, the results will be better than 
those for the case of MM. This gives us four true combinations of $\tmt$-hierarchy: LO-NH, LO-IH, MM-NH and MM-IH. $\Delta\chi^2$ is calculated for
each of these four combinations, assuming the opposite hierarchy to be the test hierarchy. In the fit, we marginalize over test $\sin^2\theta_{23}$
in its $3\sigma$ range, $\ma$ and $\sin^22\theta_{13}$ in their $2\sigma$ ranges. We considered 5\% uncertainty in the matter density, $\rho$.
Priors were added for $\rho$ ($\sigma=5\%$), $\ma$ ($\sigma=4\%$) and $\sin^22\tet$ ($\sigma=5\%$, as expected by the end of Daya Bay's run).
$\Delta\chi^2$ is also marginalized over the uncorrelated systematic uncertainties (5\% on signal and 5\% on background) in the setups, so as to obtain a
$\Delta\chi^2_{{\footnotesize \textrm{min}}}$ for every $\dcp$(true).

\begin{figure}[tp]
\centering
\includegraphics[width=12cm, height=8.0cm]{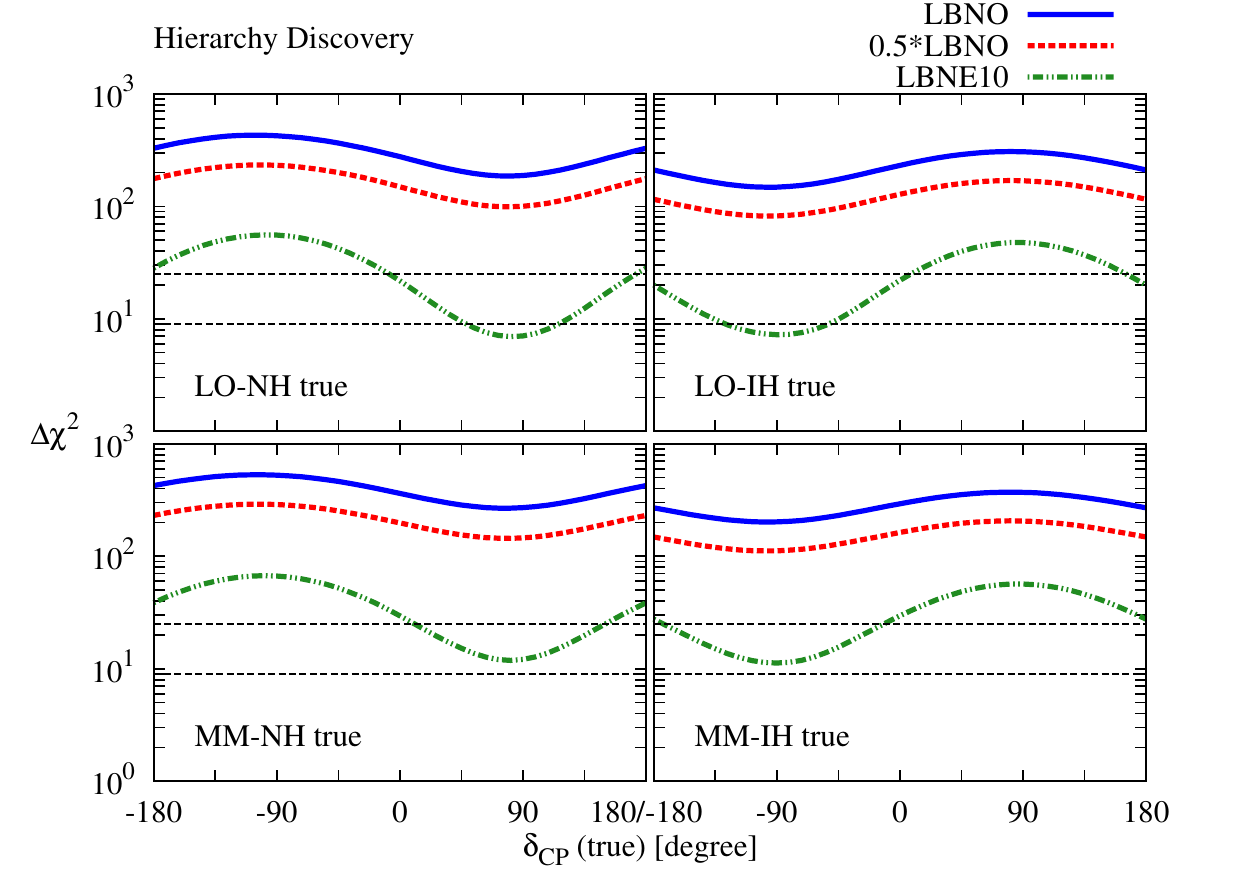}
\caption{\footnotesize{Hierarchy discovery reach for LBNO, 0.5*LBNO and LBNE10. Results are shown for the four possible true 
$\tmt$-hierarchy combinations.}}
\label{fig:hierarchy-LBNE10-LBNO}
\end{figure}

Figure~\ref{fig:hierarchy-LBNE10-LBNO} shows the discovery reach for hierarchy as a function of $\dcp$(true).
We see that even 0.5*LBNO has $\gtrsim 10 \sigma$ hierarchy discovery for all values of $\dcp$(true) and for all four
$\tmt$-hierarchy combinations. The potential of LBNO is even better. The LBNO baseline is close to bimagic which gives it a particular
advantage~\cite{Raut:2009jj,Dighe:2010js}. For LBNE10, a $5\sigma$ discovery of hierarchy is possible for only $\sim50\%$ of the $\dcp$(true),
irrespective of the true $\tmt$-hierarchy combination. For the unfavorable hierarchy-$\dcp$ combinations, {\it i.e.}
NH with $\dcp$ in the upper half plane or IH with $\dcp$ in the lower half plane, the performance of LBNE10 suffers. 
In particular, for LO and the worst case combinations ((NH, $90^\circ$) and (IH, $-90^\circ$)), {\it LBNE10 will not be
able to provide even a $3\sigma$ hierarchy discrimination}. Therefore, LBNE10 must increase their statistics, if NO$\nu$A data indicate that 
the unfavorable combinations are true. The discovery potential for all three set-ups will be better if $\tmt$ happens to lie in
HO. But, we checked that, even then, a $5\sigma$ discovery is not possible with LBNE10 for $\sim30\%$ of the upper half plane of $\dcp$ for 
HO-NH true and $\sim70\%$ of the lower half plane of $\dcp$ for HO-IH true.

We next consider the discovery reach of the same set-ups for excluding the wrong octant. We consider the true values of 
$\sin^2\tmt=0.41(\textrm{LO})$ and $\sin^2\tmt=0.59(\textrm{HO})$, so that we have the following four true combinations of octant and hierarchy:
LO-NH, LO-IH, HO-NH and HO-IH. $\Delta\chi^2$ is calculated for each of these four combinations, assuming test $\sin^2\tmt$ values from
the other octant. For LO (HO) true, we consider the test $\sin^2\tmt$ range from 0.5 to 0.67 (0.34 to 0.5). Rest of the marginalization procedure 
(over oscillation parameters as well as systematic uncertainties) is the same as that in the case of hierarchy exclusion
except with another difference: the final $\Delta\chi^2$ is marginalized over both the hierarchies as the test hierarchy, 
to obtain $\Delta\chi^2_{{\footnotesize \textrm{min}}}$.

\begin{figure}[tp]
\centering
\includegraphics[width=12cm, height=8.0cm]{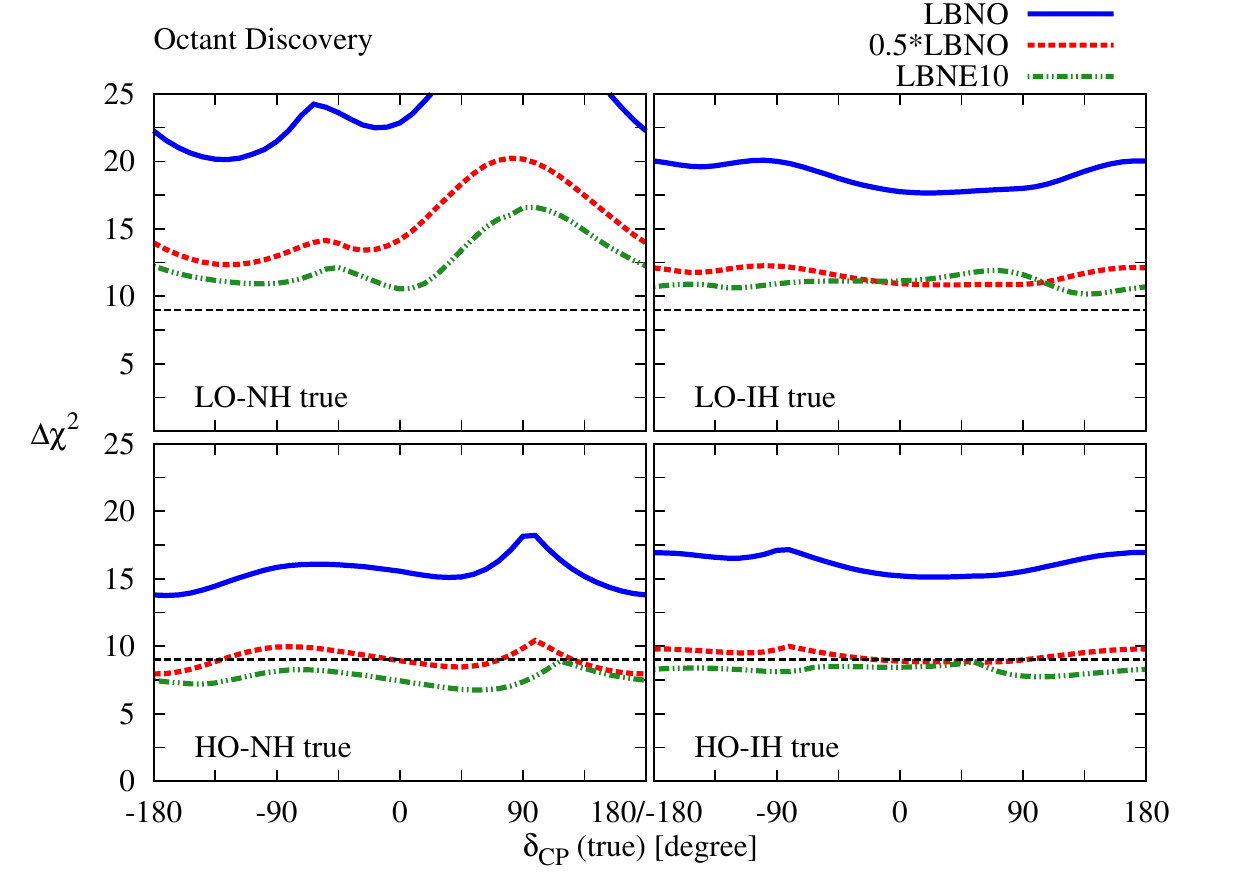}
\caption{\footnotesize{Octant resolving capability for LBNO, 0.5*LBNO and LBNE10. Results are shown for the four possible true 
octant-hierarchy combinations.}}
\label{fig:octant-LBNE10-LBNO}
\end{figure}

Figure~\ref{fig:octant-LBNE10-LBNO} shows the discovery reach for octant as a function of $\dcp$(true).
It can be seen that for (LO/HO)-IH true, the sensitivities of LBNE10 and 0.5*LBNO are quite similar whereas they are
somewhat better for 0.5*LBNO if (LO/HO)-NH are the true combinations. For LO-(NH/IH), both LBNE10 and 0.5*LBNO
have more than $3\sigma$ discovery of octant while for HO-(NH/IH), the $\Delta\chi^2_{{\footnotesize \textrm{min}}}$ 
varies from $\sim$ 6 to $\sim$ 11. However, with full LBNO, we have more than $3.5\sigma$ discovery of octant for all octant-hierarchy
combinations. A $5\sigma$ discovery of octant is possible only for LO-NH true for $\dcp(\textrm{true})\in(\sim20^\circ,\sim150^\circ)$.

\begin{figure}[tp]
\centering
\includegraphics[width=0.49\textwidth]{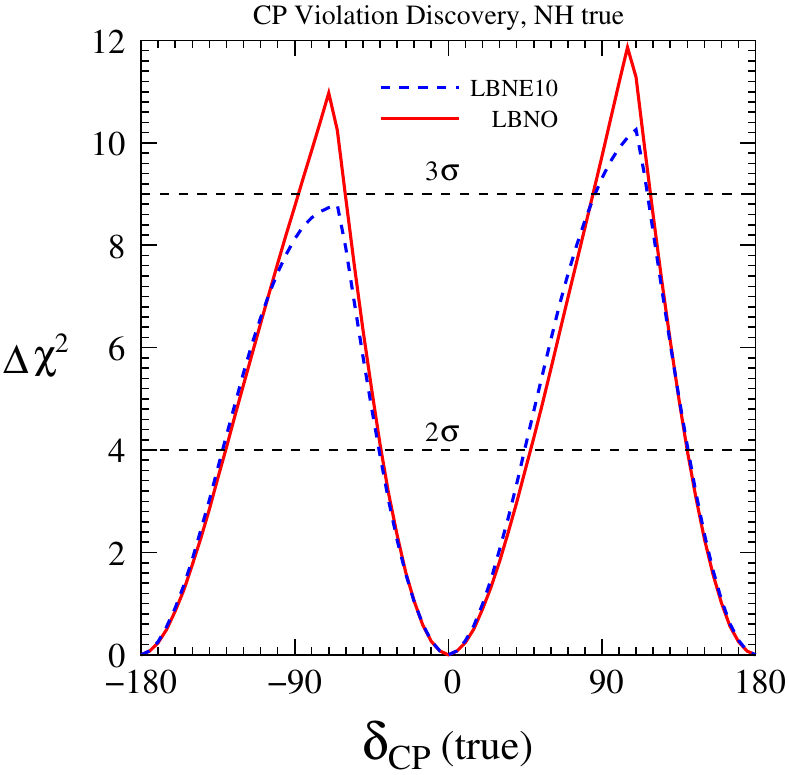}
\includegraphics[width=0.49\textwidth]{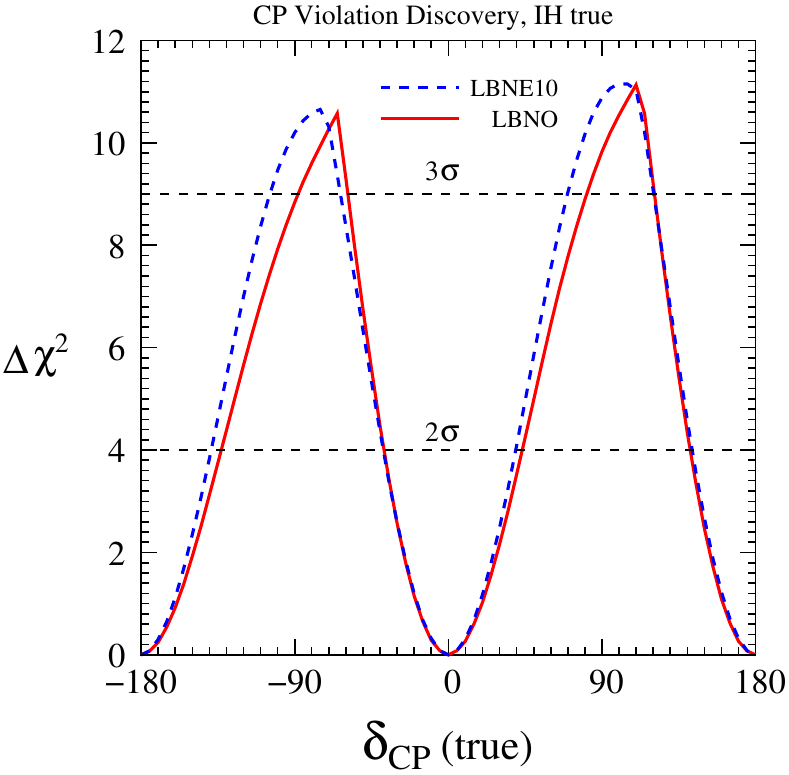}
\caption{\footnotesize{Left panel (right panel) shows the $\Delta\chi^2$ for the CP violation discovery as a function of true 
value of $\dcp$ assuming NH (IH) as true hierarchy.}}
\label{fig:cp-LBNE10-LBNO}
\end{figure}

\underline{\bf{CP violation discovery with LBNE10 and LBNO:}}
In their first phases, both LBNE10 and LBNO will have very minimal CP violation reach. Figure~\ref{fig:cp-LBNE10-LBNO} depicts the
CP violation discovery reach for LBNE10  and LBNO. In the left panel (right panel), we have considered NH (IH) as true hierarchy.
In Table~\ref{tab:CPV-compare-LBNE10-LBNO}, we mention the fraction of $\dcp$ values for which CP violation can be detected
for these two experimental setups. Both LBNE10 and LBNO have CP violation reach for around 50\% values of true $\dcp$ at $2\sigma$ 
confidence level. At $3\sigma$, their CP violation reach is quite minimal: only for 10 -- 20\% of the entire range.
  
\begin{table}[tp]
\begin{center}
\begin{tabular}{|c|c|c|} \hline\hline
\multirow{2}{*}{Setups} & \multicolumn{2}{c|}{{\rule[4mm]{0mm}{4mm}Fraction of $\dcp$(true)}}
\cr\cline{2-3}
& $2\,\sigma$ & $3\,\sigma$ \cr
\hline
LBNE10 (5 years $\nu$ + 5 years $\bar\nu$) & 0.52 (0.56) & 0.09 (0.26) \cr
\hline
LBNO     (5 years $\nu$ + 5 years $\bar\nu$)     & 0.51 (0.54) & 0.17 (0.19) \cr
\hline\hline
\end{tabular}
\caption{\footnotesize{Fractions of $\dcp$(true) for which a discovery is possible for CP violation. The numbers without (with) parentheses 
correspond to NH (IH) as true hierarchy. The results are presented at $2\sigma$ and $3\sigma$ confidence level.}}
\label{tab:CPV-compare-LBNE10-LBNO}
\end{center}
\end{table}

\subsection{T2HK, CERN-MEMPHYS, and ESS LINAC Proposals}
\label{subsec:t2hk-cern-memphys}

The T2HK proposal~\cite{Abe:2011ts} plans to use a 1.66 MW superbeam from the upgraded J-PARC proton synchrotron facility 
directed to a 1 megaton water Cherenkov detector (with fiducial mass of 560 kilotons) located at a distance of 295 $\mathrm{km}$ 
from the source at an off-axis angle of $2.5^\circ$. A proposed location for this detector is about 8 km south of Super-Kamiokande 
at an underground depth of 1,750 meters water equivalent. The main purpose of this experiment is to achieve an unprecedented 
discovery reach for CP violation. The water Cherenkov detector provides an excellent energy resolution for sub-GeV low multiplicity 
final state events. Both the high power narrow band beam and the megaton-size detector play an important role to have large numbers 
of $\nue$ and $\anue$ appearance events at the first oscillation maximum. This experiment plans to have 1.5 years of neutrino run 
and 3.5 years of anti-neutrino run with one year given by 10$^7$ seconds. The baseline of this experiment is too short to have any
matter effect. Therefore, its mass hierarchy discovery reach is very limited. The expected accuracy in the determination of CP phase
is better than $20^\circ$ at $1\sigma$ confidence level assuming that mass hierarchy is known. For $\stch$ = 0.1, the CP violation can 
be established at $3\sigma$ C.L. for 74\% of the $\dcp$ values provided we fix the hierarchy in the fit. This CP coverage reduces to 
55\% if we marginalize over both the choices of hierarchy in the fit~\cite{Abe:2011ts}. 

Like T2HK proposal, there is a superbeam
configuration is under consideration in Europe~\cite{Campagne:2006yx,Baussan:2012wf,Edgecock:2013lga} using the
CERN to Fr{\'e}jus baseline of 130 km. The aim of this proposal is to send a 4 MW high power superbeam from CERN towards a
440 kilotons MEMPHYS water Cherenkov detector~\cite{Agostino:2012fd} located 130 km away at Fr{\'e}jus.
Using this setup, CP violation can be established for roughly 60\% values of $\dcp$ at $3\sigma$ 
confidence level~\cite{Agostino:2012fd} assuming two years of neutrino run and eight years of anti-neutrino run, with 
uncorrelated systematic uncertainties of 5\% on signal and 10\% on background.

Another interesting possibility to explore leptonic CP violation using a very intense, cost effective, and high performance neutrino beam 
from the proton linac of the ESS facility currently under construction in Lund, Sweden has been studied recently in detail in 
references~\cite{Baussan:2012cw,Baussan:2013zcy}. A high power superbeam from this proton linac in conjunction with
a megaton-size water Cherenkov detector located in the existing mines at a distance of 300 to 600 km from the ESS facility
can discover leptonic CP violation at 5$\sigma$ confidence level for 50\% of the $\dcp$ values~\cite{Baussan:2013zcy}.

\subsection{Neutrino Factory}
\label{subsec:nufact}

The term ``neutrino factory''~\cite{Geer:1997iz,DeRujula:1998hd,Bandyopadhyay:2007kx,Choubey:2011zzq} has been associated to
describe neutrino beams created by the decays of high energy muons (obtained via pion decay) which are circulated in a storage ring 
with long straight sections. The decay of muons in these straight sections produces an intense, well known and pure beam of 
$\numu$ and $\anue$. If $\mu^+$ are stored, $\mu^+\rightarrow e^+\nue\anumu$ decays generate a beam consisting of equal numbers 
of $\nue$ and $\anumu$. The most promising avenue to explore CP violation, neutrino mass hierarchy, and octant of $\tmt$ at a
neutrino factory is the sub-dominant $\nue \to \numu$ oscillation channel which produces muons of the opposite charge (wrong-sign muons) 
to those stored in the storage ring and these can be detected with the help of the charge identification capability of a magnetized iron 
neutrino detector (MIND)~\cite{Cervera:2000kp,Cervera:2010rz}. In the most recent analysis~\cite{Cervera:2010rz} of MIND, low energy neutrino 
signal events down to $1\,\mathrm{GeV}$ were selected with an efficiency plateau of $\sim$~60\% for $\numu$ and $\sim$~80\% for $\anumu$ 
events starting at $\sim$~$5\,\mathrm{GeV}$, while maintaining the background level at or below 10$^{-4}$. For the neutral current background, 
the impact of migration is non-negligible and it is peaked at lower energies. This feed-down is the strongest effect of migration and thus has 
potential impact on the energy optimization, since it penalizes neutrino flux at high energies, where there is little oscillation but a large 
increase in fed-down background. 

\begin{figure}[tp]
\centering
\includegraphics[width=13.0cm, height=10.0cm]{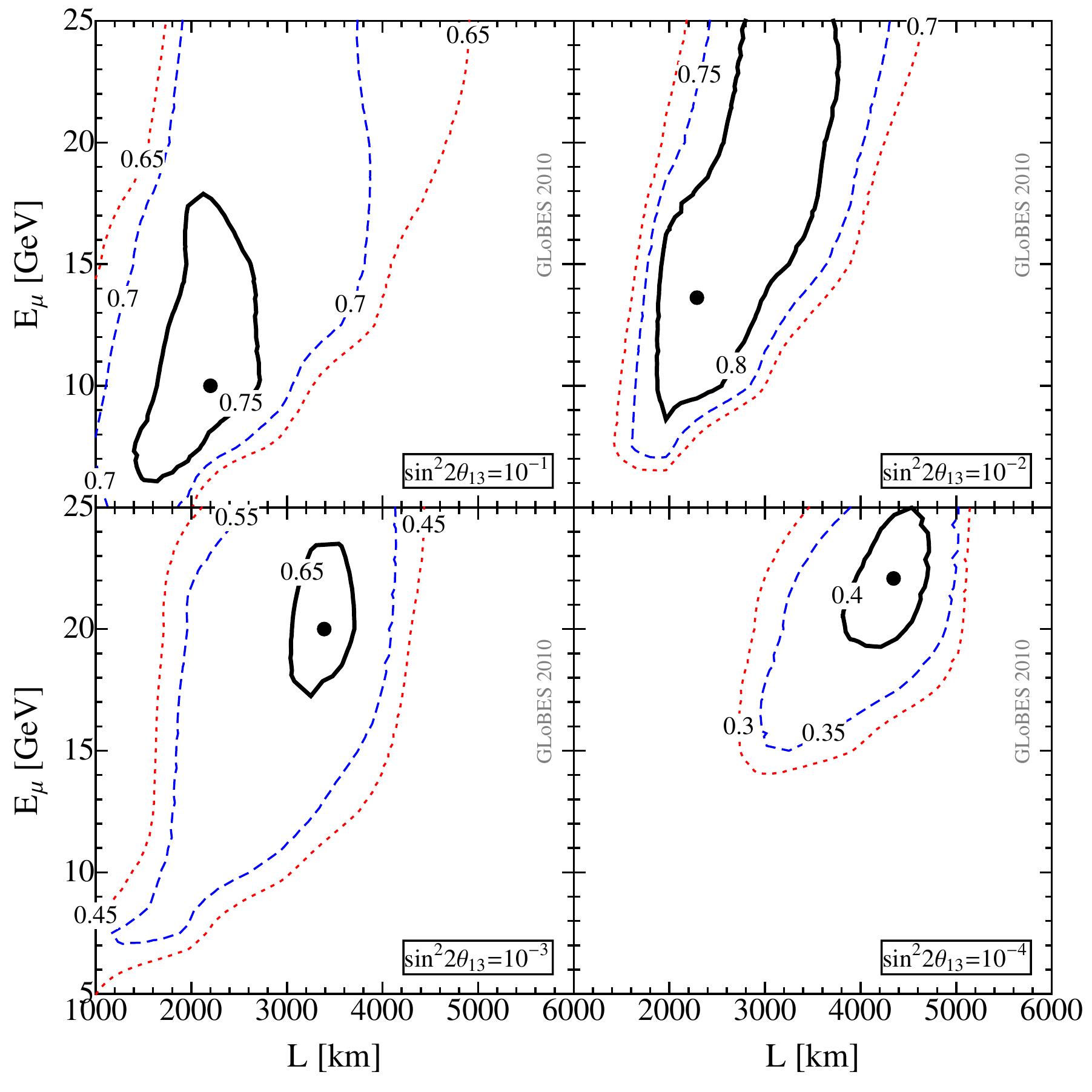}
\caption{\footnotesize{Fraction of $\dcp$(true) for which CP violation will be discovered at $3\sigma$ C.L. as a function of $L$ and 
$E_\mu$ for the single baseline Neutrino Factory. The different panels correspond to different true values of $\stch$, as given there.
Here we consider $5 \times 10^{21}$ useful muon decays in total with a 50 kilotons MIND detector. The optimal performance
is marked by a dot: (2200,10.00), (2288,13.62), (3390,20.00) and (4345,22.08) with regard to their best reaches of the fraction
of $\dcp$(true) at: 0.77, 0.84, 0.67 and 0.42. This figure has been taken from~\cite{Agarwalla:2010hk}.}}
\label{fig:LvsE}
\end{figure}

In light of recently discovered moderately large value of $\tet$, shorter baselines and lower energies are preferred to achieve high 
performance in exploring CP violation. Even $E_\mu$ as low as 5 to $8\,\mathrm{GeV}$ at the Fermilab-Homestake baseline of about 
$1300\,\mathrm{km}$ is quite close to the optimal choice (see upper left panel of Figure~\ref{fig:LvsE}), which means that the MIND 
detector approaches the magnetized totally active scintillator detector performance of the low energy Neutrino 
Factory (LENF)~\cite{Geer:2007kn}. With the present baseline choice of the neutrino factory ($E_\mu$ = 10 GeV and $L$ = 2000 km) and 
a 50 kilotons MIND detector, CP violation can be established for around 80\% values of $\dcp$ at $3\sigma$ C.L. considering
$5 \times 10^{21}$ useful muon decays in total.

\subsection{Beta-beams}
\label{subsec:beta-beam}

Zucchelli~\cite{Zucchelli:2002sa} put forward the novel idea of a 
beta-beam~\cite{Mezzetto:2003ub,Volpe:2006in,Agarwalla:2009xc,Benedikt:2011za}, which is based on the
concept of creating a pure, well understood, intense, collimated beam of $\nu_{e}$ or $\bar\nu_{e}$ through the beta-decay
of completely ionized radioactive ions. Firstly, radioactive nuclides are created by impinging a target by accelerated protons. 
These unstable nuclides are collected, fully ionized, bunched, accelerated and then stored in a decay ring.
The decay of these highly boosted ions in the straight sections of the decay ring produces the so-called beta-beam.
It has been proposed to produce $\nue$ beams through the decay of
highly accelerated $^{18}$Ne ions ({$^{18} _{10}$Ne}  $\rightarrow$  $^{18} _{9}$F  $+$  $e^+ +  {\nue}$)
and ${\anue}$ from $^6$He ({$^6 _2$He}  $\rightarrow$  $^6 _3$Li  $+$  $e^- +  {\anue}$)~\cite{Zucchelli:2002sa}. 
More recently, $^{8}$B ({$^{8} _{5}$B}  $\rightarrow$  $^{8} _{4}$Be  $+$  $e^+ +  {\nue}$)
and $^{8}$Li ({$^8 _3$Li}  $\rightarrow$  $^8 _4$Be  $+$  $e^- +  {\anue}$)~\cite{Rubbia:2006pi,Rubbia:2006zv,Mori:2005zz}
with much larger end-point energy have been suggested as alternate sources since these ions can yield higher energy $\nue$ and
$\anue$ respectively, with lower values of the 
Lorentz boost $\gamma$~\cite{Agarwalla:2006vf,Agarwalla:2007ai,Agarwalla:2008ti,Coloma:2007nn,Donini:2006dx}.
Details of the four beta-beam candidate ions can be found in Table~\ref{tab:ions}.
It may be possible to store radioactive ions producing beams with both polarities in the same ring. This will enable running
the experiment in the $\nue$ and $\anue$ modes simultaneously. In the low $\gamma$ design of beta-beams, the standard luminosity
taken for the $^{18}$Ne and $^6$He are $1.1 \times 10^{18}$ ($\nu_e$) and $2.9\times 10^{18}$ ($\bar{\nu}_e$) 
useful decays per year, respectively.

\begin{table}[tp]
\begin{center}
\begin{tabular}{||c||c||c||c||c||c||c||} \hline \hline
   Ion & $\tau$ (s) &
$E_0$ (MeV)
   & $f$& Decay fraction & Beam \\
\hline
  $^{18} _{10}$Ne &   2.41 & 3.92&820.37&92.1\%& $\nu_{e}$    \\
  $^6 _2$He   &   1.17 & 4.02&934.53&100\% &$\bar\nu_{e}$    \\
\hline
 $^{8} _5$B& 1.11 & 14.43&600872.07&100\%&$\nu_{e}$    \\
 $^8 _3$Li& 1.20 &13.47 &425355.16& 100\% & $\bar\nu_{e}$    \\
\hline \hline
\end{tabular}
\caption{\footnotesize{Beta decay parameters: lifetime $\tau$,
electron total end-point energy $E_0$, $f$-value and decay fraction for various ions. This table has been taken from~\cite{Agarwalla:2009xc}.}}
\label{tab:ions}
\end{center}
\end{table}

Within the EURISOL Design Study~\cite{Benedikt:2011za}, the $\gamma$ = 100 option with $^6$He and $^{18}$Ne ions
has been studied quite extensively. The energy spectrum of the emitted neutrinos from these radioactive ions with $\gamma$ = 100
suits well the CERN to Fr{\'e}jus baseline of 130 km. Compared to superbeam, the main advantage of using beta-beam is that it is an
extremely pure beam with no beam contamination at the source. Combining beta-beam with superbeam, we can study the T-conjugated
oscillation channels~\cite{Donini:2004hu,Donini:2004iv} and this combined setup can provide an excellent reach for 
CP violation discovery. 

\section{Summary and Conclusions}
\label{sec:conclusion}

The discovery of neutrino mixing and oscillations provides strong evidence that neutrinos are massive and leptons flavors are mixed
with each other which leads to physics beyond the Standard Model of particle physics. With the recent determination of $\tet$, for the first 
time, a clear and comprehensive picture of the three flavor leptonic mixing matrix has been established. This impressive discovery has crucial 
consequences for future theoretical and experimental efforts. It has opened up exciting prospects for current and future long-baseline 
neutrino oscillation experiments towards addressing the remaining fundamental questions, in particular the type of the neutrino 
mass hierarchy, the possible presence of a CP-violating phase in the neutrino sector, and the correct octant of $\tmt$ 
(if it turns out to be non-maximal establishing the recent claims). In this paper, we have made an attempt to review the phenomenology 
of long-baseline neutrino oscillations with a special emphasis on sub-leading three-flavor effects, which will play a crucial role in resolving 
these unknowns in light of recent measurement of a moderately large value of $\tet$. We have discussed in detail the physics reach of
current generation long-baseline experiments: T2K and NO$\nu$A which have very limited reach in addressing these unknowns for only
favorable ranges of parameters. Hence, future facilities are indispensable to cover the entire parameter space at unprecedented confidence
level. A number of high-precision long-baseline neutrino oscillation experiments have been planned/proposed to sharpen our understanding 
about these tiny particles. In this review, we have discussed in detail the physics capabilities of few of such proposals based on superbeams,
neutrino factory, and beta-beam.

\subsubsection*{Acknowledgments}
SKA would like to thank all his collaborators with whom he has worked on long-baseline neutrino oscillation physics.
SKA acknowledges the support from DST/INSPIRE Research Grant [IFA-PH-12], Department of Science and Technology, India.

\bibliographystyle{apsrev}
\bibliography{references}

\end{document}